\documentclass[%
    fontsize=10pt,
]{article}
\usepackage[T1]{fontenc}
\usepackage[latin9]{inputenc}
\usepackage[a4paper]{geometry}
\usepackage{array}
\usepackage{textcomp}
\usepackage{multirow}
\usepackage{amsmath}
\usepackage{amssymb}
\usepackage{graphicx}
\usepackage{hyperref}
\usepackage{gensymb}
\usepackage[numbers]{natbib}
\usepackage{fancyhdr}
\usepackage{authblk}
\usepackage[normalem]{ulem}
\usepackage{colortbl}
\usepackage{graphicx}
\usepackage[english]{babel}

\setcitestyle{square}

\providecommand{\tabularnewline}{\\}
\begin{document}

\title{Meshless Methods for Large Deformation Elastodynamics}

\author{M. R{\"o}thlin \thanks{Corresponding Author: mroethli@ethz.ch}}
\author{H. Klippel}
\author{K. Wegener}
\affil{Institute of Machine Tools and Manufacturing (IWF), Department of Mechanical and Process Engineering, ETH Zurich}



\maketitle

\section{Abstract}

Meshless methods are a promising candidate to reliably simulate materials 
undergoing large deformations. Unlike mesh based methods like the FEM, meshless 
methods are not limited in the amount of deformation they can reproduce since 
there are no mesh regularity constraints to consider. However, other numerical
issues like zero energy modes, the tensile instability and disorder of the 
discretization points due to the deformation may impose limits on the deformations
possible. It is thus worthwhile to benchmark a wide array of these methods since
a proper review to this end has been missing from the literature so far. 
In the interest of reproducibility, the complete source code of all methods 
considered is made public. It can be retrieved from \url{https://github.com/mroethli/mfree_iwf}.

\section{Introduction}
Meshless methods are a rather recent development in computational science. Even 
though there are methods that could reasonably be called ``meshless'' before the
inception of the SPH in 1977 by Ginghold, Monaghan and Lucy 
\cite{gingold1977smoothed,lucy1977numerical}, the SPH was the first method that 
truly discretized a continuum, to solve a wide array of partial differential 
equations. Although the method was conceived to solve problems in stellar 
dynamics, it was quickly adapted in a wide array of fields including 
computational fluid dynamics \cite{takeda1994numerical, monaghan1994simulating}, electromagnetics 
\cite{price2003smoothed,price2004smoothed,price2005smoothed} and solid dynamics 
\cite{libersky1991smooth,allahdadi1993high}, up to very involved processes like metal cutting \cite{ruttimann2010simulation} \cite{ruttimann2013simulation}. Presenting an exhaustive account of
the history of SPH is beyond the scope of this paper, instead the reader is 
referred to the excellent reviews by Monaghan \cite{monaghan2005smoothed} 
and Price \cite{price2012smoothed}.

Research on meshless methods has been an active field since the development
of the SPH, and numerous improvements have been proposed to the method. Despite 
its age, the overview by Belytschko \cite{belytschko1996meshless} can still be considered
quite comprehensive.

Meshless methods seem to be the perfect means to simulate solids undergoing very
large deformations. The current standard way of performing such computations is
to rely on Lagrangian FE methods. These methods, while well explored and widely
available in simulation tools both open source and proprietary, need numerically
expensive remeshing techniques to guarantee an accurate solution of the equations
to be solved and to resolve topology changes. However, there is a surprising
shortage of literature applying meshless methods, be it SPH or one of it's 
extensions, to problems exhibiting very large deformations in solid mechanics.

One of the earliest papers demonstrating very large deformations using meshless
methods is by Liu and co-workers \cite{jun1998explicit}. They apply the Reproducing
Kernel Particle Method (RKPM) to a weak formulation of the constitutive equations of
elasto-dynamics. Employing a modified Mooney-Rivlin material model \cite{fried1988note} 
they are able to simulate global deformations of up to 1000\%. In \cite{chen1998analysis}
a similar procedure is applied to simulate 2D metal forming processes.

In the area of the Corrected Smoothed Particle Method (CSPM) there are mainly 
the works by Mohammadi and colleagues \cite{ostad2008field,ostad2012stabilized}.
They apply CSPM to the strong form of the elastic equations while applying a 
novel field smoothing technique in order to stabilize the solution. More recently, CSPM was 
applied to model impact problems \cite{eghtesad2012study}, extending the purely
elastic model to a fully elastic-plastic-thermally coupled material. 

Meanwhile, the virtual surgery community was very active in developing Total Lagrangian
Explicit Dynamic (TLED) Algorithms to rapidly simulate soft tissues, some of them
meshless \cite{zhang2014three,horton2010meshless}. These so called Element Free
Galerkin (EFG) methods where also applied to bulk metal forming processes in a 
number of works \cite{nianfei20093d,li2001element}, including
involved physics like friction and contact modeling. Other authors like Wang 
\cite{wang2007development} explored means to dynamically adapt the resolution
of the discretization in areas of interest.

Another strategy borrowed from vortex methods \cite{cottet2000vortex} is to periodically
regularize (or ``remesh'', not to be confused with remeshing algorithms in FEM)
the particle distribution. Those so called rSPH methods where investigated by
Koumoutsakos and colleagues to simulate internal organs in \cite{hieber2008lagrangian}
and \cite{hieber2004remeshed}. It is noteworthy that their algorithms are amongst
the very few ones that actually address the issue of the smoothed boundary in SPH
methods. For this purpose they developed a Lagrangian Level Set method in 
\cite{hieber2005lagrangian}, making their method one of the more elaborate ones
in the area, both in terms of algorithmic as well as implementation complexity.

Another field which is very interested in solving large deformation problems with
rapid computational speed is the computer graphics industry. Even though the focus
of their work is in physical plausibility, as opposed to physical \emph{accuracy},
it is certainly worthwhile to mention works like \cite{muller2004point} by Muller
et al. Their idea is to reduce the order of the model PDEs by performing part of
the spatial derivatives analytically. Like in EFG they employ Moving Least Squares (MLS) 
basis functions to ensure linear completeness. Later, Solenthaler and Parajola \cite{solenthaler2007unified} 
extended the model in \cite{muller2004point} by a wide range of effects like 
melting, solidification etc. Finally, in \cite{becker2009corotated} a
computationally economic way is discussed to employ a co-rotated stress formulation
for computer graphics applications. 

So far all the works discussed are using some successor to SPH. However, there 
are also efforts to simulate large deformations with methods that are still close
to the original SPH formulation. In \cite{cleary2012modelling} a SPH formulation 
very close to the original SPH solid mechanic papers \cite{libersky1991smooth,allahdadi1993high} 
is used to simulate metal forging processes exhibiting rather large deformations. 
However, one of the most promising approaches is probably
to formulate the governing equations using a total Lagrangian framework instead of
an updated one. This approach was pursued in the doctoral thesis of Reveles 
\cite{reveles2007development}. In \cite{vidal2007stabilized} it is argued that
an updated Lagrangian formalism is preferable to a total Lagrangian one because
total Lagrangian methods can not resolve situations where initially neighboring
discretization points are separated, for example by a punch in stamping.
They use a weak form algorithm based on nodal integration and propose a correction 
term involving the difference of two approximations to a Hessian matrix. Another 
updated Lagrangian method is found in a classic paper by Gray and Monaghan \cite{gray2001sph}.
Instead of enhancing the accuracy of differential operators they opt to introduce
an artificial stress term, enabling moderately large deformations in tensile specimen.
In \cite{vignjevic2000treatment} so called stress points are introduced to 
de-collocate the integration of field variables, successfully suppressing zero
energy modes. Finally, the authors of \cite{parshikov2002smoothed} realized that
the particle interactions in SPH can be recast into a Riemann problem. Using this
paradigm they observed a significant improvement in solution quality. This is the
so called Godunov SPH approach. 

Although this overview has not been an exhaustive one, it still covers the bulk
of the work considering large deformations using meshless methods, at least for solid dynamics. A comparative
study regarding the performance of these methods has so far been missing from the
literature. The closest work probably being the review by Rabczuk \cite{rabczuk2004stable}
et al. However, they only cover total Lagrangian methods and  the large deformation problems they show all refer to CFD applications. This work aims to fill this gap, improving upon some of the methods
discussed with some novel ideas. 

\section{The SPH and the Role of the Kernel Function}
A formal, mathematically rigorous discussion of the SPH is not part of the scope
of this paper. For this purpose, the reader is again referred to the review \cite{price2012smoothed}
or such standard texts as \cite{li2007meshfree}. Nonetheless, a short summary of SPH approximation
theory is given in the following. A good approach to SPH is
using the sampling property of the Dirac delta function
\begin{equation}
f(\uline{x})=\int_{-\infty}^{+\infty}f(\uline{x}')\delta(\uline{x}'-\uline{x})dx'\label{eq:1}
\end{equation}
As it stands, this property can't be used to approximate functions. To enable 
a means of approximating the function $f$ at some locations $\uline{x}_{i}$ given
samples at some other, potentially randomly distributed locations $\uline{x}_{j}$
the Dirac delta function $\delta$ is replaced with a mollified Kernel function
$W$: 
\begin{equation}
<f(\uline{x})> =\int_{-\infty}^{+\infty}f(\uline{x}')W^h(\uline{x}'-\uline{x})dx'\label{eq:2}
\end{equation}
Where $h$ is a parameter controlling the amount of mollification. That is, controlling to size of domain of $W^h$. This expression is only approximatively true, hence the approximation operator 
$<\cdot>$ on the left hand side. In order for the error introduced in the expression above to be bound, $W$ needs to fulfill some properties:
\begin{alignat}{2}
\int_{-\infty}^{\infty} & W^h(\uline{x})dr=1\\
 & W^h(\uline{x})\rightarrow\delta(\uline{x})\qquad &  & h\rightarrow0\\
 & W^h(\uline{x})\in C^{k}\qquad &  & k\geq1
\end{alignat}
Where $C^{k}$ is the manifold of $k$ differentiable functions.  
Another important property for the practicability of a kernel is that its domain
needs to be small in terms of multiples of $h$ or at least exponentially decaying
such that it can be clamped without measurable loss of accuracy. In this paper
only one kernel, namely the cubic spline kernel is employed. This classic kernel
is used throughout most of the SPH literature and given as
\begin{align}
W(q) = c/h^2 \begin{cases} 
1 - 1.5q^2 + 0.75q^3 & 0 \leq q \leq 1\\
0.25(2-q)^3 & 1 \leq q \leq 2 \\
0 & \textrm{otherwise}
\end{cases}
\end{align}
\begin{figure}[htp]
\noindent \centering{}%
\includegraphics[width=0.5\textwidth]{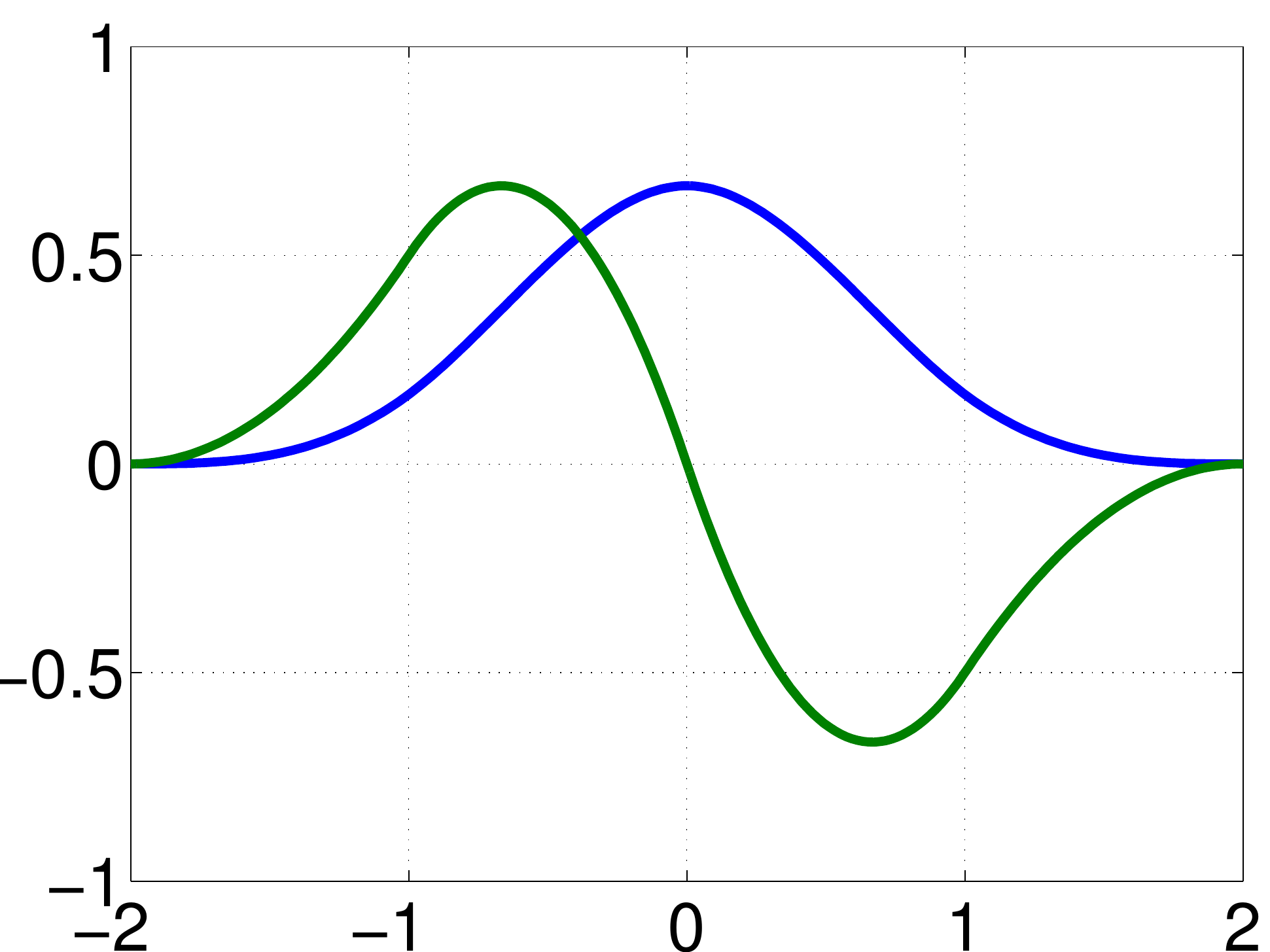}
\caption{The cubic spline kernel in blue with its first derivative in green.}
\label{fig:cubic_spline}
\end{figure}
With $c=10/7\pi$ in 2 dimensions and $q = ||\uline{x}' - \uline{x}||/h$. See figure \ref{fig:cubic_spline} for a plot. 
In order to implement 
the SPH the integral in equation \eqref{eq:2} needs to be 
discretized using some numerical quadrature scheme. In SPH this is usually done by 
nodal integration (a so called Riemann approach or trapezoidal rule):
\begin{equation}
<f(\uline{x}_{i})>=\sum_{j=1}^{N}f_{j}W^h(\uline{x}_{i}-\uline{x}_{j})\omega_{j}\label{eq:3}
\end{equation}
Where $\omega_j$ is the integration weight dictated by the trapezoidal rule. Physically,
this is the volume of discretization point, or ``particle'' in SPH terminology, with 
index $j$:
\begin{align}
\omega_j = m_j/\ \varrho_j 
\end{align}
Where $m$ is the mass of the particle and $\varrho$ its density. In the following, 
$W^h(\uline{x}_{i}-\uline{x}_{j})$ will be shortened to $W_{ij}$ to improve legibility.
This is the most basic SPH approximation formula. To approximate a derivative 
one simply multiplies with nabla: 
\begin{equation}
<\nabla\cdot f(\uline{x}_{i})>=\sum_{j=1}^{N}f_{j}\nabla\cdot W_{ij}\omega_{j}\label{eq:4}
\end{equation}
The derivative of a function can thus be approximated using the derivative of the
kernel function. It is important to note that the expression above is not unique.
By subtracting a null term to the continuous version of \eqref{eq:4} another 
expression is obtained. 
\begin{align}
<\nabla f(\uline{x})> &= \int_{-\infty}^{+\infty}f(\uline{x}')\nabla \cdot W_{ij} dx' - f(\uline{x})\underbrace{\int_{-\infty}^{+\infty}\nabla W_{ij} dx'}_{0}\\
&= \int_{-\infty}^{+\infty}(f(\uline{x}')-f(\uline{x}))\nabla W_{ij} dx' \\
&\approx\sum_{j=1}^{N}(f_{j}-f_{i})\nabla W_{ij} \omega_{j} \label{eq:5}
\end{align}
The null term on the first line above is zero because $W$ is symmetric. Hence, 
$\nabla \cdot W$ is anti-symmetric, and the integral of an anti-symmetric function
over the whole domain is always zero. This procedure of adding null terms to the
continuous equations, yielding non zero differences in the discretized equations
enables an infinite number of SPH approximation equations. Another popular approach
is to consider the expression $\nabla  f /\ \varrho$, expand it using the product	
rule and isolate a term containing $\nabla f$ \footnote{Note: $\nabla f / \varrho$ is extracted
instead of $\nabla f$ in anticipation of the momentum equation, which naturally contains a division
by $\varrho$}:
\begin{align}
\nabla \left( \frac{f}{\varrho} \right) &= \frac{\varrho \cdot (\nabla f)}{\varrho^2} - \frac{f \cdot (\nabla \varrho)}{\varrho^2} \\
\frac{(\nabla f)}{\varrho} &=  \nabla \left( \frac{f}{\varrho} \right) + \frac{f \cdot (\nabla \varrho)}{\varrho^2} \\
\end{align}
Discretization of the two nabla operators on the right using \eqref{eq:4} yields:
\begin{align}
\nabla \left( \frac{f}{\varrho} \right) + \frac{f \cdot (\nabla \varrho)}{\varrho^2} &\approx \sum_{j=1}^{N} \frac{f_j}{\varrho_j} W_{ij} \frac{m_j}{\varrho_j} + \frac{f_i}{\varrho_i^2} \cdot \sum_{j=1}^{N} \varrho_j \nabla W_{ij} \frac{m_j}{\varrho_j} \\
&= \sum_{j=1}^{N} \frac{f_j}{\varrho_j^2} W_{ij} m_j + \sum_{j=1}^{N} \frac{f_i}{\varrho_i^2} \nabla W_{ij} m_j \\
&= \sum_{j=1}^{N} \left( \frac{f_j}{\varrho_j^2} + \frac{f_i}{\varrho_i^2} \right) \nabla W_{ij} m_j \label{eq:6}
\end{align}
These different approximators exist for specific reasons: Version \eqref{eq:5} is
zero order consistent. That is, the derivative of a constant function (i.e. zero)
is obtained exactly. This can readily be seen since in that case $f_i=f_j$ and the
term in the bracket becomes zero. Version \eqref{eq:6} is symmetric about
indices $i,j$. These properties are important regarding the conservation of physical
quantities of the PDE system to be discretized. For example, it is readily understood
that in order for a discretization to be Galilean invariant it has to be zero order
complete, since a constant displacement (or velocity) field, i.e. a translation mode,
has to be approximated exactly. 

A researcher wanting to use SPH to simulate the evolution of a system of PDEs has thus
to be careful which spatial derivative operator is discretized using which SPH equation. 
Beside the quite ad-hoc method of just trying out popular options until the conservation 
proofs required start to work out, there are other options like starting from the 
Lagrangian of the system \cite{price2012smoothed} or using the variational 
approach given by Bonet et al. \cite{bonet1999variational}. Other researchers opt 
to use non-conservative approaches \cite{reveles2007development, hieber2008lagrangian}. 
While one might argue that these approaches are conceptually unsound a wide range
of phenomena has been simulated successfully using such approaches. One should also 
keep in mind that methods that conserve all properties on paper are still dispersive
due to round off errors after implementation, or may contain stabilization methods
that disperse the energy of spurious modes, and can thus not be considered 
``conservative'' either. 

\subsection{Corrected SPH Kernels}

So far all interpolation formulae presented are zero order complete at best. This seems
quite poor, even the most basic FE elements (i.e. linear ones) are first order 
complete. Furthermore, for an interpolation scheme to represent rigid body motions
exactly it seems necessary that the scheme is, in fact, first order complete. This
can easily be seen considering a pure rotation mode about the origin:
\begin{align}
\uline{v} = \begin{bmatrix}[r] -\alpha \cdot y \\ \alpha \cdot x\end{bmatrix}
\end{align}
which shows that a rotation mode is in fact linear along the coordinate axes. Thus,
it is imperative that linear functions and their gradient are approximated exactly. \par
A multitude of procedures to elevate the order of completeness of SPH approximators
exist. In this work, two such methods are employed:
\begin{itemize}
\item The Randles-Libersky correction, first mentioned in \cite{randles1996smoothed} but only properly derived in \cite{libersky1997recent}. It will be shown to be equivalent to CSPM.
\item The Reproducing Kernel Particle Method \cite{liu1995reproducing}
\end{itemize}
In the following these two methods are derived using notation consistent with 
this work and some properties of the methods at hand are discussed.

\subsubsection{The Randles Libersky Correction} \label{sec:randles_libersky}

The derivation starts with version \eqref{eq:5} of the SPH approximators:
\begin{align}
<\nabla  f > &= \sum_{j=1}^{N}(f_{j}-f_{i})\nabla W_{ij} \frac{m_j}{\varrho_j}
\end{align}
The right hand side is now enhanced with some yet to be determined tensor of second order $\uuline{B}$:
\begin{align}
<\nabla  f > &= \left[\sum_{j=1}^{N}(f_{j}-f_{i})\nabla  W_{ij} \frac{m_j}{\varrho_j} \right]\cdot\uuline{B}
\end{align}
A linear function $f=\uuline{A}\cdot \uline{x} + \uline{b}$ with $\nabla  f = \uuline{A}$ and $\uuline{A}$ being the diagonal coefficient matrix
is plugged in. Note that the approximation operator $<.>$ is omitted below, i.e. the relation is demanded to be exact.
\begin{align}
\uuline{A} &= \left[\sum_{j=1}^{N}((\uuline{A}\cdot \uline{x}_i + \uline{b})-(\uuline{A}\cdot \uline{x}_j + \uline{b})) \otimes \nabla  W_{ij} \frac{m_j}{\varrho_j} \right]\cdot\uuline{B} \\
\uuline{A} &= \uuline{A} \left[\sum_{j=1}^{N}(\uline{x}_i - \uline{x}_j) \otimes \nabla  W_{ij} \frac{m_j}{\varrho_j} \right]\cdot\uuline{B} \\
\uuline{1} &= \left[\sum_{j=1}^{N}(\uline{x}_i - \uline{x}_j) \otimes \nabla W_{ij} \frac{m_j}{\varrho_j} \right]\cdot\uuline{B} \\
\end{align}
The procedure above can be understood as the fitting of a correction tensor to the 
kernel $W_{ij}$ such that the gradients of linear functions are retrieved exactly. 
The correction tensor $\uuline{B}$ can be defined in compact form:
\begin{align}
\uuline{B} &= \left[\sum_{j=1}^{N}(\uline{x}_i - \uline{x}_j) \otimes \nabla W_{ij} \frac{m_j}{\varrho_j} \right]^{-1} \label{cspmsys}
\end{align}
Necessitating a matrix inversion at each particle location. It is customary \cite{reveles2007development,rabczuk2004stable}
to replace $W_{ij}$ by it's Shephard Interpolant \cite{shepard1968two}:
\begin{align}
\tilde{W}_{ij} = \frac{W_{ij}}{\sum_{j=1}^{N} W_{ij} \frac{m_j}{\varrho_j}} \label{eq:7}
\end{align}
This well known interpolant exactly reproduces constants and is derived by considering
a Taylor series approximation which is immediately cut after the constant term
\begin{align}
f(\uline{x}') = f(\uline{x}_i) + \mathcal{O}(\uline{x}'-\uline{x}_i)
\end{align}
Multiplying with a kernel and integrating over the whole domain:
\begin{align}
\int W(\uline{x}'-\uline{x}_i) f(\uline{x}) d\uline{x}' &= \int W(\uline{x}'-\uline{x}_i) f(\uline{x}_i) d\uline{x}' + \mathcal{O}(\uline{x}'-\uline{x}_i) \\
&= f(\uline{x}_i) \int W(\uline{x}'-\uline{x}_i) d\uline{x}' + \mathcal{O}(\uline{x}'-\uline{x}_i) 
\end{align}
Solving for $f(\uline{x}_i)$:
\begin{align}
f(\uline{x}_i) &= \frac{\int W(\uline{x}'-\uline{x}_i) f(\uline{x}') d\uline{x}'}{\int W(\uline{x}'-\uline{x}_i) d\uline{x}'} \label{eq:8}
\end{align}
It is evident that \eqref{eq:7} contains a discretized version of \eqref{eq:8}.

It is worthwhile to contrast the Randles Libersky Correction to the so called 
Corrected Smoothed Particle Method by Chen et al. \cite{chen1999improvement}.
To derive the CSPM another Taylor series expansion is employed, this time only cutting
after the linear element.
\begin{eqnarray}
f(x',y') & = f(x_i,y_i)+(x'-x_i)\frac{\partial f}{\partial x}|_{(x_i,y_i)}+(y'-y_i)\frac{\partial f}{\partial y}|_{(x_i,y_i)}
\end{eqnarray}
Note that the derivation is done in 2D for brevity. The extension to three
or more dimensions is tedious but straight forward. The Taylor series expansion above is now, in turn, multiplied by
$\frac{\partial}{\partial x}W$, $\frac{\partial}{\partial y}W$, then integrated
over the whole domain, giving the following two equations
\begin{align}
\int \frac{\partial W}{\partial x} f(x',y') d\uline{x}' & = \int \frac{\partial W}{\partial x}f(x_i,y_i)d\uline{x}+\int (x'-x_i) \frac{\partial W}{\partial x} \frac{\partial f}{\partial x}|_{(x_i,y_i)} d\uline{x}'+ \int (y'-y_i)\frac{\partial W}{\partial x}\frac{\partial f}{\partial y}|_{(x_i,y_i)}d\uline{x} \\
\int \frac{\partial W}{\partial y} f(x',y') d\uline{x}' & = \int \frac{\partial W}{\partial y}f(x_i,y_i)d\uline{x}+\int (x'-x_i) \frac{\partial W}{\partial y} \frac{\partial f}{\partial x}|_{(x_i,y_i)}d\uline{x}'+ \int (y'-y_i)\frac{\partial W}{\partial y}\frac{\partial f}{\partial y}|_{(x_i,y_i)}d\uline{x}
\end{align}
Reorganizing these terms:
\begin{align}
\int \frac{\partial W}{\partial x} (f(x',y') - f(x_i,y_i)) d\uline{x}' & = \frac{\partial f}{\partial x}|_{(x_i,y_i)} \int (x'-x_i) \frac{\partial W}{\partial x} d\uline{x}' + \frac{\partial f}{\partial y}|_{(x_i,y_i)}\int (y'-y_i)\frac{\partial W}{\partial x}d\uline{x}'+\cdots \\
\int \frac{\partial W}{\partial y} (f(x',y') - f(x_i,y_i)) d\uline{x}' & = \frac{\partial f}{\partial x}|_{(x_i,y_i)} \int (x'-x_i) \frac{\partial W}{\partial y} d\uline{x}' + \frac{\partial f}{\partial y}|_{(x_i,y_i)}\int (y'-y_i)\frac{\partial W}{\partial y}d\uline{x}'+\cdots \\
\end{align}
Discretizing employing the usual nodal integration:
\begin{align}
\sum_{j=1}^{N} \frac{\partial W}{\partial x} (f(x_j,y_j) - f(x_i,y_i)) \frac{m_j}{
\varrho_j} & = \frac{\partial f}{\partial x}|_{(x_i,y_i)} \sum_{j=1}^{N} (x_j-x_i) \frac{\partial W}{\partial x} \frac{m_j}{
\varrho_j}+ \frac{\partial f}{\partial y}|_{(x_i,y_i)}\sum_{j=1}^{N} (y_j-y_i)\frac{\partial W}{\partial x}\frac{m_j}{
\varrho_j} \\
\sum_{j=1}^{N} \frac{\partial W}{\partial y} (f(x_j,y_j) - f(x_i,y_i)) \frac{m_j}{
\varrho_j} & = \frac{\partial f}{\partial x}|_{(x_i,y_i)} \sum_{j=1}^{N} (x_j-x_i) \frac{\partial W}{\partial y} \frac{m_j}{
\varrho_j}+ \frac{\partial f}{\partial y}|_{(x_i,y_i)}\sum_{j=1}^{N} (y_j-y_i)\frac{\partial W}{\partial y}\frac{m_j}{
\varrho_j} \\
\end{align}
Writing this in compact matrix form reveals that the CSPM corrector is in fact
the familiar Randles Libersky correction:
\begin{align}
\underbrace{\left[\sum_{j=1}^{N}(\uline{x}_i - \uline{x}_j) \otimes \nabla W_{ij} \frac{m_j}{\varrho_j} \right]}_{\uuline{B}^{-1}} <\nabla  f >_i &= \left[ \sum_{j=1}^{N}(f_{j}-f_{i})\nabla W_{ij} \frac{m_j}{\varrho_j} \right]
\end{align}
it is thus not entirely clear why the term CSPM was coined or why it spawned it's own string
of papers \cite{chen1999completeness,chen1999corrective,chen2001corrective,chen2000generalized}.
There is however, as was just shown, a difference in mathematical motivation between the 
two correctors: 
\begin{itemize}
\item Randles and Libersky introduce a correction tensor to exactly reproduce
a linear Ansatz function 
\item The correction tensor by Chen, Beraun and Jih is the 
result of multiple simultaneous Taylor expansions 
\end{itemize}
The latter approach might be more
attractive because it extends to higher order derivatives in systematic fashion
\cite{chen1999corrective}. On the other hand, the CSPM approach does not allow
for replacing $W$ by $\tilde{W}$ (the Shephard interpolant) since the approximator 
is unique. Also, Vidal et al. used an approach very close to Randles Liberskys 
idea to come up with a corrector for the Hessian \cite{vidal2007stabilized}, showing
that their idea too extends to higher order derivatives, albeit maybe in less methodical fashion. \par
In this paper Randles and Libersky's version, that is, including the Shephard interpolant,
is used henceforth. It will be denoted by $\overset{\triangle}W$.

\subsubsection{The RKPM} \label{sec:RKPM}

Randles and Libersky introduced a second order tensor to exactly reproduce an 
Ansatz function while Chen and coworkers used Taylor series expansions to arrive at
the same correction tensor (CSPM). Another idea, the idea of RKPM that is, is to introduce
a correction \emph{function} to the kernel estimate, this time yielding a truly
different corrector. Starting from \eqref{eq:3}, $\psi$ being the correction function 
mentioned:
\begin{align}
<f(\uline{x}_{i})>&=\sum_{j=1}^{N}f_{j}\underbrace{\psi(\uline{x}_{i},\uline{x}_{j})\cdot W(\uline{x}_{i}-\uline{x}_{j})}_{K(\uline{x},\uline{x}')}\omega_{j}
\end{align}
To ensure linear completeness the following Ansatz for $\psi$ is chosen:
\begin{align}
\psi(\uline{x}_{i},\uline{x}_{j}) &= \left[ C_0(\uline{x}) + \uline{C}_1(\uline{x})\cdot(\uline{x}-\uline{x}') \right]
\end{align}
with (unknown) coefficient $C_0$ and coefficient vector $\uline{C}_1$. To determine
these coefficients, one employs the moment conditions:
\begin{align}
1 &= \int K(\uline{x},\uline{x}') d\uline{x}' \\ \label{eq:mom1}
0 &= \int (\uline{x}-\uline{x}')\cdot K(\uline{x},\uline{x}') d\uline{x}'
\end{align}
It can be shown that any kernel that fulfills the first $n$ moment conditions is
$n$-th order complete (if the moment conditions start being numbered at zero, 
which is customary). 

It is convenient to recall the definition of the first two moments of the 
kernel function $W$ about some point $\uline{x}$ to shorten some notation later on:
\begin{align}
m_{0} & =\int W(\uline{x}-\uline{x}')d\uline{x}'\\
\uline{m}_{1} & =\int (\uline{x}-\uline{x}')W(\uline{x}-\uline{x}')d\uline{x}'\\
\uuline{m}_{2} & =\int (\uline{x}-\uline{x}')\otimes(\uline{x}-\uline{x}')W(\uline{x}-\uline{x}')d\uline{x}'
\end{align}
The first condition \eqref{eq:mom1} is rewritten by expanding K(\uline{x},\uline{x}'):
\begin{align}
1 & =\int \left[C_{0}(\uline{x})+\uline{C}_{1}(\uline{x})(\uline{x}-\uline{x}')\right]W(\uline{x}-\uline{x}')d\uline{x}'\\
 & =C_{0}(\uline{x})\int W(\uline{x}-\uline{x}')d\uline{x}'+\uline{C}_{1}(\uline{x})\int (\uline{x}-\uline{x}')W(\uline{x}-\uline{x}')d\uline{x}'\\
 & =C_{0}(\uline{x})\; m_{0}+\uline{C}_{1}(\uline{x})\;\uline{m}_{1}
\end{align}

The same for the second condition: 
\begin{align}
0 & =\int (\uline{x}-\uline{x}')\left[C_{0}(\uline{x})+\uline{C}_{1}(\uline{x})(\uline{x}-\uline{x}')\right]W(\uline{x}-\uline{x}')d\uline{x}'\\
 & =C_{0}(\uline{x})\int (\uline{x}-\uline{x}')W(\uline{x}-\uline{x}')d\uline{x}'+\uline{C}_{1}(\uline{x})\int (\uline{x}-\uline{x}')\otimes(\uline{x}-\uline{x}')W(\uline{x}-\uline{x}')d\uline{x}'\\
 & =C_{0}(\uline{x})\;\uline{m}_{1}+\uline{C}_{1}(\uline{x})\;\uuline{m}_{2}
\end{align}
This system of equations can now be written in compact form:
\begin{align}
\left[\vphantom{\begin{matrix}m_{0} & \uline{m}_{1}^{T}\\
\uline{m}_{1} & \uuline{m}_{2}
\end{matrix}}\begin{matrix}1\\
\uline{0}
\end{matrix}\right] & =\left[\vphantom{\begin{matrix}m_{0} & \uline{m}_{1}^{T}\\
\uline{m}_{1} & \uuline{m}_{2}
\end{matrix}}\begin{matrix}C_{0}\\
\uline{C}_{1}
\end{matrix}\right]\left[\begin{matrix}m_{0} & \uline{m}_{1}^{T}\\
\uline{m}_{1} & \uuline{m}_{2}
\end{matrix}\right]\\
\uline{P} & =\uline{C}(\uline{x})\uuline{M}(\uline{x})\label{rkpmsys}
\end{align}
This approximator is exact for linear and constant functions since the coefficients
$c \in \uline{C}(\uline{x})$ were fit to fulfill the first two moment conditions. To 
approximate gradients using RKPM the system above undergoes consecutive derivation
with respect to the coordinates desired. Starting from $K(.)$:
\begin{align}
\frac{\partial}{\partial x}K(\uline{x},\uline{x}')=\left(\frac{\partial}{\partial x}\psi(\uline{x},\uline{x}')\right)W(\uline{x}-\uline{x}')+\psi(\uline{x},\uline{x}')\left(\frac{\partial}{\partial x}W(\uline{x}-\uline{x}')\right)
\end{align}
The term $\frac{\partial}{\partial x}\psi(\uline{x},\uline{x}')$ contains the
derived coefficients $\frac{\partial}{\partial x}\uline{C}(\uline{x})$.
To obtain them, system \eqref{rkpmsys} is derived:
\begin{align}
\frac{\partial}{\partial x}\uline{P} & =\frac{\partial}{\partial x}(\uline{C}(\uline{x})\uuline{M}(\uline{x}))\\
0 & =\left(\frac{\partial}{\partial x}\uline{C}(\uline{x})\right)\cdot\uuline{M}(\uline{x})+\uline{C}(\uline{x})\cdot\left(\frac{\partial}{\partial x}\uuline{M}(\uline{x})\right)
\end{align}
This procedure is repeated for $\frac{\partial}{\partial y}$ (and $\frac{\partial}{\partial z}$ in 3D)
to arrive at a system of equations which can in turn be solved for the unknowns 
$\nabla \cdot \uline{C}(\uline{x})$. RKPM corrected kernel functions are denoted by $\overset{\circ}{W}$ 
in this paper.

\subsection{Choosing smoothing length $h$} \label{chosing_h}

A distinctive feature of the SPH is the presence of a completely unphysical parameter
$h$. Such a parameter is not present in methods like FDM, FEM, FVM and the like 
and can also not be equated to the mesh size in FEM. However, its choice
is imperative in solution quality since it determines the number of neighbors
available for any SPH approximator. In corrected SPH methods the solution 
might even stop completely if not enough neighbors are available since in that 
situation the correction matrices in \eqref{rkpmsys} and \eqref{cspmsys} may become singular, and
can thus not be inverted.

\begin{figure}
\noindent \centering{}%
\includegraphics[width=0.7\textwidth]{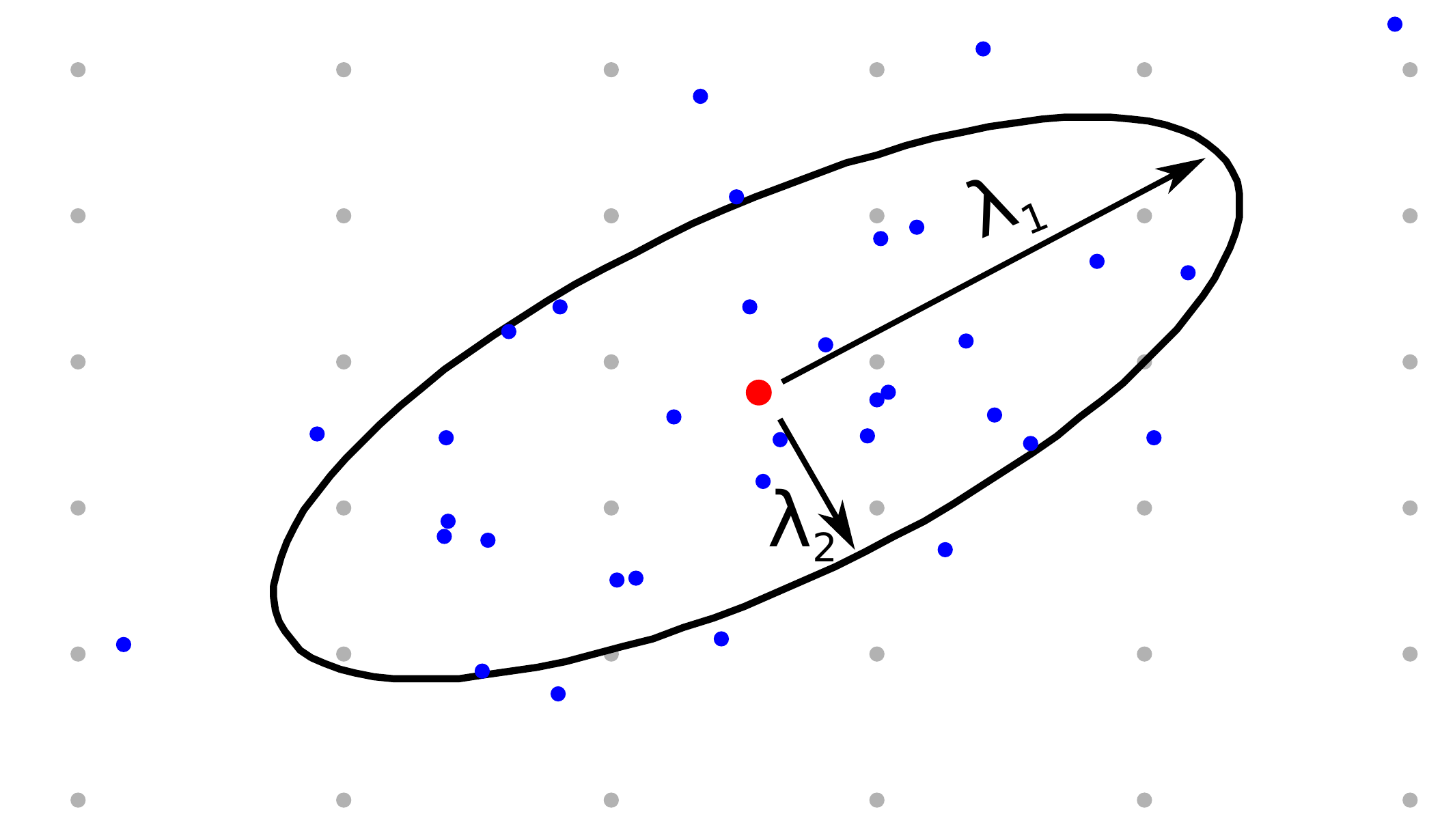}
\caption{Sketch of the proposed algorithm. Original neighborhood in gray, current positions in blue. The covariance matrix of the map from the gray positions to the blue positions is investigated. The eigenvalues $\lambda$ and the corresponding eigenvectors of said covariance matrix yield radii and major axis of a confidence ellipse shown above. It can be observed that the ellipse follows the heavy shearing of the transformation from the original to current positions, while some outlier particles are removed.}
\label{fig:ellipse}
\end{figure}

\begin{figure}[htp]
\begin{tabular}{c}
\includegraphics[width=\textwidth]{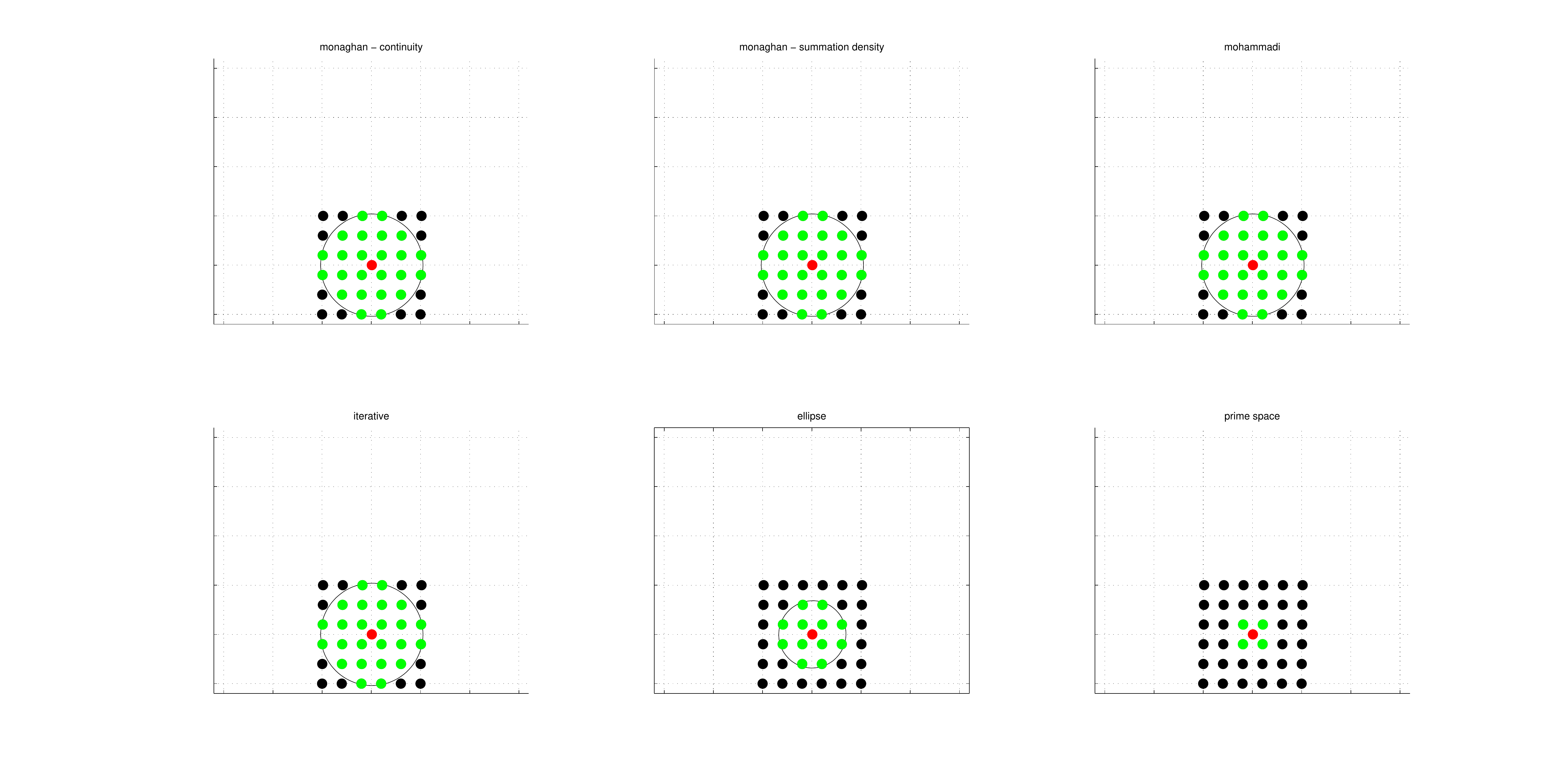} \tabularnewline
\includegraphics[width=\textwidth]{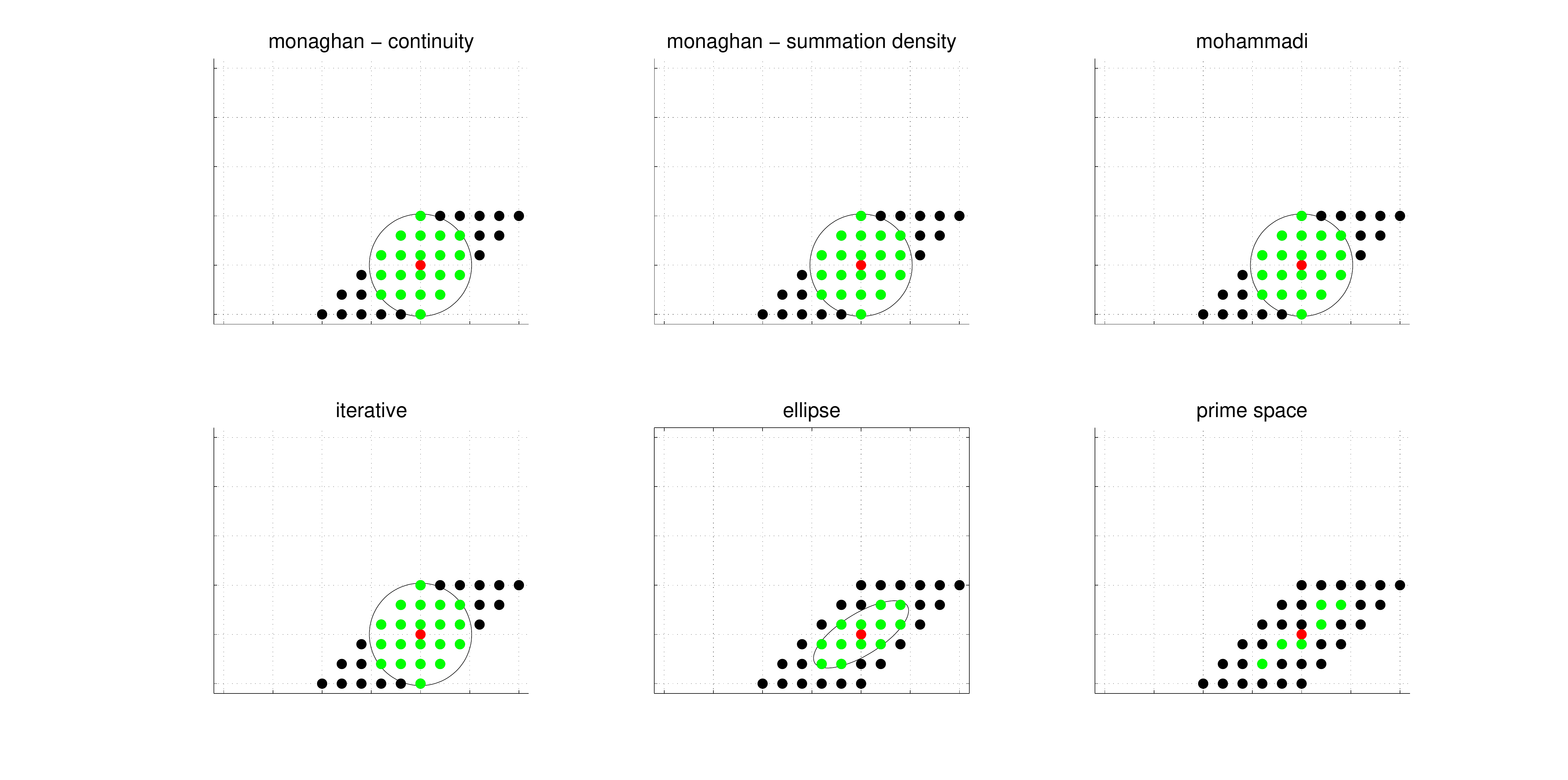} \tabularnewline
\includegraphics[width=\textwidth]{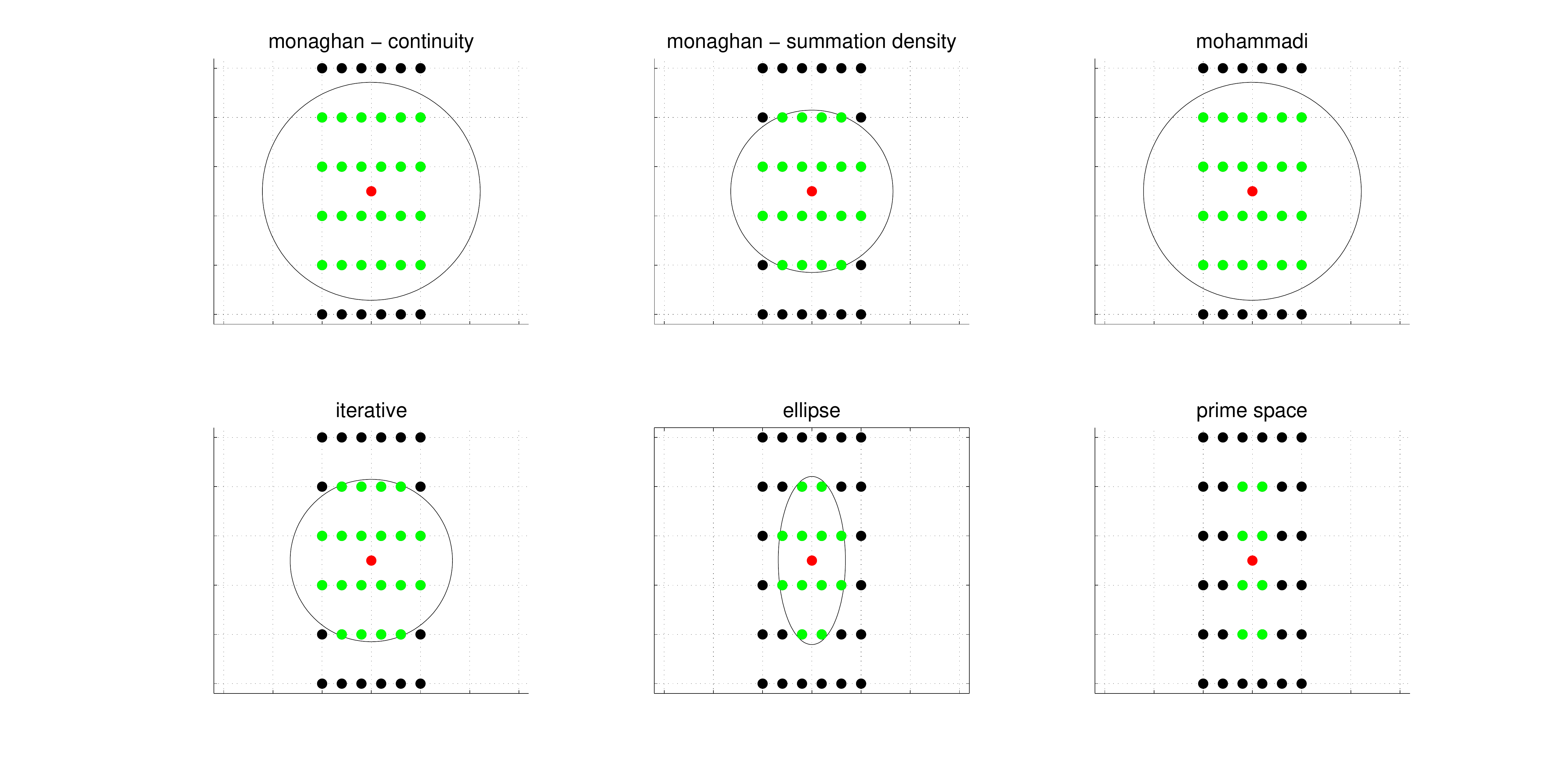} \tabularnewline
\end{tabular}
\caption{Benchmark of some algorithms regarding the choice of smoothing length.
Initial conditions on top, shear case in the middle, tensile case at the bottom.}
\label{fig:smooth_len}
\end{figure}

It is thus clear that the smoothing length needs to be adapted throughout the simulation
to reflect changes in the local density of particles. Thus, $h$ should become small in 
areas where there is a lot of particles to ensure reasonable runtimes and large 
in areas that are sampled sparsely to ensure maintenance of solution quality. The earliest idea to 
ensure this \cite{gingold1977smoothed} is to set $h$ proportional to the density
estimator:
\begin{align}
h \propto \frac{1}{<\varrho>^{\frac{1}{\eta}}} \label{prop_dens}
\end{align}
$\eta$ being the dimension of the problem at hand. It was noticed simultaneously  
by Monaghan \cite{monaghan2002sph} and Springel, Helmquist \cite{springel2002cosmological} 
that $h$ and the quantity of
\begin{align}
<\varrho> = \sum_{j=1}^{N} W_{ij} \cdot m_j \label{sum_density}
\end{align}
are interdependent. Hence, they apply a fixed point iteration to find $h$. There
is a multitude of other options of choosing $h$. A standard approach is to 
use \eqref{prop_dens} but use the continuity equation \eqref{ulcont} to come
up with $<\varrho>$ instead of \eqref{sum_density}. Other authors like Mohammadi \cite{ostad2012stabilized} prefer to use an equation of the form:
\begin{align}
h^{n+1} = \sqrt{\frac{\varrho}{\varrho^{n+1}}}
\end{align}
with $n+1$ being the new time step, $n$ being the old one to obtain $h$. All of 
these approaches are isotropic, i.e. the local anisotropy of the distribution of
particles is not taken into consideration. In a deformation field that distorts the particles
heavily in one direction but compresses them in the other there may thus be
neighborhoods that feature an excessive number of particles along one direction but
a too low one in the other. An interesting idea is presented in \cite{randles1999neighbors,randles2000normalized}. For a particle at $\uline{x}_i$ particles from a candidate set are mapped to a prime space using the transofrmation 
\begin{align}
\uline{x}_j' = \frac{\uline{x}_{íj}}{\chi ||\uline{x}_{ij}||}
\end{align}
where $\chi$ is a measure of local anisotropy. The closest neighbors are mapped farthest away from $\uline{x}_i$ with regard to local anisotropy $\chi$. The final neighborhood is then found using the convex hull algorithm. However, their algorithm needs some tedious corrections at the boundary of the particle set
and was found to perform rather poor in some situations (see figure \ref{fig:smooth_len}).
Another simple idea borrowed from descriptive statistics is to compute the local covariance matrix
\begin{align}
\uuline{H}_i = \frac{1}{N-1}\sum_{j=1}^N \uline{x}_{ij} \otimes \uline{x}_{ij}
\end{align}
Assuming that the points $\uline{x}_j$ are bivariate (or trivariate in 3D) normally
distributed about $\uline{x}_i$ one can do an Eigenvalue-analysis of $\uuline{H}$ to obtain a 
confidence ellipse. If $\lambda_1, \lambda_2$ are the Eigenvalues of $\uuline{H}$
and $\uline{e}_1, \uline{e}_2$ the Eigenvectors then the confidence ellipse
has half axis along $\uline{e}_i$ with length $\sqrt{\lambda_i}$. Since 
$\sqrt\lambda_i = \sigma_i$, with $\sigma_i$ being the standard deviations of
the normal distribution assumed, one can fit a confidence ellipse by scaling
$\sqrt\lambda_i$ by a suitable $\chi^2$ value, e.g. $\sqrt{5.991}$ for $95\%$). A sketch of the procedure is found in figure \ref{fig:ellipse}.
A similar algorithm could not be found in literature.
However, since the idea is very simple, it is doubtful that the idea 
is truly original. In fact, the approach may be regarded as a heavily simplified version of the tesnor formulation approach given in \cite{owen1998adaptive}, \cite{shapiro1996adaptive}.

Figure \ref{fig:smooth_len} compares all the algorithms mentioned in this section. 
No physics simulation was done, particle movement was induced by an external velocity
field. Shear and tensile situations were investigated. Generally speaking, the two
classical approaches identify an excess number of neighbors due to the isotropic 
expansion of the neighborhood. Also, the choice of density estimator seems to be
way more important than the specific equation for $h$. The ellipse fit identifies
a neighborhood set preferable to the prime space algorithm \cite{randles2000normalized}
at least in the shear case situation: The neighborhood in the shear case 
identified by the prime space algorithm is not symmetric about the query point 
in red and quite sparse. In the tension case it is less clear which neighborhood is
preferable. Some additional neighbors are identified by the ellipse fit while the
prime space algorithm follows the anisotropy of the deformation more closely. The
additional neighbors may stabilize the solution, but at the same time impact the 
runtime of the algorithm negatively. 

\section{Governing Equations and Material Modeling} \label{governing}

Conservation of mass and momentum in a solid, considering an updated Lagrangian 
frame, can be expressed using the following two PDEs:
\begin{align}
\dot{\varrho} &= -\varrho \nabla \uline{v}  \label{ulcont} \\  
\dot{\uline{v}} &= \frac{1}{\varrho} \nabla \uuline{\sigma} \label{ulmoment}
\end{align}
where \eqref{ulcont} is the continuity equation and \eqref{ulmoment} is 
the balance of momentum without body forces, containing the Cauchy stress tensor $\uuline{\sigma}$. The following convention is adapted in this paper: if a derivative operator like $\nabla(.)$ is without index it refers to the \textit{current} configuration. If the derivative operator refers to the initial configuration it is written with upper index $0$. The usual convention for $x$ being coordinates in the current configuration and $X$ being coordinates in the initial configuration is adapted too.
The same set of equations can be expressed in a total Lagrangian frame: 
\begin{align}
\varrho &= \varrho_0 \cdot J \label{tlcont} \\ 
\dot{\uline{v}} &= \frac{1}{\varrho_0} \nabla^0 \uuline{P} \label{tlmoment}
\end{align}
The $J$ in \eqref{tlcont} denotes the determinant of the deformation gradient $\uuline{F}$.
Instead of the Cauchy stress tensor \eqref{tlmoment} contains the nominal stress 
tensor \uuline{P} (the transpose of the first Piola-Kirchhoff stress tensor). Beside the balance equations above, constitutive laws are needed to describe the underlying material behaviour.

The choice of constitutive equations depends on the material to be simulated. The
scope of this work contains linear elastic materials undergoing large deformations
only. That is, the stress strain curve shall be linear, i.e. Hooke's law, but
geometrical non-linearities, i.e. rotations, need to be resolved correctly. 

A simple proposition for a constitutive update could look like this:
\begin{align}
\dot{\uuline{\sigma}} &= \underline{\underline{\underline{\underline{C}}}}:\dot{\uuline{\varepsilon}} \label{hookes} \\
\dot{\uuline{\varepsilon}} &= \frac{1}{2}(\uuline{L} + \uuline{L}^T) \\
L_{ij} &= \frac{\partial v_i}{\partial x_j} 
\end{align}
with $C$ being the fourth order Hooke's elasticity tensor. Some materials may ask for different models regarding the deviatoric and hydrostatic (pressure) stress. This is especially relevant in J2 plasticity which is adequate for a wide range of metals. 
It is thus preferable to decompose $\uuline{\sigma}$ into a deviatoric
and isotropic part:
\begin{align}
\uuline{\sigma} &= -p\uuline{I} + \uuline{S} \label{cauchy_decomposition}
\end{align}
In case of an isotropic, homogeneous material Hooke's law can be restated conveniently using the Lam{\'e} parameters:
\begin{align}
\underline{\underline{\sigma}} = \underbrace{\frac{E}{1+\nu}}_{2G}\underline{\underline{\varepsilon}} + \underbrace{\frac{\nu}{(1-2\nu)(1+\nu)}}_{\lambda}\underline{\underline{I}}\cdot tr(\underline{\underline{\varepsilon}}) \label{eq:hooke_lame}
\end{align}
This directly yields an expression for $\underline{\underline{S}}$:
\begin{align}
\underline{\underline{S}} &= 2\mu\cdot dev(\underline{\underline{\varepsilon}}) + \underbrace{dev(\lambda \cdot tr(\underline{\underline{\varepsilon}}))}_{0} \\
&= 2\mu\cdot (\underline{\underline{\varepsilon}} - 1/3 tr(\underline{\underline{\varepsilon}}))
\end{align}
This can readily be derived in time: 
\begin{align}
\underline{\underline{\dot{S}}} &= 2\mu\cdot (\underline{\underline{\dot{\varepsilon}}} - 1/3 tr(\underline{\underline{\dot{\varepsilon}}})) \label{hookedev}
\end{align} 
This covers $\uuline{S}$ in \eqref{cauchy_decomposition}, leaving $p$ to be defined.
A popular choice is 
\begin{align}
p = c_0^2(\varrho_0-\varrho) \label{eos1}
\end{align}
This is used for example in the first SPH paper regarding solids \cite{libersky1991smooth} 
but also later works \cite{hieber2008lagrangian}. It is also present in commercial
simulation packages like \verb+LSDYNA+ \cite{hallquist2006ls}. It is not clear
where this equation of state has been used first. An attempt at deriving it follows.
Hookes law can alternatively be written as 
\begin{align}
\uuline\sigma = 3K\cdot vol(\uuline\varepsilon) + 2G\cdot dev(\uuline\varepsilon)
\end{align}
Thus:
\begin{align}
vol(\uuline\sigma) &= 3K\cdot vol(\uuline\varepsilon) \\
\uuline{I} \cdot p &= \uuline{I}\cdot3K\cdot \frac{1}{3} tr(\uuline{\epsilon}) \\
p &= K\cdot tr(\uuline{\epsilon}) \label{eq:presscomp}
\end{align}
Assuming $|\varepsilon_{ii}| \ll 1$ it holds that:
\begin{align}
tr(\uuline\varepsilon) = \frac{V-V_0}{V_0}
\end{align}
Inserting into \eqref{eq:presscomp}
\begin{align}
K\left(\frac{V-V_0}{V_0}\right) = p
\end{align}
Using the mass conservation $m=V_0\cdot\varrho_0=V\cdot\varrho$ this is transformed into:
\begin{align}
p &= \frac{K}{\varrho}(\varrho_0 - \varrho) \label{eos2} \\
&= c^2(\varrho_0-\varrho)
\end{align}
Note the subtle difference in \eqref{eos1} and \eqref{eos2}. The authors of this paper
were only able to identify a single work which uses \eqref{eos2}, namely \cite{cleary2012modelling}.
The motivation to use this equation of state is not entirely clear. It seems
that the pressure term is artificially limited to small deformation elasticity.
Nonetheless, to stay consistent with existing SPH literature it is employed in 
this work too. 
\subsection{Objective Stress Rates}

\begin{figure}[htp]
\begin{tabular}{ccc}
\includegraphics[width=0.3\textwidth]{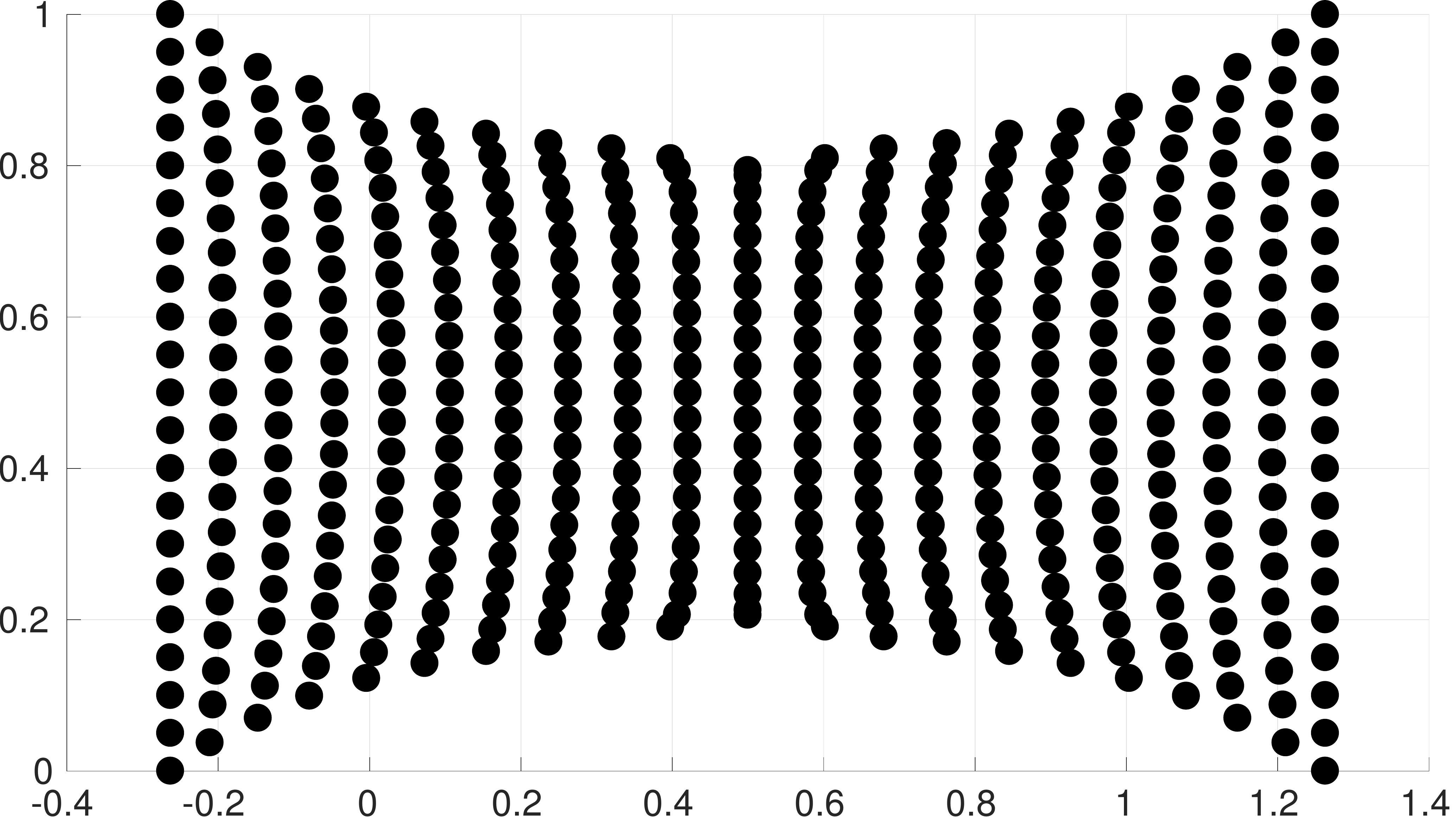} &
\includegraphics[width=0.3\textwidth]{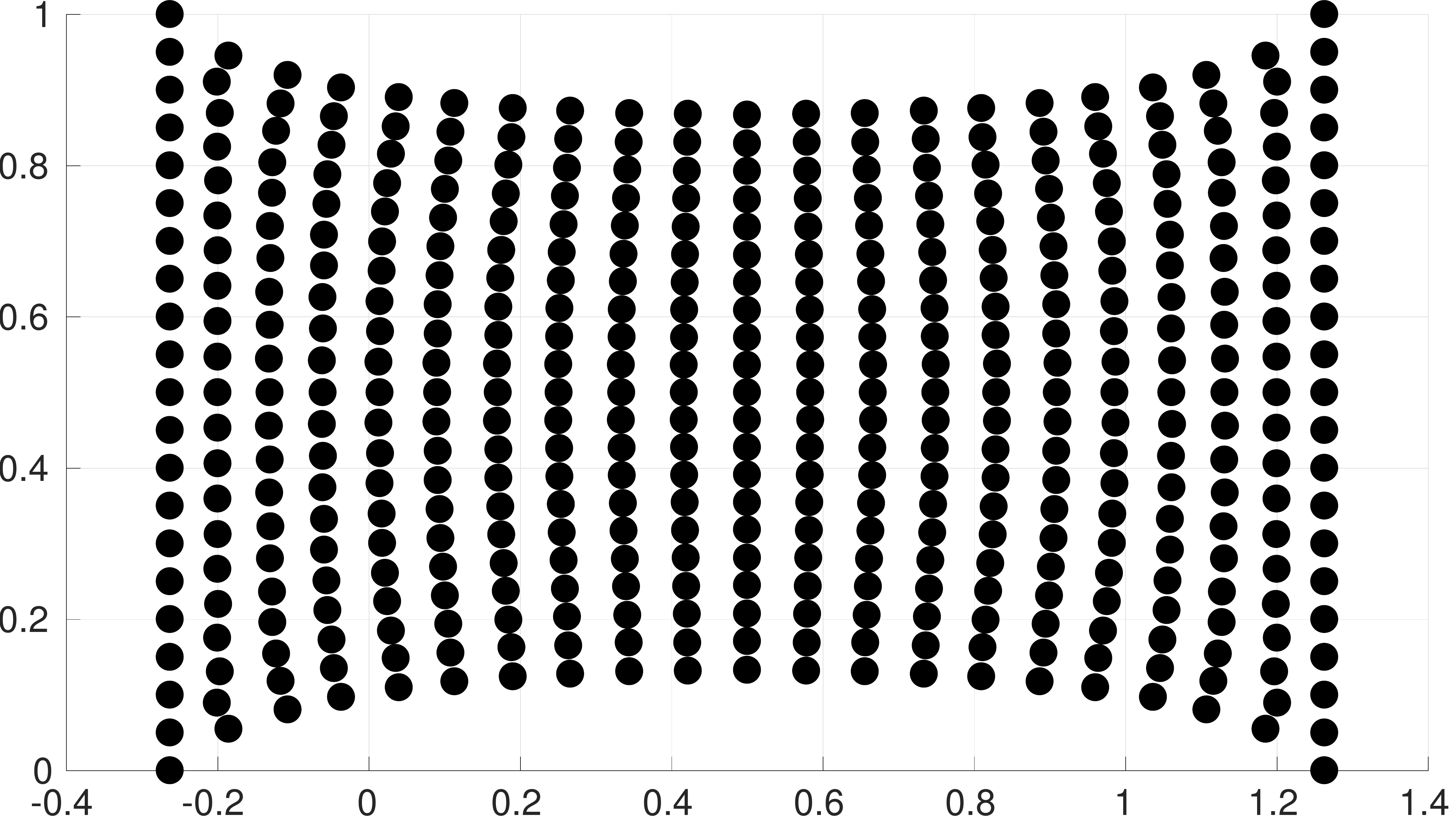} &
\includegraphics[width=0.3\textwidth]{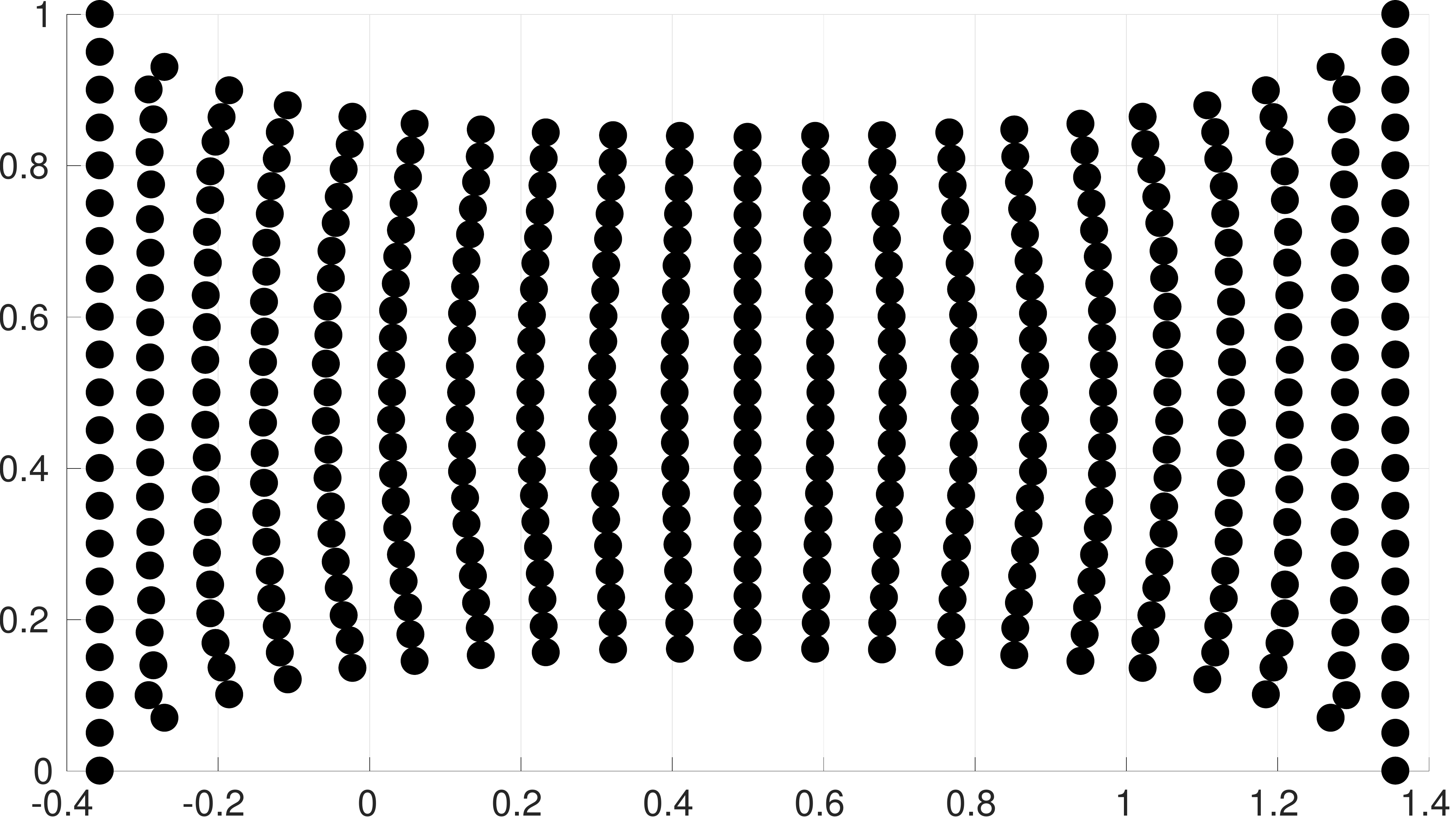} 
\end{tabular}
\caption{From left to right: state based model at failure, rate based model at the
same time step, rate based model at failure. About 40\% more extension can be resolved
using the rate based model.}
\label{fig:ratevsstate}
\end{figure}

Another topic not discussed so far is frame invariance. It is clear that the 
Cauchy stress tensor transforms correctly under rotation:
\begin{align}
\uuline{\sigma_r} = \uuline{Q}\cdot\uuline\sigma\cdot\uuline{Q}^T
\end{align}
However, the same is not true for the rate of the Cauchy stress tensor:
\begin{align}
\dot{\uuline{\sigma_r}} &= \uuline{\dot{Q}}\cdot\uuline\sigma\cdot\uuline{Q}^T + \uuline{Q}\cdot\dot{\uuline\sigma}\cdot\uuline{Q}^T + \uuline{Q}\cdot\uuline\sigma\cdot\uuline{\dot{Q}}^T \\
&\neq \uuline{Q}\cdot\dot{\uuline\sigma}\cdot\uuline{Q}^T
\end{align}
The same holds for the deviatoric part $\uuline{S}$ of $\uuline{\sigma}$. To address
this problem, an objective stress rate, a corotational derivative of the stress instead of the time derivative must be used. The most popular
choice, both in SPH literature and in general, is probably the Jaumann-Rate \cite{jaumann1911geschlossenes}:
\begin{align}
\mathring{\uuline{S}} &= \dot{\uuline{S}} + \uuline{S}\cdot\uuline\omega - \uuline\omega\cdot\uuline{S} \\
\uuline\omega &= \frac{1}{2}\left(\uuline{L} - \uuline{L}^T\right)
\end{align}
Well known alternatives include the Green Naghdi \cite{green1971some,naghdi1961time} 
and the Truesdell \cite{truesdell1966continuum} rate. An overview over objective 
stress rates is given in \cite{wegener1991berechnung}. Consequently, Hooke's Law \eqref{hookedev}
is now plugged into $\mathring{\uuline{S}}$:
\begin{align}
2\mu\left(\uuline{\dot\varepsilon} - \frac{1}{3}\cdot\uuline{I}\cdot tr(\uuline{\dot\varepsilon})\right) &= \dot{\uuline{S}} + \uuline{S}\cdot\uuline\omega - \uuline\omega\cdot\uuline{S} \label{eq_deviatoric}
\end{align}
Rearranging for $\uuline{\dot{S}}$ combined with \eqref{eos1} yields the complete
constitutive update:
\begin{align}
\dot{\uuline{S}} &= 2\mu\left(\uuline{\dot\varepsilon} - \frac{1}{3}\cdot\uuline{I}\cdot tr(\uuline{\dot\varepsilon})\right) - \uuline{S}\cdot\uuline\omega + \uuline\omega\cdot\uuline{S} \label{update_ul_start} \\
p &= c_0^2(\varrho_0-\varrho) \\
\uuline{\sigma} &= -p\uuline{I} + \uuline{S} \label{update_ul}
\end{align}
Since it is easy to convert between $\uuline{\sigma}$ and $\uuline{P}$:
\begin{align}
J\cdot\uuline{\sigma} = \uuline{P}\cdot\uuline{F}^T \label{cauchy_to_nominal}
\end{align}
this material model is applicable to both total Lagrangian as well as updated
Lagrangian frameworks. Slightly inconvenient is that the material model contains
derivatives with regard to the current configuration $x$, namely the velocity. 
A total Lagrangian model should only feature derivatives with respect to $X$, the
reference configuration, though. To this end, the following identity can be used:
\begin{align}
\uuline{L} = \uuline{\dot{F}}\;\uuline{F}^{-1} \label{ftol}
\end{align}
The proof for this identity is very short:
\begin{align}
\uuline{L}\cdot\uuline{F} &= \dot{\uuline{F}} \\
\frac{\partial v_i}{\partial x_k} \frac{\partial x_k}{\partial X_j} &= \frac{\partial v_i}{\partial X_j} \\
\frac{\partial v_i}{\partial X_j} &= \frac{\partial v_i}{\partial X_j}
\end{align}
It might be argued that a material model of the form:
\begin{align}
\uuline{\sigma} &= \underline{\underline{\underline{\underline{C}}}}:\uuline{E} \label{stress_state} \\
\uuline{E} &= \frac{1}{2}\left[(\nabla^0 \uline{u})^T + (\nabla^0 \uline{u}) + (\nabla^0 \uline{u})^T \cdot (\nabla^0 \uline{u}) \right] \label{green_st_venant} \\
\uline{u} &= \uline{x} - \uline{X} \label{update_tl}
\end{align}
is more natural for total Lagrangian algorithms. While this may be true it was 
found in preliminary studies that ``rate based'' models like \eqref{update_ul_start}-\eqref{update_ul}
are more stable in a lot of situations. Such a situation is given in figure \ref{fig:ratevsstate}.
The total Lagrangian algorithm in section \ref{sec:a3} was run twice, once with a rate based
model as in \eqref{stress_state} and once using a model like \eqref{update_tl}, i.e. ``state based''. 
The latter model severely overestimates the contraction when compared to a FEM simulation 
and fails due to excess oscillatory modes about 40\% earlier. 
While a formal proof of this can't be provided, it makes sense on an
intuitive level. Say a spurious mode, like a zero energy mode with amplitude $A$
is imposed on $\uline{u}$ at a single time step. Using \eqref{update_tl} $\uuline{\sigma}$ is affected with
the same amplitude $A$. Alternatively, let $\uline{v}$ be disturbed with the same
spurious mode using \eqref{update_ul}. The deviatoric stress tensor $\uuline{S}$ and
thus $\uuline{\sigma}$ will only be affected by a mode with amplitude $c\cdot\delta t \cdot A$,
the constant $c$ being dependent on the time stepper used. It seems clear that
spurious modes need to be present in the solution over multiple time steps to significantly
affect the stress tensor using \eqref{update_ul}. Over these multiple time steps
those spurious frequencies may be canceled out or suppressed by stabilization 
measures like the artificial viscosity, hour glassing control or smoothing schemes. 

\subsection{Spatial Discretization and Boundary Conditions}
The canonical way of discretizing system \eqref{ulcont},\eqref{ulmoment} using SPH is
as follows:
\begin{align}
<\dot{\uline{v}}>_i &= \sum_{j=1}^{N} \left( \frac{\uuline{\sigma_j}}{\varrho_j^2} + \frac{\uuline{\sigma_i}}{\varrho_i^2} \right) \nabla W_{ij} m_j \label{disc31} \\
<\dot{\varrho}>_i &= \varrho_i \sum_{j=1}^{N} \left( \uline{v}_j - \uline{v}_i \right) \cdot \nabla W_{ij} \frac{m_j}{\varrho_j} \label{disc32} \\
<\uuline{L}>_i &= \sum_{j=1}^{N} \left( \uline{v}_j - \uline{v}_i \right) \otimes \nabla W_{ij} \frac{m_j}{\varrho_j} \label{disc33} 
\end{align}
Where the last equation is needed for the material model. This way the system conserves mass, since there is no evolution equation 
for $m$, i.e. the mass of any particle is simply never changed. It is also Galilean
invariant, since a pure translation field implies $\uline{v}=\uline{v}_i=\uline{v}_j$.
Thus, both the right hand sides for \eqref{disc32} and \eqref{disc33} remain zero. 
No pressure is induced due to \eqref{eos1} and no deviatoric stresses due to \eqref{eq_deviatoric}. Finally, the
system conserves linear momentum since \eqref{disc31} is anti-symmetric about the indices $i,j$:
The linear momentum of a discrete system can be written as
\begin{align}
\uline{P} = \sum_{i=1}^N m_i\uline{v}_i
\end{align}
The change in linear momentum is thus:
\begin{align}
\dot{\uline{P}} = \sum_{i=1}^N m_i\dot{\uline{v}}_i
\end{align}
inserting into \eqref{disc31}:
\begin{align}
\dot{\uline{P}} = \sum_{i=1}^N \sum_{j=1}^{N} \left( \frac{\uuline{\sigma_j}}{\varrho_j^2} + \frac{\uuline{\sigma_i}}{\varrho_i^2} \right) \nabla W_{ij} m_j m_i
\end{align}
rearranging the sum:
\begin{align}
\dot{\uline{P}} = \frac{1}{2}\sum_{i=1}^N \sum_{j=1}^{N} \left[ \left( \frac{\uuline{\sigma_j}}{\varrho_j^2} + \frac{\uuline{\sigma_i}}{\varrho_i^2} \right) \nabla W_{ij} m_j m_i + \left( \frac{\uuline{\sigma_i}}{\varrho_i^2} + \frac{\uuline{\sigma_j}}{\varrho_j^2} \right) \nabla W_{ji} m_i m_j \right]
\end{align}
since $\nabla W_{ij} = -\nabla W_{ji}$ the term above becomes zero and liear momentum
is conserved. However, in corrected SPH $\nabla W_{ij} \neq -\nabla W_{ji}$. This is quite troublesome, since,
consequently, corrected SPH does not conserve linear momentum. However, it is necessary
to use at least linear order complete, i.e. corrected SPH, to correctly resolve rotations.
Otherwise, \eqref{disc32}, \eqref{disc33} will not be constant for a linear velocity field,
inducing unphysical stresses. Thus, an SPH scheme which is both conserving linear 
momentum and able to exactly resolve rigid body motion seems impossible, 
at least on the strong form. A sentiment which is also discussed in 
\cite{price2003smoothed}. 

Beside this lack of consistency and other concerns like lack of integrability and lack or lack of stability mentioned in  \cite{belytschko1998completeness}
an often mentioned sore point in SPH in particular, and meshless methods in general
is the the lack of a canonical way to apply boundary conditions, essential or 
otherwise. Essential boundary conditions are often applied by just overwriting the
current state variables of the particles affected, ignoring any consistency with
approximator \eqref{eq:3}. Consistency regarding \eqref{eq:3} can be restored using
the procedure outlined in \cite{jun1998explicit} if required. 

More critical is probably that systems \eqref{ulcont},\eqref{ulmoment} and \eqref{tlcont},\eqref{tlmoment}
are only valid in an unbound domain. In a bound domain the following constraint
has to be met on traction free boundaries:
\begin{align}
\uuline\sigma\cdot\uline{n} = \uline{0} \label{stress_free}
\end{align}
If \eqref{disc31} is used, this constraint is approximately true since the stress
field will tend to go to zero at the boundary due to the lack of zero order completeness of 
\eqref{disc31}, see \cite{randles2000normalized}. Vice versa, if any corrected scheme is
employed, \eqref{stress_free} can not be ignored anymore, even though there are
some works that do, e.g. \cite{reveles2007development}. There are a number of
ways to meet this constraint:
\begin{itemize}
\item by treating \eqref{stress_free} as a system of linear equations. 
The system is underdetermined, i.e. there are an infinite number of solutions. In \cite{randles2000normalized} the system is solved by rotation of the stress
tensor along $\uline{n}$, then zeroing the in plane stresses and stresses along $\uline{n}$,
while the authors in \cite{ostad2008field} solve \eqref{stress_free} by elimination 
of variables.
\item by seeding a thin layer of ghost particles along the domain, manipulating their
stress tensors such that \eqref{stress_free} holds, see \cite{hieber2008lagrangian}.
\item by adding a body force of magnitude $-\uuline\sigma\cdot\uline{n}$ to boundary particles as in
\cite{hieber2004remeshed}.
\end{itemize}
Probably the cleanest option and one not discussed so far is to use the weak form
of systems \eqref{ulcont},\eqref{ulmoment} or \eqref{tlcont},\eqref{tlmoment}, respectively.
Taking \eqref{ulmoment} as an example, the weak form can be found by multiplying the
equation with a so called test function $\delta\uline{v}$ and integrating over the 
whole current domain:
\begin{align}
\int  \delta\uline{v} \varrho \dot{\uline{v}} d\Omega &=  \int \delta\uline{v} \nabla \uuline{\sigma} d\Omega \label{weak1}
\end{align}
The test function will later be approximated using a shape function:
\begin{align}
\delta\uline{v}_i = \uline{v}_i N_i(\uline{x})
\end{align}
if $N_i(\uline{x}_j) = 0$ one obtains FEM methods. If $N_i(\uline{x}_j) \neq 0$
EFG methods are obtained \cite{belytschko1994element}. Since $N_i(.)$ is known 
analytically in either case, the goal is to manipulate \eqref{weak1} such that the
spatial derivative term $\nabla$ is only applied to $\delta\uline{v}$. To this end, the term
on the right hand side is expanded using the product rule:
\begin{align}
\int \delta\uline{v} \left(\nabla \uuline{\sigma}\right) d\Omega &= \int \left[ \nabla\left( \delta\uline{v} \uuline{\sigma}\right) - \left( \nabla \delta\uline{v}\right) \uuline{\sigma} \right] d\Omega \label{weak2}
\end{align}
The Gauss theorem can be used on the first term on the right hand side of the 
above to yield:
\begin{align}
\int \left( \nabla \delta\uline{v} \uuline{\sigma}\right) d\Omega  = \int \delta\uline{v} \left( \uuline{\sigma} \cdot \uline{n} \right) d\Gamma
\end{align}
which contains \eqref{stress_free}. Thus, if no boundaries with prescribed tractions $\uline{t}$ are present,
this term can just be omitted from any future derivations. Otherwise $\uuline{\sigma}\cdot\uline{n} = \uline{t}$:
\begin{align}
\int \left( \nabla \delta\uline{v} \uuline{\sigma}\right) d\Omega  = \int \delta\uline{v} \left( \uuline{\sigma} \cdot \uline{n} \right) d\Gamma = \int \delta\uline{v} \cdot \uline{t} d\Gamma
\end{align}
The only term left containing $\nabla\uuline\sigma$ has now been replaced, which
was the goal at hand. Inserting \eqref{weak2} back into \eqref{weak1}:
\begin{align}
\int  \delta\uline{v} \varrho \dot{\uline{v}} d\Omega &= \int \left[ \nabla\left( \delta\uline{v} \uuline{\sigma}\right) - \left( \nabla \delta\uline{v}\right) \uuline{\sigma} \right] d\Omega \\
&= \int \delta\uline{v} \cdot \uline{t} d\Gamma - \int \left( \nabla \delta\uline{v}\right) \uuline{\sigma} d\Omega \label{weak_final_ul}
\end{align}
This derivation of the weak form for the updated Lagrangian momentum equation
was not complete, since body forces where ignored. It was also not entirely
rigorous since the issue of non-continuous stresses inside the body has been left out.
For a complete review consider \cite{belytschko2013nonlinear}.

\section{Implemented Algorithms}

It is not possible to exhaustively implement all the meshless algorithms for 
dynamic solid mechanics published in the literature. Still, a broad selection 
was attempted. As was demonstrated in the last section, there are updated Lagrangian
(UL) and total Lagrangian (TL) formulations for the problem at hand, as well
as strong and weak formulations. It seems imperative to implement at least one
algorithm of each combination for a meaningful review. There also needs to be a
baseline against which each algorithm's performance can be compared to.\par
The final list of algorithms implemented for this paper is thus:
\begin{itemize}
\item Reference Algorithms
    \begin{itemize}
        \item Updated Lagrangian FEM
        \item Total Lagrangian FEM
    \end{itemize}
\item Updated Lagrangian, Strong Form
    \begin{itemize}
        \item Gray and Monaghans stress corrected algorithm in \cite{gray2001sph}
        \item Godunov SPH according to Parshikov et al. \cite{parshikov2002smoothed}
    \end{itemize}    
\item Total Lagrangian, Strong Form    
    \begin{itemize}
        \item Reveles Total Lagrangian algorithm presented in \cite{reveles2007development}
        \item An algorithm based on the elastic potential force instead of the standard momentum equation, inspired by \cite{muller2004point}
        \item An algorithm inspired by the corotated formulation in \cite{becker2009corotated}
    \end{itemize}    
\item Updated Lagrangian, Weak Form
    \begin{itemize}
        \item An algorithm similar to the stress point algorithm in \cite{vignjevic2000treatment}
    \end{itemize}
\item Total Lagrangian, Weak Form
    \begin{itemize}
        \item The explicit RKPM given in \cite{jun1998explicit}
    \end{itemize}    
\end{itemize}
All of these works use a wide array of material models and time steppers. Since
the purpose of this work is not to implement the original works as close as possible,
but to compare their respective algorithmic performance,
both of these aspects where changed in most cases. Wherever possible, the constitutive
model given in \eqref{governing} is employed. In all cases the time stepper was 
replaced by a fourth order Runge-Kutta method.

There are also works which would make sense in this comparison, but could not
be reimplemented. One of them is \cite{ostad2012stabilized}. It seems to
be the only work published that applies corrected SPH directly to the strong form.
Another interesting approach is the one given in \cite{hieber2008lagrangian}. The remeshing idea presented therein seems to
be a strong paradigm to overcome tensile instabilities and avoids the whole problem
of choosing the smoothing length $h$ outlined in section \eqref{chosing_h}. However,
both of these works omit crucial implementation details and their results could
not be replicated here.

In the following a short description and discussion of each algorithm is given,
along with what was changed to the original implementation

\subsection{Algorithm 1: Gray and Monaghans Stress Corrected SPH} \label{sec:a1}
Monaghans paper \cite{gray2001sph} uses a number of correctors to overcome the tensile instability problem
present in most meshless methods. Two of these correctors, namely the artificial viscosity
\cite{monaghan1983shock} and XSPH \cite{monaghan1989problem}, have been used 
before. However, they propose a novel corrector to the stress tensor. Their idea
is to diagonalize the stress tensor, then add an artificial term to each component
which is positive, indicating tension. The usual momentum equation \eqref{disc31} is thus replaced by 
\begin{align}
\dot{\uline{v}}_i &= \sum_{j=1}^{N} \left( \frac{\uuline{\sigma}_j}{\varrho_j^2} + \frac{\uuline{\sigma}_i}{\varrho_i^2} + \Pi_{ij} + f^n\left(\uuline{R}_i + \uuline{R}_j \right) \right) \nabla W_{ij} m_j 
\end{align}
Where $\Pi$ is the artificial viscosity and $\uuline{R}$ are the newly proposed artificial stress terms. $f$ is a term depending on the particle spacing between $i$ and $j$, that is, $f$ increases the influence of the artificial stress terms if $i$, $j$ are close and vice versa. To find $\uuline{R}$, $\uuline{\sigma}$ needs to be diagonalized. That is, the nominal stress is computed. In 2D this can easily be achieved by the following procedure:

The nominal stress $\bar{\uuline\sigma}$ 
can be found using:
\begin{align}
2\theta = \frac{2\sigma_{xy}}{\sigma_{xx} - \sigma_{yy}}
\end{align}
And:
\begin{align}
\bar{\sigma}_{xx} &= c^2\sigma_{xx} + 2sc\sigma_{xy} + s^2\sigma_{yy} \\
\bar{\sigma}_{yy} &= s^2\sigma_{xx} + 2sc\sigma_{xy} + c^2\sigma_{yy}
\end{align}
with $s=\sin{\theta}$ and $c=\cos{\theta}$. If $\bar{\sigma}_{xx} > 0$ then:
\begin{align}
\bar{R}_{xx} = -\varepsilon \frac{\bar{\sigma}_{xx}}{\varrho^2}
\end{align}
the same holds for $\bar{\sigma}_{yy} > 0$. This modified stress is then rotated
back into the reference frame:
\begin{align}
R_{xx} &= c^2\bar{R}_{xx} + s^2\bar{R}_{yy} \\
R_{yy} &= s^2\bar{R}_{xx} + c^2\bar{R}_{yy} \\
R_{xy} &= sc(\bar{R}_{xx} - c^2\bar{R}_{yy})
\end{align}
The velocity gradient and continuity equation are discretized
using the usual approaches \eqref{disc32} \eqref{disc33}. The advection equation
\begin{align}
\uline{\dot{x}}_i = \uline{v}_i
\end{align}
is replaced with
\begin{align}
\uline{\dot{x}}_i = \uline{v}_i +\varepsilon \sum_{j=1}^N \frac{m_j}{\varrho_i + \varrho_j}(\uline{v}_j - \uline{v}_i)W_{ij}
\end{align}
i.e. the XSPH correction, as mentioned. 

The authors show that their algorithm is stable in a wide array of test cases. It
is not unconditionally stable under tension though, however in the tensile test
they investigate numerical fracture, i.e. tensile instability, occurs only beyond
rupture strength of the material at hand. An argument that this is true for 
all, or at least most materials is not given.

Another issue is the number of unphysical parameters in this algorithm, XSPH and 
artificial viscosity need four tunable parameters, their artificial
stress involves another two, raising the total number up to six. This is problematic,
since in a strict sense these algorithms would need to be tuned using experimental 
data. In \cite{johnson1996normalized} it was shown that the selection of artificial viscosity parameters
may alter the physics simulated quite severly. 

\subsection{Algorithm 2: Godunov SPH}  \label{sec:a2}

The idea behind Godunov SPH is the realization that in any SPH approximation formula
containing a gradient, for example \eqref{disc32},
\begin{align}
\dot{\varrho}_i &= \sum_{j=1}^{N} \left( \uline{v}_j - \uline{v}_i \right) \cdot \nabla W_{ij} \frac{m_j}{\varrho_j} \label{eq:godu1}
\end{align}
the gradient $\nabla W_{ij}$ can be rewritten as 
\begin{align}
\nabla W_{ij} = \frac{\uline{x}_{ij}}{|\uline{x}_{ij}|} \frac{\partial W_{ij} }{\partial x_{ij} }
\end{align}
by extracting the inner derivative. Reinserting into \eqref{eq:godu1}:
\begin{align}
\dot{\varrho}_i &= \sum_{j=1}^{N} \underbrace{\left( \uline{v}_j - \uline{v}_i \right) \cdot \frac{\uline{x}_{ij}}{|\uline{x}_{ij}|}}_{A} \frac{\partial W_{ij} }{\partial x_{ij} } \frac{m_j}{\varrho_j}
\end{align}
in geometrical terms the term marked $A$ means that velocities $\uline{v}_i$, $\uline{v}_j$ 
are projected onto the inter particle distance $\uline{x}_{ij}$ and can thus be regarded
as scalar (the length of their protection) initial values, piecewise constant, with a single 
discontinuity somewhere along line $\uline{x}_{ij}$. This, combined with a conservation
law, i.e. \eqref{ulcont} for the physical quantity projected is called a Riemann problem.
A wide array of Riemann solvers regarding such problems exists, both analytical and numerical \cite{puri2014approximate}.
The observation that SPH particle interactions can be recast into Riemann problems
was first made in \cite{monaghan1997sph}. The principle was then first applied to 
solid mechanics in \cite{parshikov2002smoothed}. 

The reasoning why this procedure should be beneficial to the solution quality is not made explicit in the original paper, however it is commonly argued, for example in \cite{connollysecond} that algorithms constructed this
way do not require any artificial damping terms like artificial viscosity. Certainly
a remarkable feature, especially considering algorithm \ref{sec:a1}. Additionally,
it was shown in \cite{mehra2012tensile} that using Godunov SPH can be beneficial
in impact problems. A second order accurate version regarding the Riemann problem
of this algorithm exists \cite{connollysecond} \cite{connolly2012godunov}, but 
there a time operator splitting procedure is required. It is not clear how to integrate
this procedure into a standard time stepper like the 4th order Runge Kutta method used throughout 
this paper and was thus not implemented. 

In it's structure the algorithm is very similar to \ref{sec:a1} but devoid of 
any stabilization terms. The standard continuity and momentum equations are replaced
by versions containing the Riemann solutions
\begin{align}
\dot{\varrho}_i &= -2\sum_{j=1}^N\frac{m_j\varrho_i}{\varrho_j h} \left( v_i^R - v_{ij}^{*R} \right) \frac{\partial W_{ij} }{\partial x_{ij} } \\
\dot{\uline{v}}_i &= -2\sum_{j=1}^N\frac{m_j}{\varrho_i\varrho_j h} \uline{\sigma}_{ij}^R \frac{\partial W_{ij} }{\partial x_{ij} }
\end{align}
where the superscript $()^R$ denotes a quantity that was projected onto the line
$\uline{x}_{ij}$. The velocity becomes thus a scalar whereas the stress becomes
a vector quantity. The superscript $()^*$ denotes a Riemann solution. In accordance
to the paper an acoustic approximation was chosen. For example for the velocity:
\begin{align}
v_{ij}^{*R} = \frac{v_j^R\varrho_j c_j + v_i^R\varrho_i c_i - p_j + p_i}{\varrho_j c_j + \varrho_i c_i}
\end{align}
for the exact details on how to find the projections $()^R$ and the solution for 
the Riemann problem regarding the momentum equation the reader is referred to the
original publication \cite{parshikov2002smoothed} or the source code provided with this
paper. 

\subsection{Algorithm 3: Total Lagrangian SPH on the Strong Form}  \label{sec:a3}

This algorithm is basically a corrected, total Lagrangian version of \ref{sec:a1}
and follows very closely the procedure outlined in \cite{reveles2007development}, 
except for the time stepper and material model. Consequently, system \eqref{disc31} \eqref{disc32} \eqref{disc33} 
is replaced by:
\begin{align}
\dot{\uline{v}}_i &= \sum_{j=1}^{N} \left( \frac{\uuline{P_j}}{\varrho_j^2} + \frac{\uuline{P_i}}{\varrho_i^2} + \Pi_{ij} \right) \nabla^0 \overset{\triangle}{W_{ij}} m_j \label{disctl1} \\
\varrho_i &= J\cdot\varrho_0 \label{disctl2} \\
\dot{\uuline{F}_i} &= \sum_{j=1}^{N} \left( \uline{v}_j - \uline{v}_i \right) \otimes \nabla^0 \overset{\triangle}{W_{ij}} \frac{m_j}{\varrho_j} \label{disctl3} \\
\uuline{F}_i &= \sum_{j=1}^{N} \left( \uline{u}_j - \uline{u}_i \right) \otimes \nabla^0 \overset{\triangle}{W_{ij}} \frac{m_j}{\varrho_j} + \uuline{I} \label{disctl4}
\end{align}
To find $\uuline{L}$ that goes into the material model \eqref{ftol} can be used. In
the system above, $\overset{\triangle}{W}$ is the kernel corrected by the Randles Libersky
Correction, as outlined in section \ref{sec:randles_libersky}. 

\subsection{Algorithm 4: Total Lagrangian SPH based on the elastic potential}  \label{sec:a4}

An interesting idea from the computer graphics community is to disregard the usual
momentum equations \eqref{ulmoment} \eqref{tlmoment} and instead observe that the 
strain energy density function in constitutive modeling is given by
\begin{align}
U = \frac{1}{2}(\uuline{\sigma} : \uuline{\varepsilon})
\end{align}
with $\uuline{\varepsilon}$ being some measure of strain. Generally speaking, the force \emph{per unit volume} 
due to some potential function $\omega$ at some point $\uline{x}_i$ is the 
gradient of that potential with respect to that points displacement $\uline{u}_i$ 
from it's rest state:
\begin{align}
\uline{\bar{f}}_i = -\nabla_{\uline{u}i} \omega
\end{align}
now setting $\omega = U$:
\begin{align}
\uline{\bar{f}}_i &= -1/2\nabla_{\uline{u}_i} \left(\uuline{\sigma} : \uuline{\varepsilon}\right) 
\end{align}
inserting Hooke's law for isotropic media in the form \ref{eq:hooke_lame}:
\begin{align}
\underline{\bar{f}}_i = -1/2\nabla_{\underline{u}{}_i}\left[ \left( 2\mu\cdot\underline{\underline{\varepsilon}} + \lambda \cdot tr(\underline{\underline{\varepsilon}})\cdot\underline{\underline{I}} \right):\underline{\underline{\varepsilon}} \right]
\end{align}
distributing $\underline{\underline{\varepsilon}}$, then $\nabla_{\underline{u}{}_i}$:
\begin{align}
\underline{\bar{f}}_i = -1/2\left[ \nabla_{\underline{u}{}_i} \left( 2\mu\cdot\underline{\underline{\varepsilon}}:\underline{\underline{\varepsilon}}\right) + \nabla_{\underline{u}{}_i}\left(\lambda \cdot tr(\underline{\underline{\varepsilon}})\cdot\underline{\underline{I}}:\underline{\underline{\varepsilon}}\right)  \right]
\end{align}
which equals
\begin{align}
\underline{\bar{f}}_i = -1/2\left[  4\mu\cdot\underline{\underline{\varepsilon}}:\nabla_{\underline{u}{}_i} \underline{\underline{\varepsilon}} + 2\lambda \cdot tr(\underline{\underline{\varepsilon}})\cdot\underline{\underline{I}}:\nabla_{\underline{u}{}_i}\underline{\underline{\varepsilon}} \right]
\label{eq:potential_partial}
\end{align}
While 
\begin{align}
\nabla_{\underline{u}{}_i} \left( 2\mu\cdot\underline{\underline{\varepsilon}}:\underline{\underline{\varepsilon}}\right) = 4\mu\cdot\underline{\underline{\varepsilon}}:\nabla_{\underline{u}{}_i} \underline{\underline{\varepsilon}}
\end{align}
is quite straight forward
\begin{align}
\nabla_{\underline{u}{}_i}\left(\lambda \cdot tr(\underline{\underline{\varepsilon}})\cdot\underline{\underline{I}}:\underline{\underline{\varepsilon}}\right) = 2\lambda \cdot tr(\underline{\underline{\varepsilon}})\cdot\underline{\underline{I}}:\nabla_{\underline{u}{}_i}\underline{\underline{\varepsilon}}
\end{align}
is best understood by simply expanding the components, e.g. in 2D:
\begin{align} 
\lambda\nabla_{\underline{u}{}_i}\left((\varepsilon_{xx} + \varepsilon_{yy})\varepsilon_{xx} + (\varepsilon_{xx} + \varepsilon_{yy})\varepsilon_{yy} \right) &=  \nonumber \\ \lambda\begin{bmatrix}
2\cdot(\partial/(\partial u^i_x) \varepsilon_{xx})\varepsilon_{xx} +  2\cdot(\partial/(\partial u^i_x) \varepsilon_{xx})\varepsilon_{yy} \nonumber \\ 
2\cdot(\partial/(\partial u^i_y) \varepsilon_{yy})\varepsilon_{xx} +  2\cdot(\partial/(\partial u^i_y) \varepsilon_{yy})\varepsilon_{xx}
\end{bmatrix} &= \\
2\lambda \cdot tr(\underline{\underline{\varepsilon}})\cdot\underline{\underline{I}}:\nabla_{\underline{u}{}_i}\underline{\underline{\varepsilon}}
\end{align}
Factoring out $2\cdot \nabla_{\underline{u}{}_i}\underline{\underline{\varepsilon}}$ from \ref{eq:potential_partial} yields:
\begin{align}
\underline{\bar{f}}_i &= \nonumber -\left[  2\mu\cdot\underline{\underline{\varepsilon}} + \lambda \cdot tr(\underline{\underline{\varepsilon}})\cdot\underline{\underline{I}} \right]:\nabla_{\underline{u}{}_i} \underline{\underline{\varepsilon}} \\
&= -\underline{\underline{\sigma}}:\nabla_{\underline{u}{}_i} \underline{\underline{\varepsilon}}
\label{eq:potential_start}
\end{align}
After a lot of quite involved simplification and reordering steps, see \ref{sec:app1}, the following expression is retrieved for force $\underline{f}_j$ on particle $j$ enacted by particle $i$:
\begin{align}
\underline{f}_i = -\;2 \; \sum_{j=1}^N  \underline{\underline{F}}{}_i \cdot \underline{\underline{\sigma}}{}_i \cdot \nabla \overset{\triangle}{W}_{ij} \cdot m_i/\varrho_i
\label{eq:potential_final}
\end{align}
The implementation in this paper differs in the following two aspects: A Randles-Liberksy (see \ref{sec:randles_libersky})
corrected version of the usual cubic spline kernel is employed, to keep the method
as comparable as possible to the other ones implemented. Additionally, the
interaction forces are symmetrized as proposed in \cite{becker2009corotated}. See \eqref{symm},
but without any rotations.
The algorithm at hand needs to do remarkably little work per time step, basically
the precomputed corrected kernel functions are used to obtain an approximation 
to \eqref{green_st_venant}, then \eqref{eq:potential_final} is computed for each interacting 
particle. The force can then be converted to an acceleration and is supplied to the
time integration routine. 

\subsection{Algorithm 5: Co-rotated Total Lagrangian SPH} \label{sec:a5}

Co-rotated methods are well known within a FEM context. The main idea of these methods
is to replace the standard Cauchy stress tensor $\uuline{\sigma}$ with a co-rotated
version
\begin{align}
\uuline{\hat{\sigma}} = \uuline{R}^T\uuline\sigma\;\uuline{R}
\end{align}
the rotation is usually found by way of the polar decomposition of deformation 
gradient $\uuline{F}$
\begin{align}
\uuline{F} = \uuline{R}\;\uuline{U}
\end{align}
which decomposes $\uuline{F}$ into a purely rotational part $\uuline{R}$ and a 
purely stretching part $\uuline{U}$. However, the authors in \cite{becker2009corotated} suggest
using a purely geometrical consideration to extract $\uuline{R}$ based on a 
shape matching procedure presented originally in \cite{muller2005meshless}. 
They calculate a local transformation matrix $\uuline{A}$:
\begin{align}
\uuline{A}{}_i =  \sum_{j=1}^{N} m_j W_{íj} (\uline{x}_j-\uline{x}_i) \otimes (\uline{X}_j-\uline{X}_i)
\end{align}
and extract $\uuline{R}$ by means of polar decomposition of $\uuline{A}$ instead
of $\uuline{F}$. They proceed to employ this rotation matrix to compute a co-rotated
deformation gradient $\uuline{\hat{F}}$:
\begin{align}
\uuline{\hat{F}} &= \sum_{j=1}^{N} \uline{\hat{u}}_{ij} \nabla^0 W_{ij}\frac{m_j}{\varrho_j} + \uuline{I} \\
\uline{\hat{u}}_{ij} &= \uuline{R}^{-1}(\uline{x}_j-\uline{x}_i) - (\uline{X}_j-\uline{X}_i)
\end{align}
they then use a potential based approach very similar to \ref{sec:a4} but using
the Cauchy Green strain tensor instead of the Green St. Venant:
\begin{align}
\uuline{\varepsilon} &= \frac{1}{2}(\nabla^0 \uline{u} + \nabla^0  \uline{u}^T)
\end{align}
trying to stay close to the other methods it was decided to use the standard momentum
equation for the computation of particle interaction forces $\uline{\bar{f}}$ in this paper. Since
the displacement was computed in corotated fashion, the forces computed need to account
for this rotation
\begin{align}
\uline{f}_i = \frac{-\uuline{R}{}_i\uline{\bar{f}}_{ji} + \uuline{R}{}_j\uline{\bar{f}}_{ji}}{2} \label{symm}
\end{align}
also, as demonstrated in the equation above a symmetrization procedure is added. 
This algorithm is the only one that uses a constitutive update of the form \eqref{update_tl}
since altering it to rate based models is not straight forward. Corotating the velocity
gradient $\uuline{L}$ combined with \eqref{symm} did not yield a stable result.

The remarkable feature about this algorithm is that rotations are resolved correctly,
even though only zero order complete kernel functions in conjunction with the
Cauchy Green strain tensor are employed. 

\subsection{Algorithm 6: Total Lagrangian, weak} \label{sec:a6}

This algorithm is a slightly simplified version of \cite{jun1998explicit}. Contrary 
to all algorithms discussed so far it works on the weak form of the governing
equations. Thus, \eqref{weak_final_ul} is discretized using RKPM kernel functions,
albeit in total Lagrangian fashion:
\begin{align}
\int_{\Omega_0} \left( \left( \nabla^0 \delta \uline{u} \right) \uuline{P} + \delta\uline{u} \varrho_0\ddot{\uline{u}} \right) d\Omega_0 = 0 \label{weak_final_tl}
\end{align}
traction boundaries and body forces have been assumed zero in the above formulation.
Just as in FEM, the displacement $\uline{u}$ is approximated by means of shape
functions $N(.)$ at particle $i$
\begin{align}
\uline{u} = \uline{u}_i N_i
\end{align}
And for the test functions $\delta\uline{u}$
\begin{align}
\delta\uline{u} = \delta\uline{u}_i M_i
\end{align}
In \cite{jun1998explicit} both $M$ and $N$ are set to $\overset{\circ}{W}$, which are
the RKPM corrected kernels discussed in \ref{sec:RKPM}. Note that, contrary to 
FEM, these kernel functions are not restricted to the elements containing $i$ and
$W_i(\uline{x}_j) \neq 0$ for $i \neq j$. The discretization of 
\eqref{weak_final_tl} for some particle $i$ reads:
\begin{align}
\delta \uline{u}_i \int_{\Omega_0} \left( \left( \nabla^0 \overset{\circ}{W_i} \right) \uuline{P} \right) d\Omega_0 + \delta \uline{u}_i \int_{\Omega_0} \left( \varrho_0 \overset{\circ}{W_i} \overset{\circ}{W_j} \ddot{\uline{u}}_j \right) d\Omega_0 = 0 
\end{align}
since $\delta \uline{u}_i$ is arbitrary:
\begin{align}
\underbrace{\int_{\Omega_0} \left( \left( \nabla^0 \overset{\circ}{W_i} \right) \uuline{P} \right) d\Omega_0}_{\uline{f}_{\textrm{int}}^{i}} + \ddot{\uline{u}}_j \underbrace{\int_{\Omega_0} \left( \varrho_0 \overset{\circ}{W_i} \overset{\circ}{W_j} \right) d\Omega_0}_{\uuline{M}^{i}} = 0 
\label{weak_disc_tl}
\end{align}
if one would solve this for all $i$ one would arrive at a linear system of the form
\begin{align}
\uuline{M}\uline{\ddot{u}} = -\uline{f}_{\textrm{int}}
\end{align}
of size $\mathcal{N}$ where $\mathcal{N}$ is the number of degrees of freedom. Even though $\uuline{M}$
is sparse solving such a linear system is quite costly. It is thus common to 
simplify or even diagonalize matrix $\uuline{M}$. This technique  is 
known as ``mass lumping''. Row sum mass lumping can be achieved by simplifying 
\eqref{weak_disc_tl} to: 
\begin{align}
\int_{\Omega_0} \left( \left( \nabla^0 \overset{\circ}{W_i} \right) \uuline{P} \right) d\Omega_0 + \ddot{\uline{u}}_j \int_{\Omega_0} \left( \varrho_0 \overset{\circ}{W_j} \right) d\Omega_0 = 0 
\end{align}
the physical interpretation of which is to completely neglect inertial action between
lumped masses \cite{carlos2013lecture}.

 A short digression regarding the difference between weak and strong forms in a meshless
context follows: A simple attempt at discretizing the above further with regard to the numerical 
quadrature yields:
\begin{align}
\int_{\Omega_0} \left( \left( \nabla^0 \overset{\circ}{W_i} \right) \uuline{P} \right) d\Omega_0 + \ddot{\uline{u}}_j \underbrace{\int_{\Omega_0} \left( \varrho_0 \overset{\circ}{W_j} \right) d\Omega_0}_{\approx m_j} &= 0 \\
-\sum_{j=1}^{N} \left( \left( \nabla^0 \overset{\circ}{W}_{ij} \right) \uuline{P}_j \right) \frac{m_j}{\varrho_j} &= \ddot{\uline{u}}_j m_j
\end{align}
where the usual Riemann sum approach was used. This equation is very close, but
\emph{not} equal to a SPH discretization of \eqref{tlmoment} with approach \eqref{eq:4}
\begin{align}
\sum_{j=1}^{N} \left( \left( \nabla^0 \overset{\circ}{W}_{ij} \right) \uuline{P}_j \right) \frac{m_j}{\varrho_j} &= \ddot{\uline{u}}_i \varrho_i
\end{align}
from an implementation perspective the index on the right hand side is especially
crucial. 

However, since the reference configuration never changes, a more sophisticated
approach regarding the numerical quadrature is usually chosen in algorithms like
in \cite{jun1998explicit} or similar, e.g. \cite{belytschko1994element}, namely
Gaussian quadrature. To this end, a distinct set of quadrature points $Q$ is added to 
the particle locations. Their location relative to the particles is dictated by
the rules of Gaussian quadrature, as is their integration weight $w_q$. The number
of Gauss points per element, also known as the degree of the Gauss quadrature, 
is an input parameter. In this work all simulations using this algorithm employ four
Gauss points per (rectangular) element. The final algorithm:
\begin{center}
\begin{tabular}{l | p{10cm}}

1 & Compute deformation gradient and it's rate: 

$\begin{aligned}
\underline{\underline{F}}{}_q &= \sum_{i=1}^N \nabla^0 \overset{\circ}{W}_{iq} \underline{u}_i \frac{mi}{\varrho_i} \\
\underline{\underline{\dot{F}}}{}_q &= \sum_{i=1}^N \nabla^0 \overset{\circ}{W}_{iq} \underline{v} \frac{mi}{\varrho_i}
\end{aligned}$
which, this time around, \emph{is} in fact the same as discretizing 
$\begin{aligned}
\underline{\underline{F}} &= \nabla^0 \cdot \underline{u} + \underline{\underline{I}} \\
\underline{\underline{\dot{F}}} &= \nabla^0 \cdot \underline{v}
\end{aligned}$
by \eqref{eq:4} \\

\hline

2 & employ material model \eqref{update_ul} and turn $\underline{\underline{\sigma}}$ into $\underline{\underline{P}}$ using
\eqref{cauchy_to_nominal}. \\

\hline

3 & update internal forces:

$\begin{aligned}
\underline{f}_i \leftarrow \underline{f_i} + \nabla^0  \overset{\circ}{W}_{iq} \underline{\underline{P}}_q w_q
\end{aligned}$ \\

\hline

4 & Solve system 

$\begin{aligned}
\underline{\underline{M}}\;\underline{\ddot{u}} - \underline{f}_{\textrm{int}} = 0 
\end{aligned}$

using either the full or diagonalized $\underline{\underline{M}}$. \\

\hline

5 & turn the \emph{virtual} accelerations into the actual ones

$\begin{aligned}
\underline{\ddot{u}}_i &= \sum_{j=1}^N \overset{\circ}{W}_{ij} \underline{\ddot{u}}_j \frac{m_j}{\varrho_j} 
\end{aligned}$
\end{tabular}
\end{center}

The last step is necessary since the kernel functions used do not possess the 
Kronecker delta property, while a system of linear equations is 
solved using said kernel functions as a basis \cite{nguyen2008meshless}. There 
is two things to note about this: First, a lot of implementations seem to disregard
this step, e.g. \cite{zhang2014three} \cite{li2001element}. Second, both the
XSPH scheme \cite{monaghan1989problem} and the velocity field smoothing 
procedure in \cite{ostad2008field}\cite{ostad2012stabilized} are remarkably 
similar to this step. For XSPH:
\begin{align}
\uline{\dot{x}}_i = \uline{v}_i +\varepsilon \sum_{j=1}^N \frac{m_j}{\varrho_i + \varrho_j}(\uline{v}_j - \uline{v}_i)W_{ij}
\label{xsph}
\end{align}
This is the same as the last step in the algorithm above with averaged density and a user defined parameter
$\varepsilon$. As for the velocity field smoothing:
\begin{align}
\uline{\dot{x}}_i &= \sum_{j=1}^N \overset{\triangle}{W}{}^{\bar{h}}_{ij} \uline{v}_j \frac{m_j}{\varrho_j}d \\
\bar{h} &=\varepsilon h
\end{align}
which is similar to \eqref{xsph}, except the amount of smoothing is directed by scaling of the
smoothing length instead of linear combination as in XSPH. 

It is worthwhile to finish with a note on terminology. The algorithm just described
is basically the algorithm in \cite{belytschko1994element} for elasto-dynamics 
instead of elasto-statics, and is thus thus really close to \cite{li2001element}, except
for the fact that RKPM kernels are used instead of Moving Least Squares (MLS) ones. Both \cite{belytschko1994element} and \cite{li2001element}
call the methods they present Element Free Galerkin (EFG) methods. The RKPM and MLS
shape functions differ in their construction by either using a shifted or non-shifted
basis. However, it turns out that, mathematically, the approaches are equal 
\cite{jin2001equivalence}. However, in \cite{jin2001equivalence} it is also shown
that numerically, the non-shifted, i.e. MLS/EFG approach quickly becomes unstable as
$\mathcal{P}$ grows, where $\mathcal{P}$ is the degree of polynomial basis chosen. In fact, all works
regarding EFG seem to use a linear basis. In conclusion, weak form RKPM and EFG
approaches are mathematically equivalent, although their mathematical motivation
differs, similar to the CSPM vs Randles-Liberksy corrections discussed in section 
\ref{sec:randles_libersky}.

\subsection{Algorithm 7: Updated Lagrangian, weak} \label{sec:a7}

Essentially, this algorithm is equal to \ref{sec:a6} but using an updated Lagrangian
framework instead of a total Lagrangian one. The most pressing issue in such a 
formulation is the issue of numerical quadrature in general and how to move the
quadrature points in particular. At least three options are possible:
\begin{itemize}
\item
Reseed the quadrature points after every time step to conform to Gaussian quadrature 
rules. This entails that some sort of connectivity between the particles is maintained.
Algorithms of that nature can thus not be called truly mesh-free. In fact they might
need a remeshing of the background mesh if heavy distortions are encountered
\item
Move quadrature points with an interpolated velocity and settle for Riemann quadrature
\item
Work completely without quadrature points and solve the weak form equations on the particles
\end{itemize}
In an effort to minimize zero energy modes that version three would entail, but to retain the meshless nature of the algorithm
at hand the second option was chosen. The procedure
for interpolating the velocity from the particles onto the quadrature points is quite straight forward:
\begin{align}
\uline{v}_q &= \sum_{i=1}^N \overset{\circ}{W_{iq}} \uline{v}_i \frac{m_i}{\varrho_i} \label{vel_interp}
\end{align}
otherwise the algorithm proceeds exactly as outlined in section \eqref{sec:a6}, 
using the updated Lagrangian equations instead of the total ones, rendering the
algorithm very similar to the one published in \cite{vignjevic2000treatment}. 
There are two noteworthy differences however, the authors in \cite{vignjevic2000treatment}
use a novel kernel corrector whereas this implementation relies on the RKPM correction.
Second, and probably more crucial, they employ stress points instead of quadrature points 
and operate on the strong form. 

To highlight the difference between quadrature points on the weak form and stress points on 
the strong form the listing from \ref{sec:a6} is repeated in abbreviated form, 
as if stress points were used instead of quadrature points and the strong form was 
employed in an updated Lagrangian framework:
\begin{enumerate} 
\item
\begin{align}
\uuline{\dot{L}}_q &= \sum_{i=1}^N \nabla_{x} \overset{\circ}{W_{iq}} \uline{v}_i \frac{mi}{\varrho_i}
\end{align}

\item 
employ material model \eqref{update_ul_start} \eqref{update_ul}

\item
\begin{align}
\uline{a}_i &= \sum_{q=1}^N \nabla_{x} \overset{\circ}{W_{qi}} \uuline{\sigma}_q \frac{mq}{\varrho_q}
\end{align}
\end{enumerate}

the main difference is evident in the last step by the presence of $\overset{\circ}{W_{qi}}$.
Consequently, stress points carry their own kernel functions where the kernel function
is only ever evaluated at quadrature points.

This algorithm is the only one that adapts smoothing length $h$. The novel 
procedure using the fit of a confidence ellipse outlined in section \ref{chosing_h} is employed. 

\subsection{Contact} \label{sec:contact}

\begin{figure}[htp]
\begin{tabular}{ccc}
\includegraphics[width=0.3\textwidth]{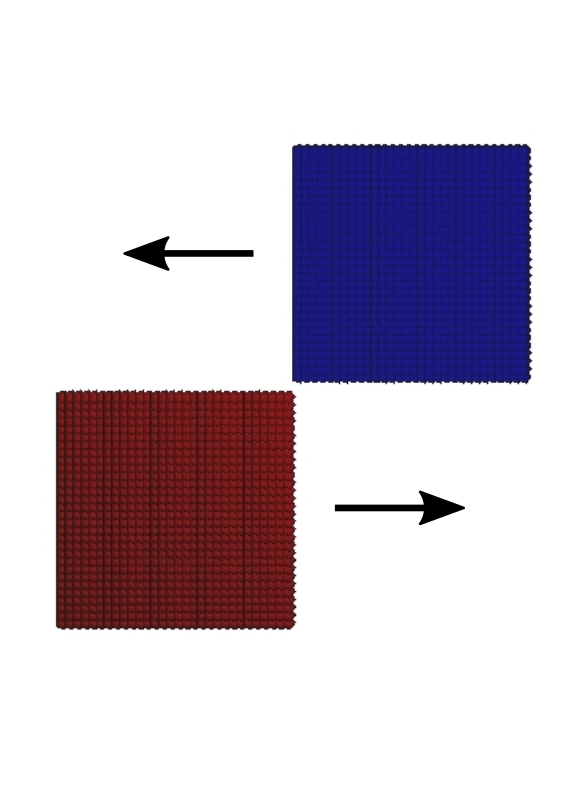}  & \includegraphics[width=0.3\textwidth]{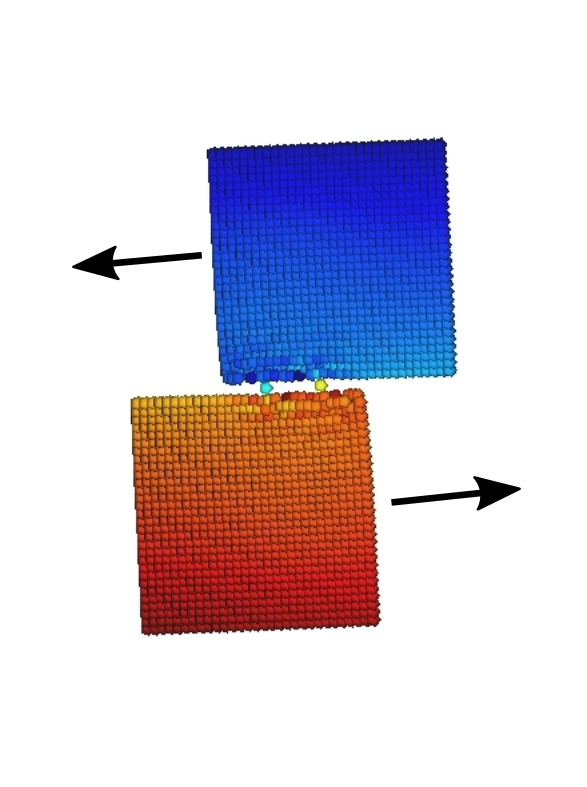} & \includegraphics[width=0.3\textwidth]{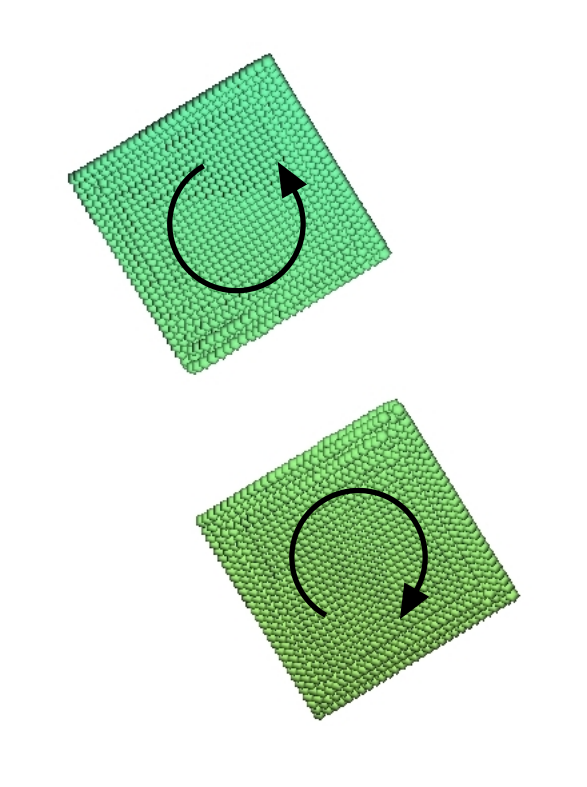} \tabularnewline
\end{tabular}
\label{fig:sliding}
\caption{Two squares are accelerated towards each other. They are slightly off set and would not touch. However, due to spurious particle interactions between the two bodies shear 
stresses develop and the two bodies deflect. In fact, the deflection is violent enough that two particles are ripped out of each of the two bodies. The coloring depicts the velocity in x direction.}
\end{figure}

In this paper a case involving contact between two bodies will be examined in 
section \ref{sec:impact}. Contact in meshless methods is an interesting topic since
the boundary of a meshless body is not well defined, due to the smooth nature of
the kernel function. Interestingly, updated Lagrangian meshless methods can be
run without any contact algorithm even if multiple bodies are present by simply 
including the particles of the other body in the computation of the momentum 
equation. However, while successfully applied for example in \cite{gray2001sph}, this 
may be problematic: Sliding contact may induce shear stresses, see \ref{fig:sliding}, 
while development of tensile stresses across body boundaries may lead to sticking
of bodies \cite{campbell2000contact}. Also, such a procedure is not readily 
applicable to total Lagrangian schemes, since Total Lagrangian Schemes compute their neighbor set only
once, on the reference configuration. 

\begin{figure}[ht]
\noindent \centering{}%
\includegraphics[width=0.3\textwidth]{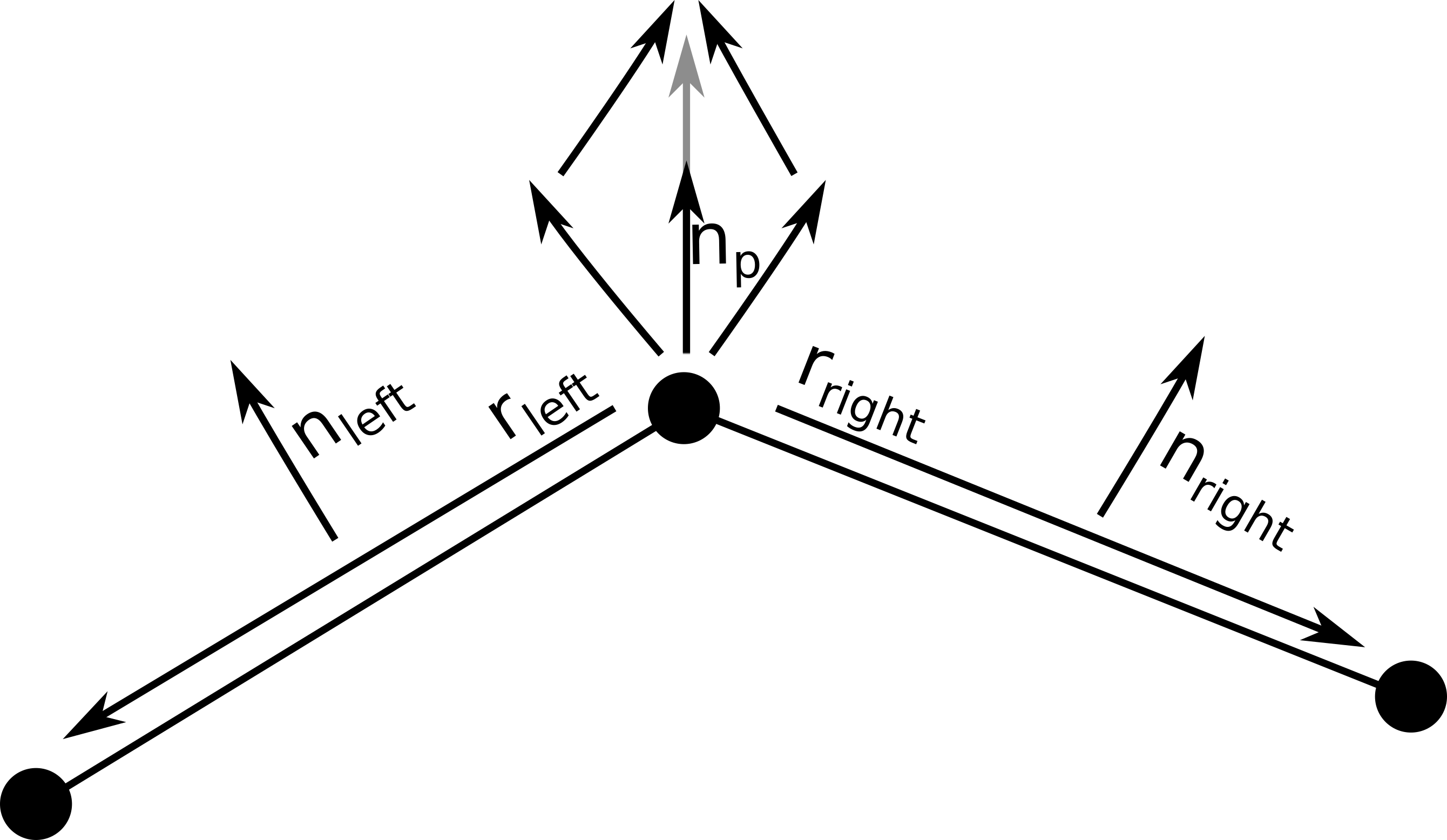}
\caption{Schematic view of the computation procedure regarding the boundary normals}
\label{fig:normcomp}
\end{figure}

This asks for proper contact conditions between two bodies. Contact algorithms
for meshless methods are not a very well explored topic in literature. There exist
certain methods for the special case of contact of an elastic (or liquid) body
with a rigid one, like the contact forces presented in \cite{monaghan2003fluid} or the ghost
particle technique employed in \cite{libersky1991smooth}. For contact between 
two elastic bodies all algorithms published are a variation of the pentalty method
\cite{vignjevic2006frictionless} \cite{campbell2000contact} \cite{li2001element}.
For this paper a very simple formulation of the penalty method was chosen, very
close to the one presented in \cite{swegle1994analysis} but for two meshless bodies instead of contact
between FEM and meshless bodies:

Let there be two bodies $\mathcal{S}$ and $\mathcal{M}$ with particles $\uline{x}_s \in \mathcal{S}$
and $\uline{x}_m \in \mathcal{M}$. $\mathcal{M}$ is the master body while $\mathcal{S}$
denotes the slave body. The boundary of the master body is extracted and a circular
tour around this boundary is formed. This circular tour $\mathcal{B} = 0 \dots p \dots N_{B}$ with $N_{B}$ equal to the
number of boundary particles is used to compute the boundary normals $\uline{n}_p$ at each boundary particle $p$. Let:
\begin{align}
\uline{r}_{\textrm{left}} &= \frac{x_m^{p-1} - x_m^{p}}{||x_m^{p-1} - x_m^{p}||} \\
\uline{r}_{\textrm{right}} &= \frac{x_m^{p+1} - x_m^{p}}{||x_m^{p+1} - x_m^{p}||}
\end{align}
Now both of these distance vectors are rotated outwards
\begin{align}
\uline{n}_{\textrm{left}} &= \uuline{R}{}_{90\degree} \uline{r}_{\textrm{left}} \\
\uline{n}_{\textrm{right}} &= \uuline{R}{}_{-90\degree} \uline{r}_{\textrm{right}}
\end{align}
Note the sign change in $\uuline{R}$ since both distance vectors $\uline{r}$ point away from $\uline{x}_m^p$.
The particle normal is then simply:
\begin{align}
\uline{n}_p = 0.5(\uline{n}_{\textrm{left}} + \uline{n}_{\textrm{right}})
\end{align}
This procedure is repeated for all boundary particles. A sketch of the procedure can be found in 
figure \ref{fig:normcomp}. Now for each particle $\uline{x}_s$
the closest particle $\uline{x}_p^c$ in $\mathcal{B}$ is extracted. If the inter particle distance  
$\uline{x}_s - \uline{x}_p^c$ is in the same direction as the particle normal $\uline{n}_p$
no contact is detected. Otherwise a penalty force is applied to $\uline{x}_s$. The 
closest particle is extracted by means of a kd-tree to cut the runtime of the 
Algorithm from $\mathcal{O}(N_B \cdot N_S)$ to $\mathcal{O}(N_B \cdot \log{N_S}) + \mathcal{O}(N_S \cdot \log{N_S})$
, where the second term stems from the construction of the kd-tree \cite{bentley1975multidimensional}.
In case a small time step is used one can probably get away with only testing
the boundary particles of the slave body. In any case, if contact is detected
the penetration depth $g$ can be measured as:
\begin{align}
g = | (\uline{x}_p^c - \uline{x}_s) | \cdot \uline{n}_p
\end{align}
and a penalty force $\uline{f}$ of magnitude
\begin{align}
\uline{f} = \alpha \uline{n}_p \frac{g\cdot m_s}{\delta t^2}
\end{align}
is applied, where $\alpha$ is a user defined parameter, i.e. the contact stiffness, $m_s$ is the mass of the slave particle and $\delta t$ the current time step. To eliminate the user 
defined parameter $\alpha$ the procedure in \cite{li2001element} is suitable. 
However, this improvement did not proof necessary, even for quite involved
contact conditions. 

\section{Benchmarks} \label{sec:bench}

In this section three benchmarks will be carried out:

\begin{enumerate}
\item Rotation of a disk
\item A tensile test
\item Collision of two rubber rings
\end{enumerate}

The rationale behind the rotation test is to demonstrate what methods actually
do conserve angular momentum and to what extent. The tensile test was chosen 
because of the well known tensile instability mode present in meshless methods
\cite{xiao2005material} \cite{swegle1995smoothed}. Even though the term ``tensile
instability'' is a misnomer \cite{li2007meshfree}, that is, the same instability mode may also be observed
in compression, a tensile test is a suitable means to investigate tensile instability
modes. Finally the rubber ring impact is interesting for a multitude of reasons. 
Involved contact conditions are combined with tensile as well as compressive stresses.
Those stresses need to be resolved across two length scales separated by one order
of magnitude; across the ring diameter on one hand, along the circumference on the other.
Large deformations are encountered. Only one material is investigated in this 
section, having numerically convenient but imaginary material parameters:
\begin{align}
E &= 1e7 & [Pa]\\
\nu &= 0.4 & [-]\\
\varrho_0 &= 1 & [kg/m^3]
\end{align}

\subsection{Rotating Disk}

\begin{figure}[ht]
\noindent \centering{}%
\includegraphics[width=\textwidth]{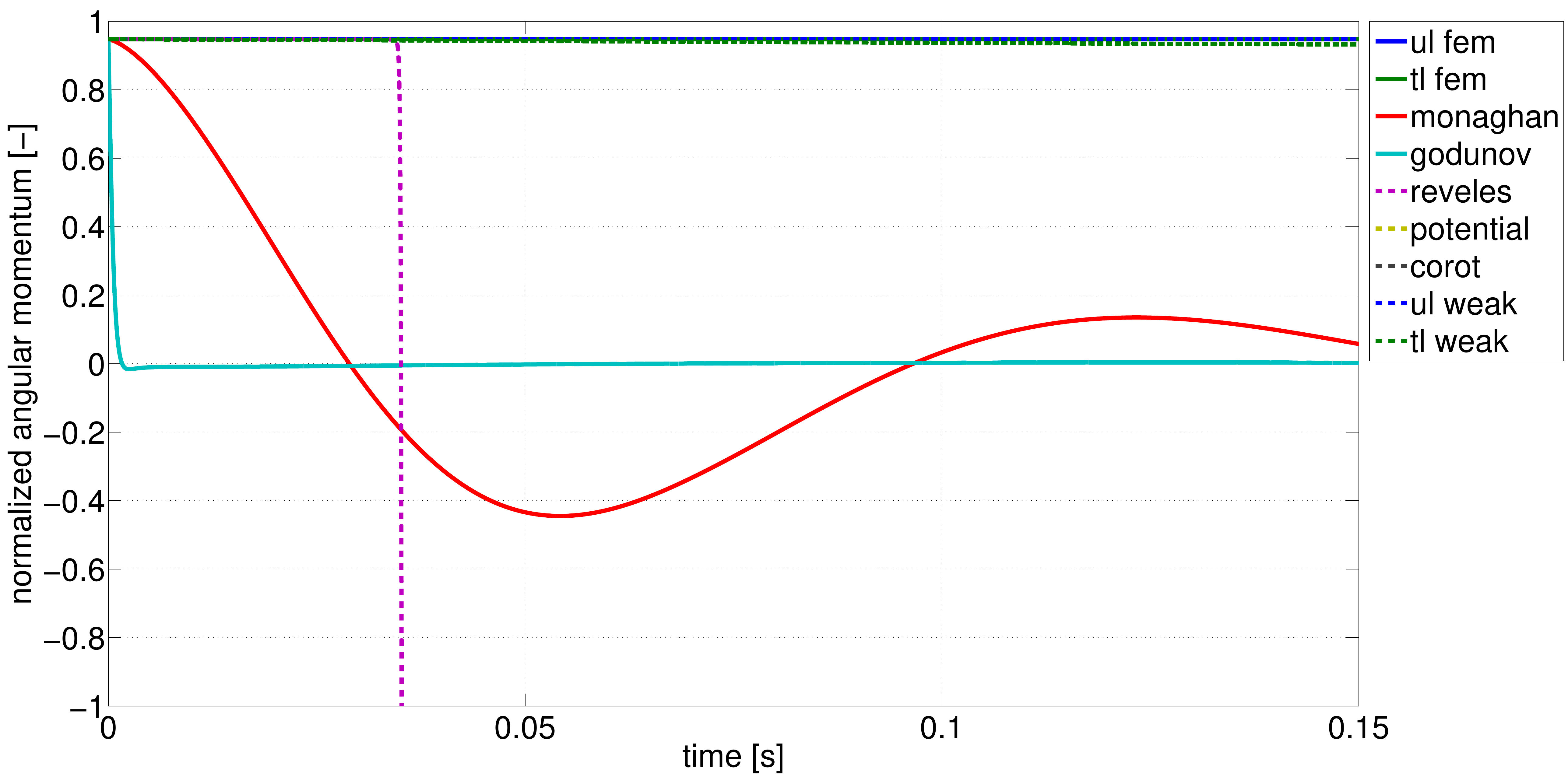}
\caption{Plot of the normalized angular momentum of each rotating disk benchmark}
\label{fig:momentum}
\end{figure}

For this benchmark the unit disk is sampled with particles on a cartesian grid. 
Each particle has an angular velocity of 50 1/s initially. No explicit boundary conditions are applied at any time. That is, the stress free boundary conditions are ignored and no correction scheme to this end, like the procedure given in \cite{ostad2012stabilized}, is employed. However, some schemes approximately fulfill the stress free boundary conditions due to lack of zero order completeness while others fulfill them due to their weak form construciton. The simulation is run until 0.15 seconds. The reasoning behind this 
is that one revolution takes about $2\pi/50 = 0.13$ seconds. This was rounded up
since the sampling of the disk is not perfect and also to allow for some numerical
damping due to round off error, while still simulating a full revolution. During the 
simulation the current angular momentum was recorded:
\begin{align}
H = \sum_{i=1}^N x_i m_i v_{yi} - y_i m_i v_{xi}
\end{align}
The analytical angular momentum, as if the unit disk was continuous instead of 
a discrete sampling that is, can be computed as:
\begin{align}
\bar{H} = \frac{1}{2}MR^2\cdot\omega
\end{align}
where $M$ is the total mass, $R$ is the radius and $\omega$ the angular velocity.
A plot of $H/\bar{H}$ can be found in figure \ref{fig:momentum}. Note that this 
ratio will not be exactly equal to one even at the outset since the discretization
underestimates the mass. 

Most methods conserve the angular momentum quite well. The
weak form algorithm in the total Lagrangian frame shows signs of slight numerical
dissipation. The Monaghan algorithm does not conserve angular momentum but completely
reverses the spinning of the disk. This is due to the lack of first order completeness of
this algorithm. Interestingly enough this can be fixed by replacing $W$ with $\overset{\triangle}W$ but not $\overset{\circ}{W}$, the latter leading to immediate divergence of the simulation. The Godunov type scheme is not linearly complete either, however in that case the rotation 
stops completely due to excess numerical dissipation. Both of these algorithms where
included in this work anyway, since they are shown to resolve the third benchmark
case, which includes large deformations, in the respective original works. 

The last noteworthy exception is the algorithm proposed by Reveles, which diverges. 
It is not clear why this is the case. The algorithm neither respects stress free 
boundaries nor does it conserve linear momentum. All other algorithms that successfully
complete a whole revolution fulfill at least one of those properties. A \verb+LSDYNA+
simulation which is suspected to implement the same algorithm fails in this situation 
too. 

\subsection{Tensile Test}\label{sec:tens}

\begin{figure}[ht]
\noindent \centering{}%
\includegraphics[width=0.5\textwidth]{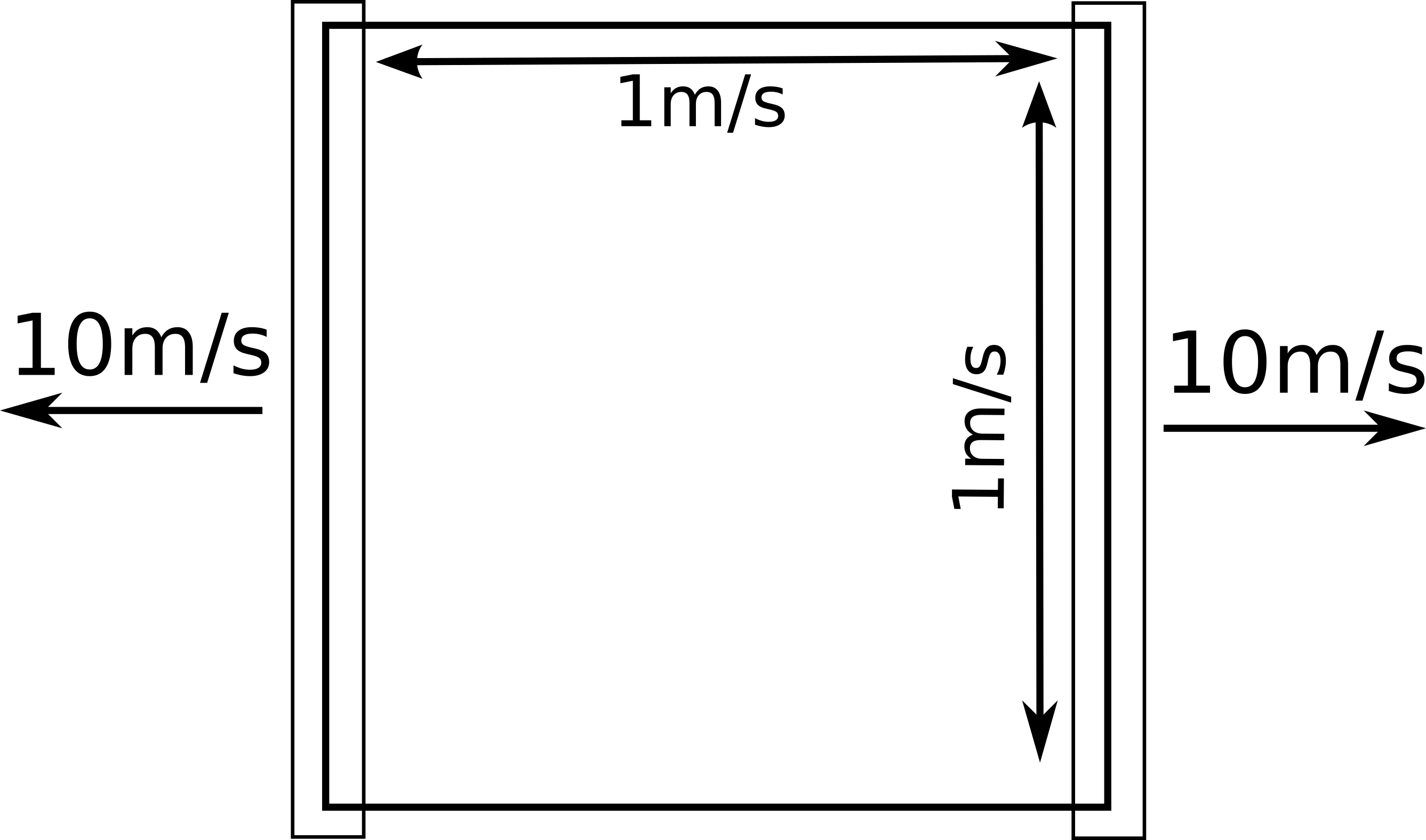}
\caption{Schematic view of the setup of the tensile benchmark}
\label{fig:tenssetup}
\end{figure}

In this section a simplified tensile test is investigated. A 1m x 1m square is sampled
regularly with particles. A single column of particles on both the left hand and
right hand side is moved with a constant velocity of 10 m/s in x direction and is 
completely fixed in y direction. A schematic view of the setup is found in figure \ref{fig:tenssetup}.
The boundary conditions are applied directly on the boundary particles on each time step, 
simply overwriting the values of the position, velocity and acceleration field variables. That is, no
consistency scheme as for example in \cite{jun1998explicit} is applied. The test
is run until the tensile specimen has a length of 1.5m, that is 50\% extension. The
reasoning being that most materials except for soft tissues, rubbers and the 
like rupture earlier. 

\begin{figure}[htp]
\noindent \centering{}%
\includegraphics[width=\textwidth]{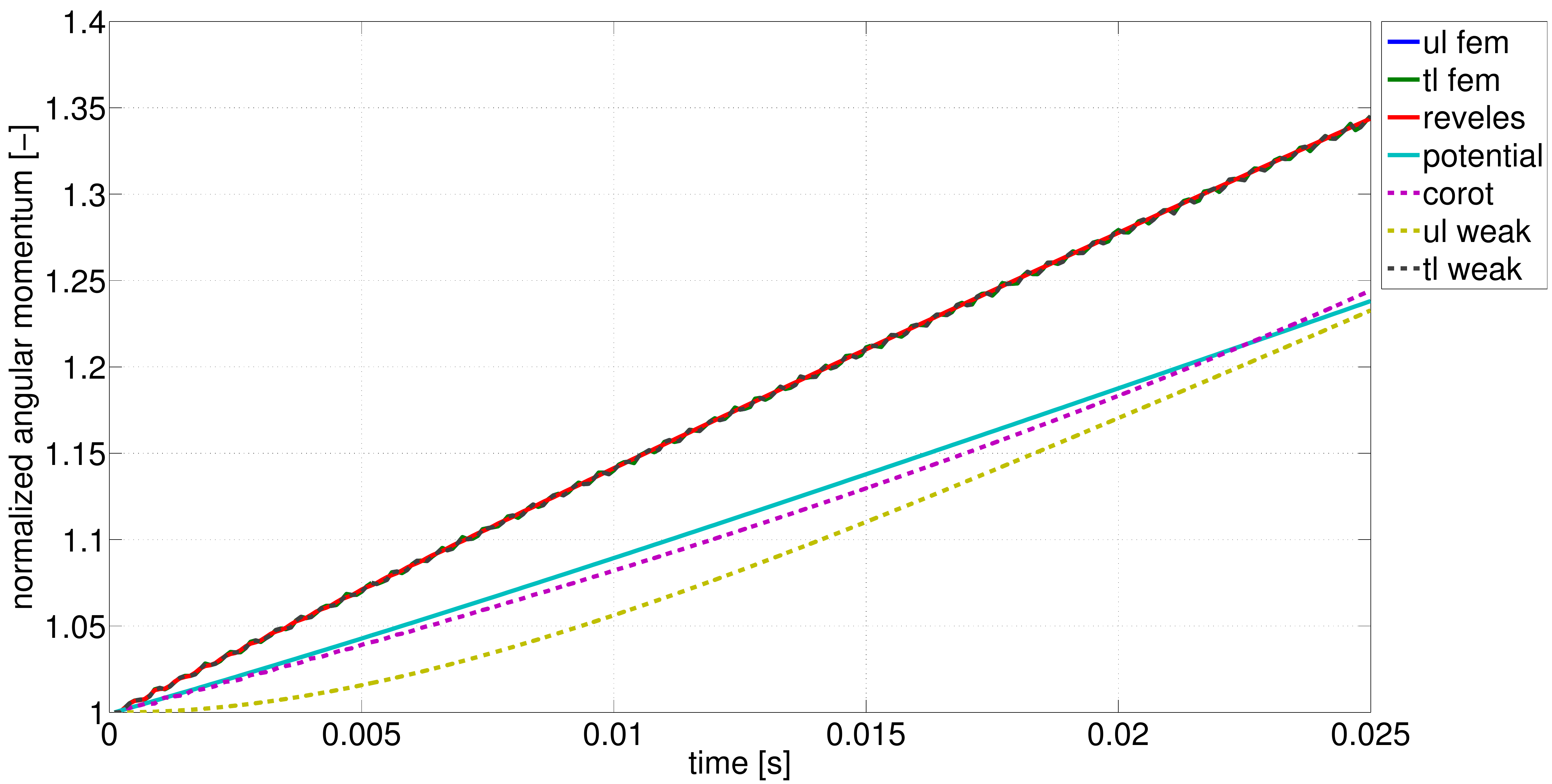}
\caption{Plot of the normalized contraction of each tensile test}
\label{fig:contraction}
\end{figure}

The normalized contraction, that is, $\bar{L} = l/l_0$ is recorded, where $l_0$ is
the initial height of the specimen, and $l$ the current height of the specimen 
where it is most contracted (usually at the middle). The results can be seen in
figure \ref{fig:contraction}. Both the Godunov as well as Monaghan algorithms 
are not able to complete this benchmark, see chapter \ref{sec:app} for some discussion regarding
this. The results of the remaining methods are quite interesting. Both Reveles
algorithm as well as the total Lagrangian weak form algorithm are very close to
the FEM solutions. The potential based as well as co-rotated algorithm converge
to another, less contracted solution. This is not very surprising because the
construction of these two algorithms is quite similar. More surprising is that the
weak form algorithm in the updated Lagrangian frame converges to this solution 
as well, even though it is way more similar to the algorithms in the other cluster in figure \ref{fig:contraction}. This certainly necessitates further investigation. One reason for this might be that
the algorithm is quite dissipative as can be seen in the next section, so it might
lose some velocity in y direction over time. 

\begin{figure}[htp]
\noindent \centering{}%

\begin{tabular}{ccc}
\includegraphics[width=0.3\textwidth]{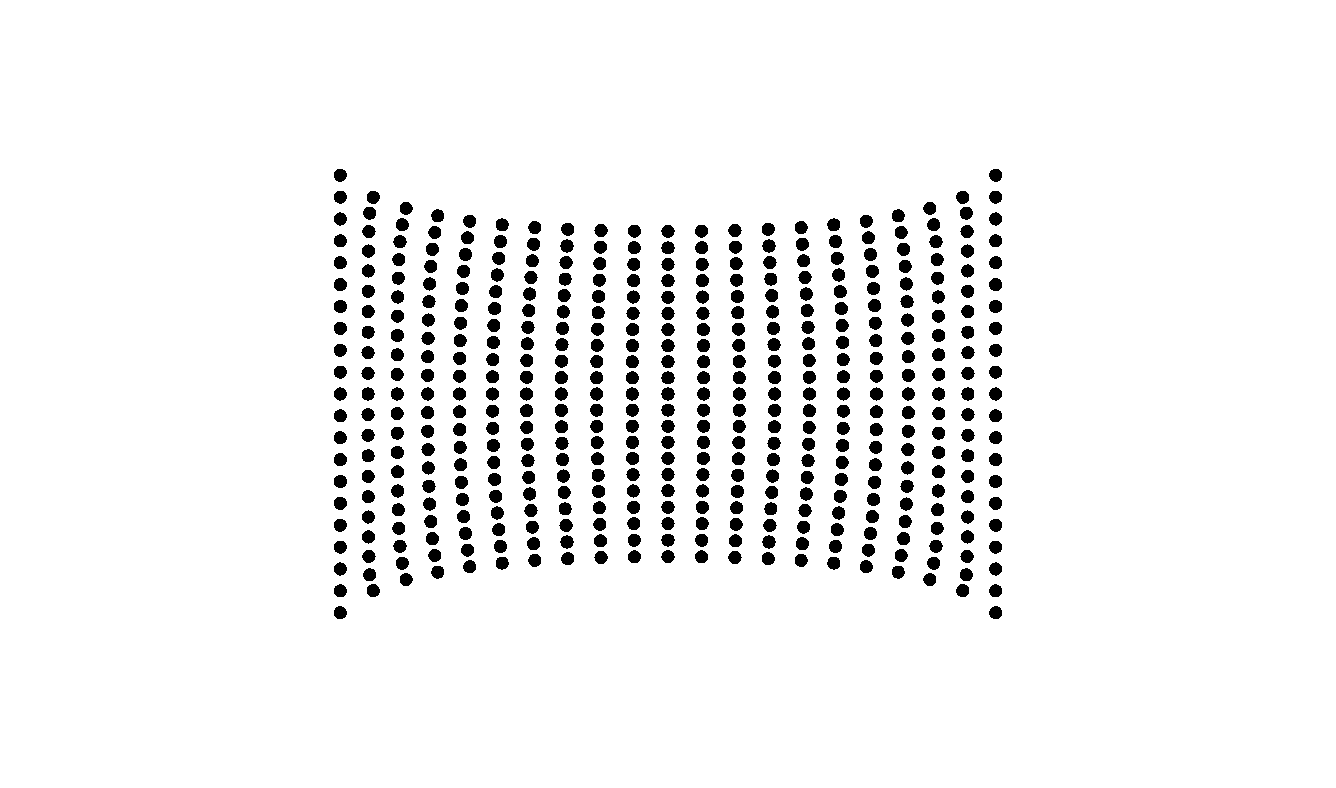}  & \includegraphics[width=0.3\textwidth]{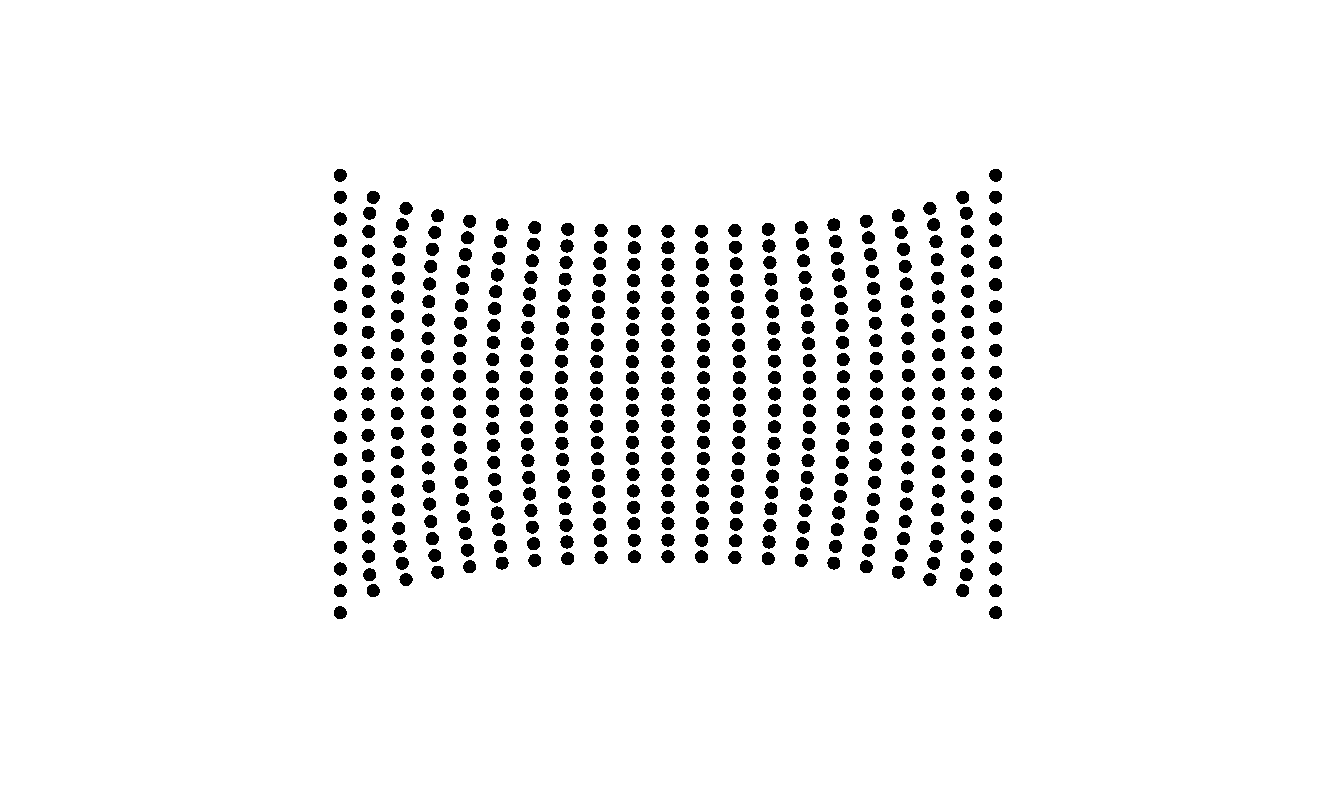} & \includegraphics[width=0.3\textwidth]{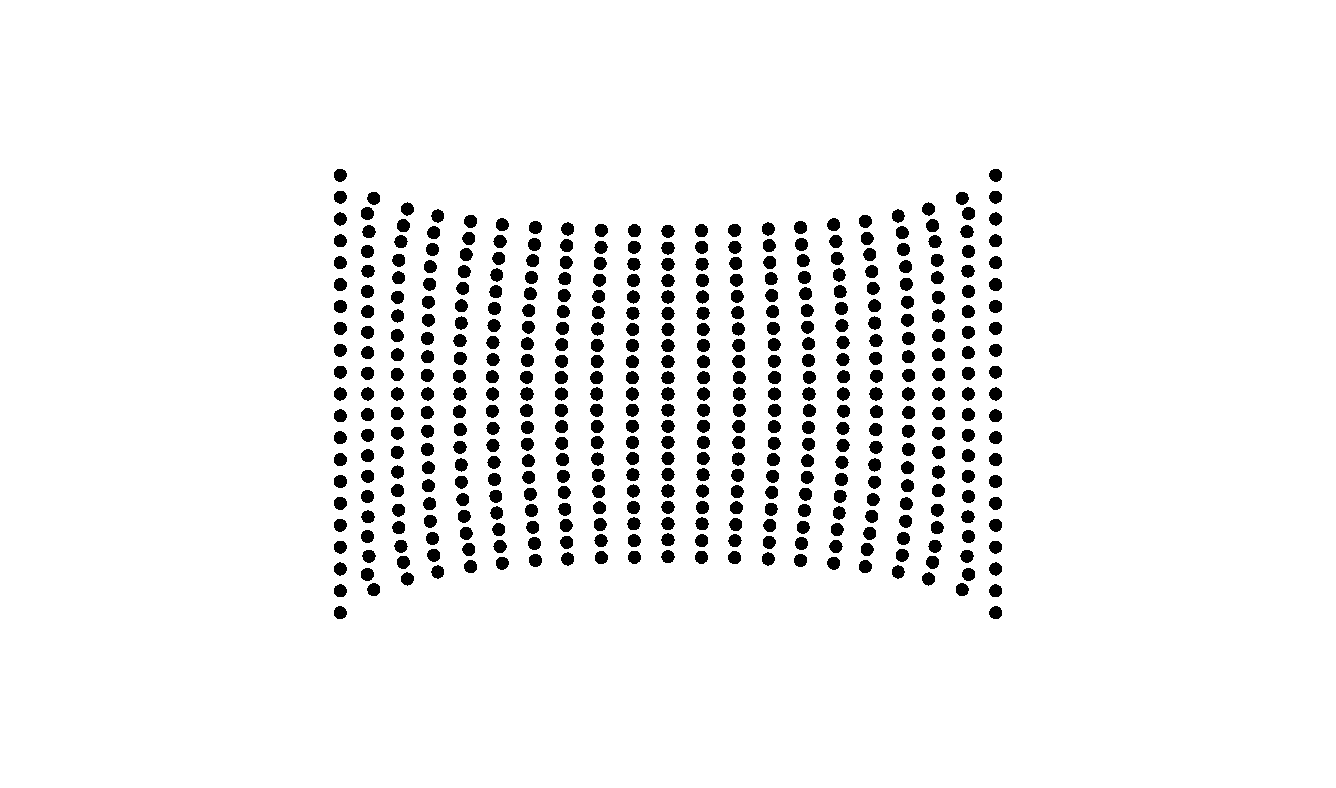} \tabularnewline
FEM UL  & FEM TL & Reveles \tabularnewline
\includegraphics[width=0.3\textwidth]{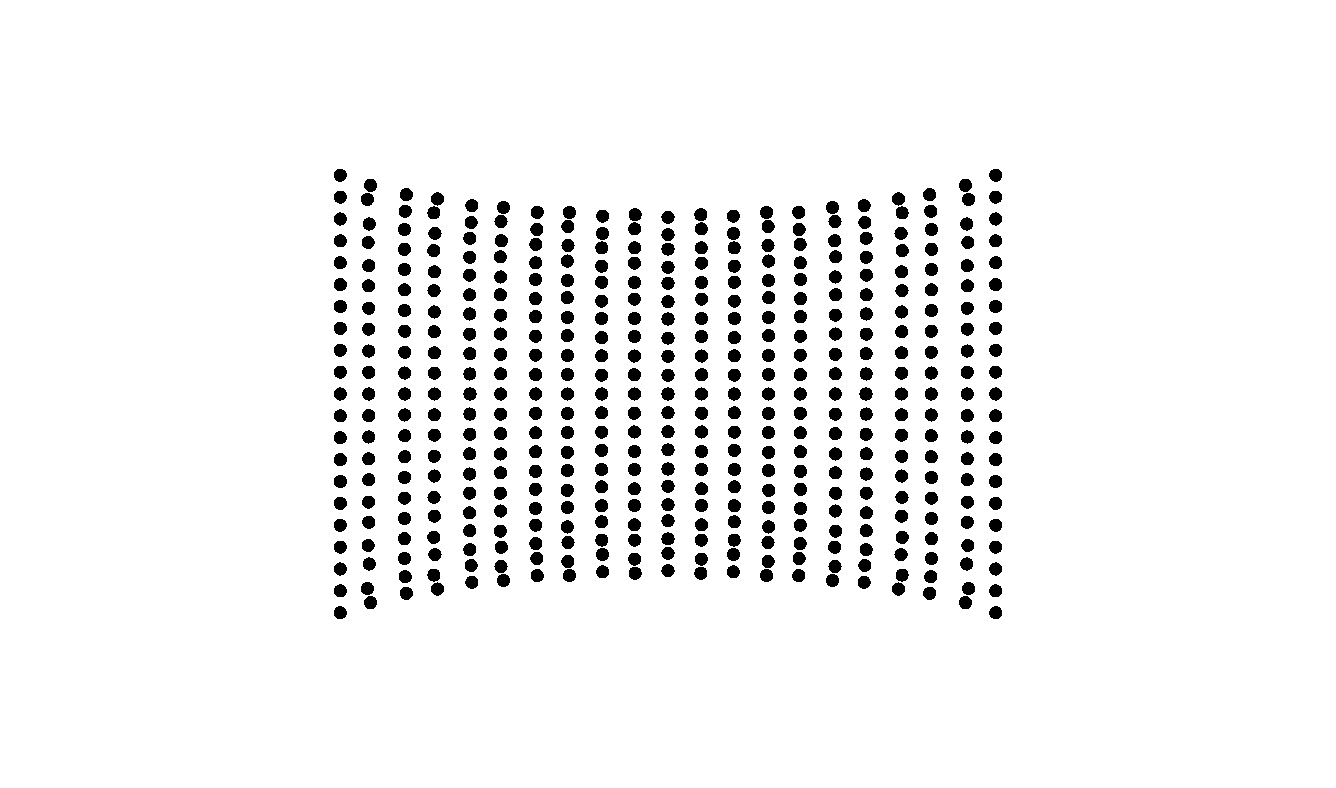}  & \includegraphics[width=0.3\textwidth]{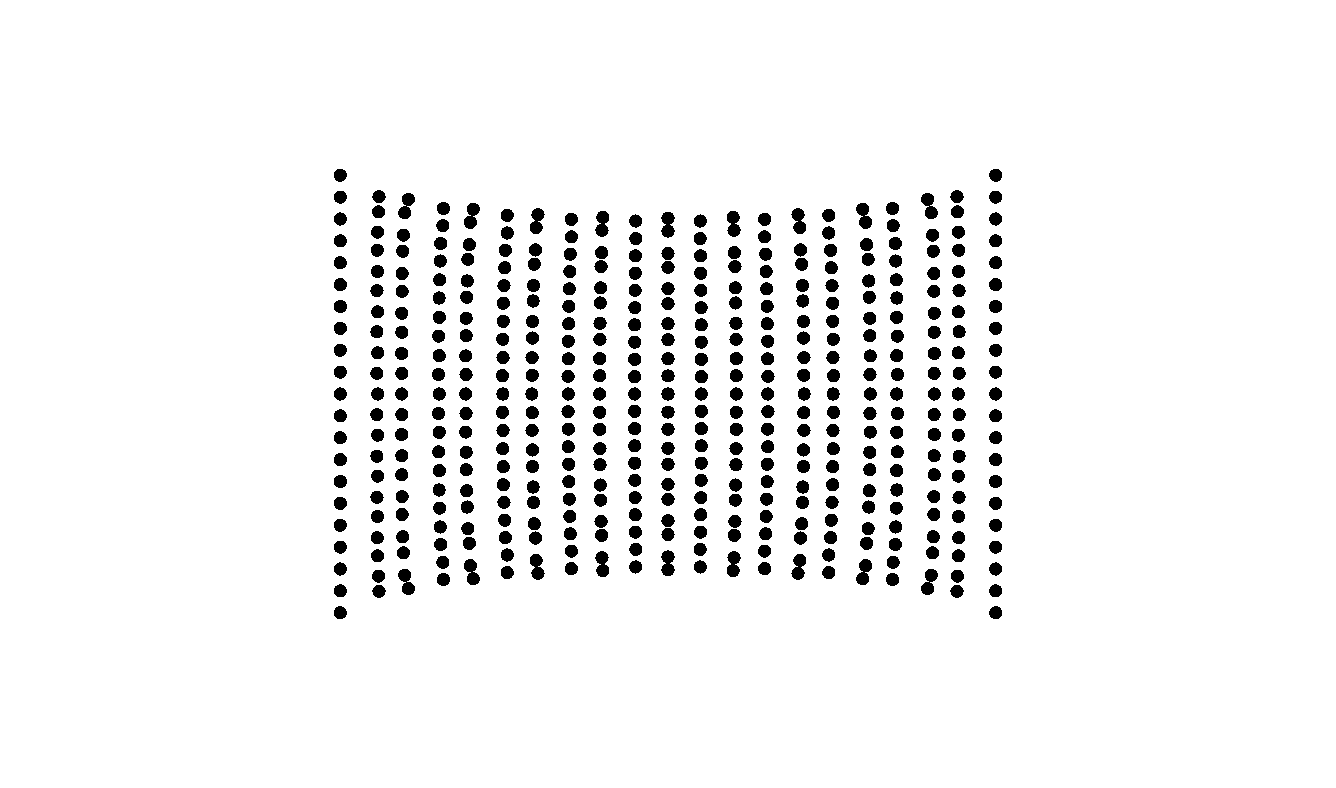} & \includegraphics[width=0.3\textwidth]{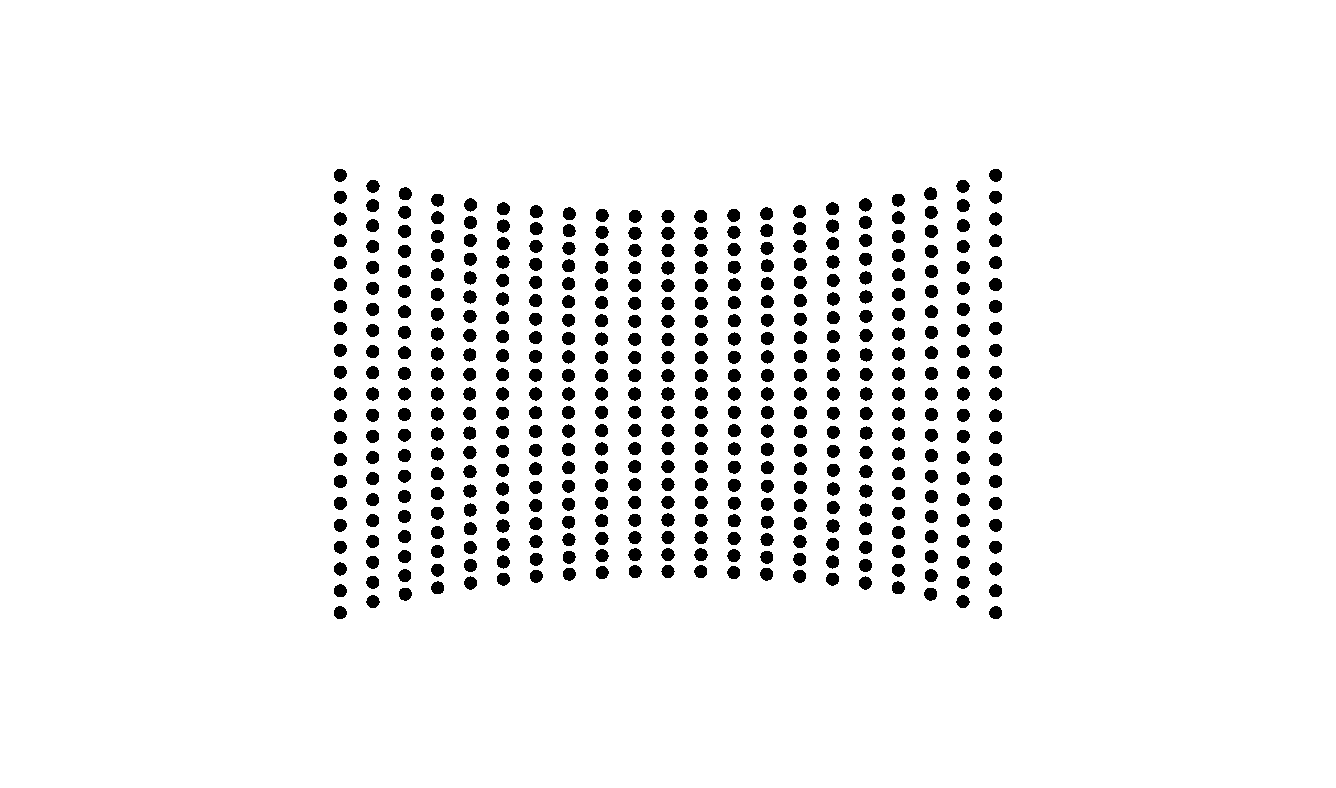} \tabularnewline
Potential  & Co-rotated & UL weak \tabularnewline
& \includegraphics[width=0.3\textwidth]{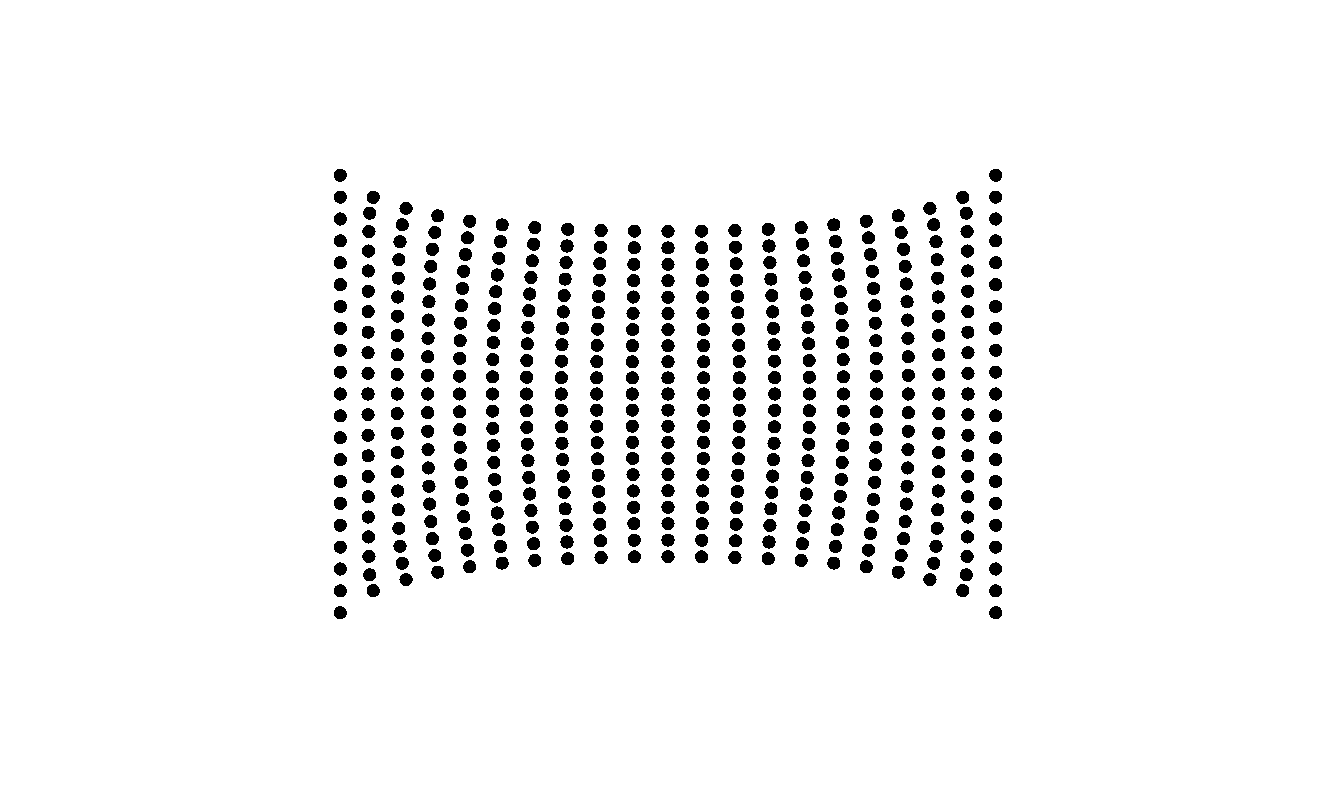}    & \tabularnewline
& TL weak   &  \tabularnewline
\end{tabular}

\caption{Final particle distribution of tensile test, all methods}
\label{fig:tensile_final}
\end{figure}

The particle distribution of each method is shown in figure \ref{fig:tensile_final}.
There is some clustering of the particles along the y axis in the potential based 
method, and more severe clustering of the particles along the x axis for the co-rotated
method. In computer graphics this is no problem, since the discretization points 
are almost never rendered on the screen directly. However, this raises some 
concerns regarding these methods in use for physics simulations. 

\subsection{Rubber Ring Impact}\label{sec:impact}

\begin{figure}[ht]
\noindent \centering{}%
\includegraphics[width=0.5\textwidth]{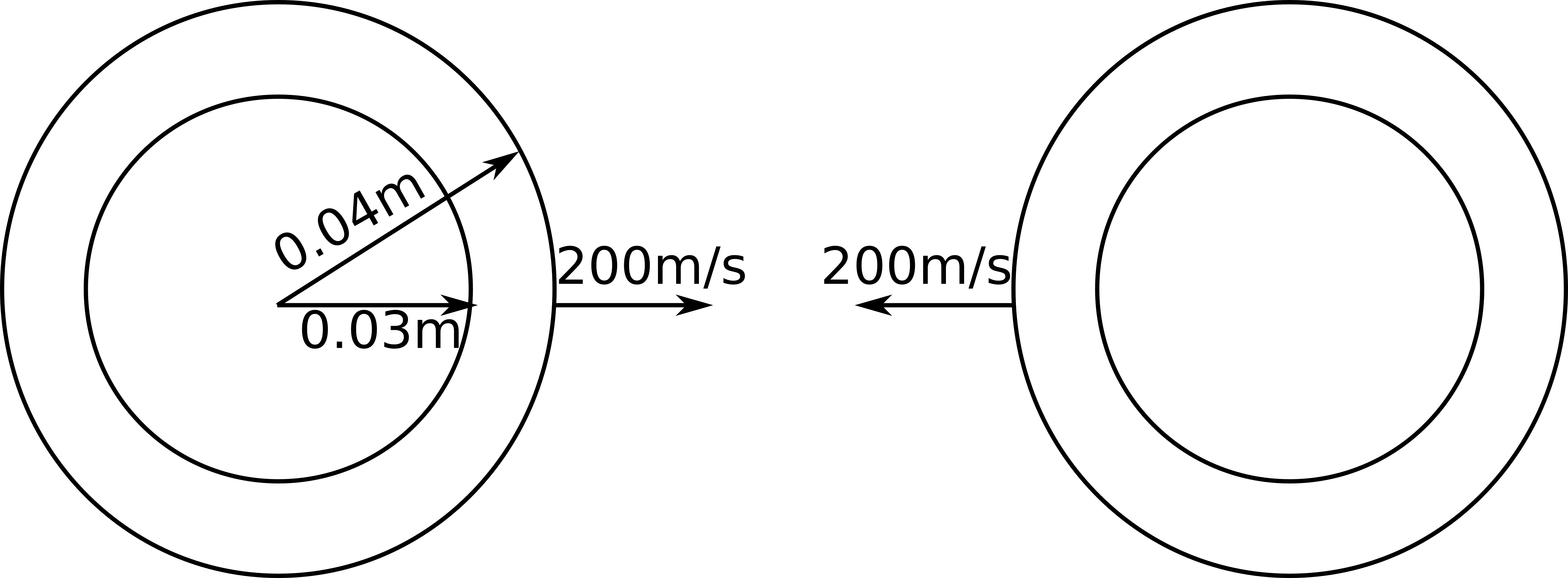}
\caption{Schematic view of the setup of the colliding rings benchmark}
\label{fig:ringsetup}
\end{figure}

This is the most involved test case investigated in this paper. Two rubber rings with circular cross section, outer
dimension 0.04 and inner dimension 0.03 meters are colliding with an initial relative
velocity of 200 m/s. A schematic view of the setup is found in figure \ref{fig:ringsetup}. The contact algorithm described in section \ref{sec:contact} 
is employed for resolving the contact between the bodies. This benchmark is quite
well known and was investigated by a number of authors, for example \cite{gray2001sph}, 
\cite{swegle1995smoothed}, \cite{vidal2007stabilized} and \cite{parshikov2002smoothed}.
The rings are expected to deform upon impact, reflect, than oscillate freely a number
of times. Since there is no friction the rings should retain all their kinetic
energy after the impact.

\begin{figure}[htp]
\noindent \centering{}%
\includegraphics[width=\textwidth]{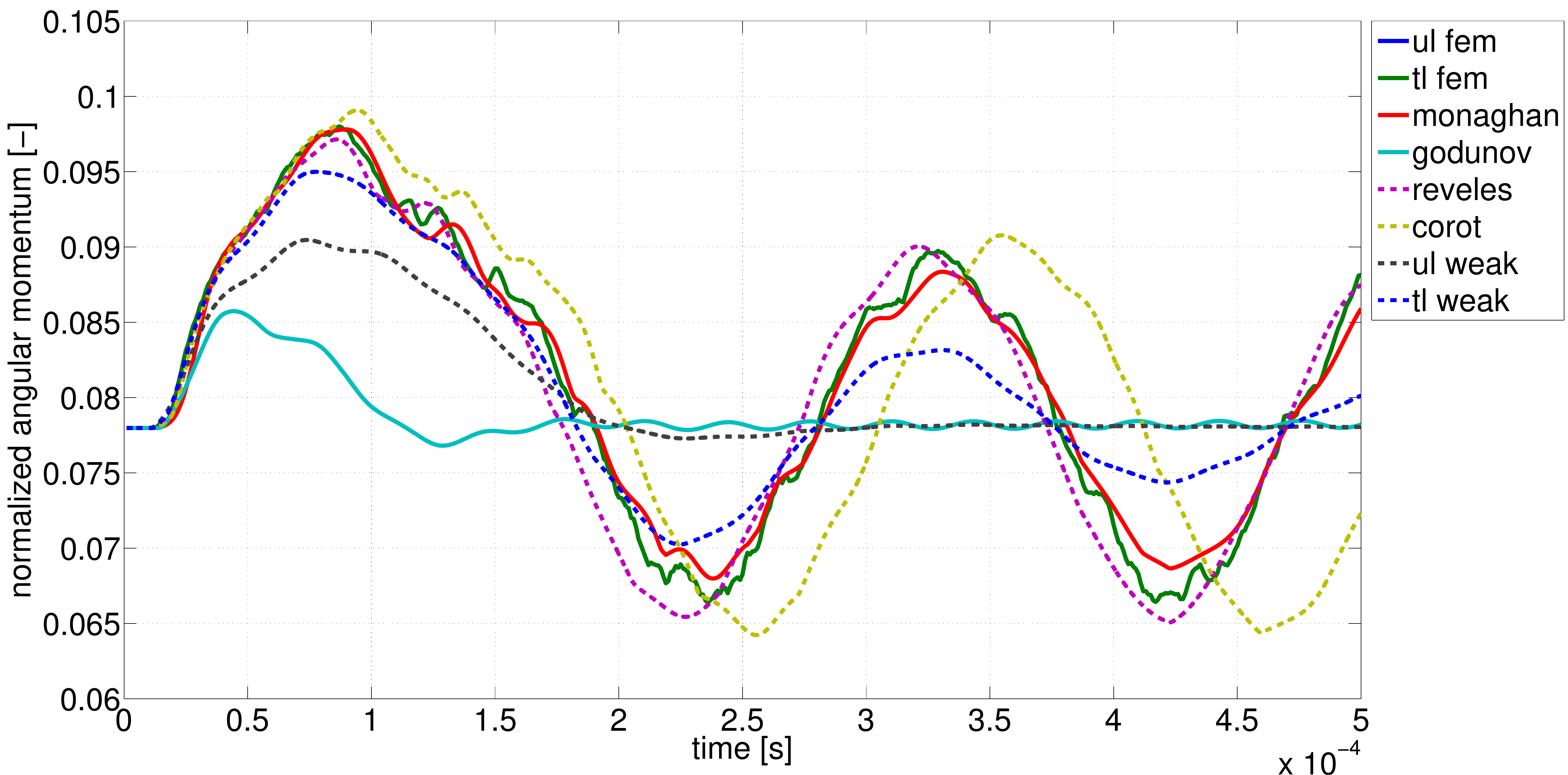}
\caption{Height of left ring in rubber ring impact simulation}
\label{fig:rings}
\end{figure}

To compare the methods at hand the height of the left rubber ring was recorded
in all simulations, see figure \ref{fig:rings}. In this simulation the variation
of results was higher than in the other test cases. This makes sense since the 
simulation is the most dynamic. and the multidimensionality of the deformation is the largest. The following results are remarkable:

\begin{itemize}
\item The UL FEM and TL FEM results are indistinguishable. Monaghan's algorithm 
is closest, followed by Reveles'. However, to get a stable result, Reveles' algorithm
had to be stabilized by quite large amounts of artificial viscosity: $\alpha = 3$, 
$\beta = 3$, $\eta = 0.1$. In light of the findings in \cite{johnson1996normalized}, where it was shown that choice of artificial viscosity constants can severely alter the physics of the problem at hand, this is quite concerning.

\item The co-rotated algorithm follows the amplitude of the FEM simulations but the free vibrations after impact are out of phase

\item The total Lagrangian weak form algorithm is in phase but suffers from additional amplitude decay

\item Godunovs algorithm dissipates almost all energy during the collision of the
rubber rings. A fact acknowledged in the original publication \cite{parshikov2002smoothed}. 

\item The updated Lagrangian weak form algorithm follows the shape of the first impact
quite closely, albeit with less energy. No further oscillations are observed. This
is probably due to excessive smoothing due to the velocity interpolation procedure 
\eqref{vel_interp}. 
\end{itemize}

\begin{figure}[htp]
\noindent \centering{}%

\begin{tabular}{cccc}
\includegraphics[width=0.23\textwidth]{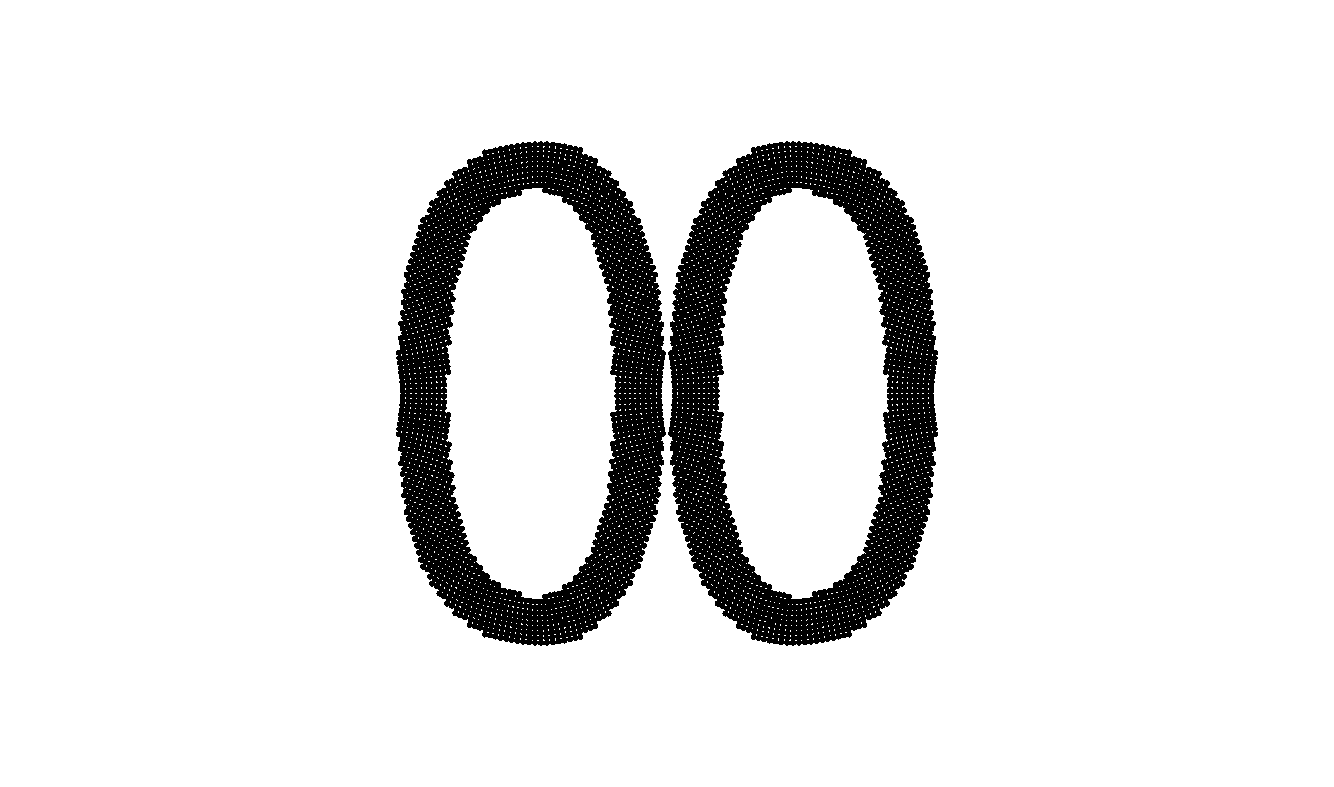}  & \includegraphics[width=0.23\textwidth]{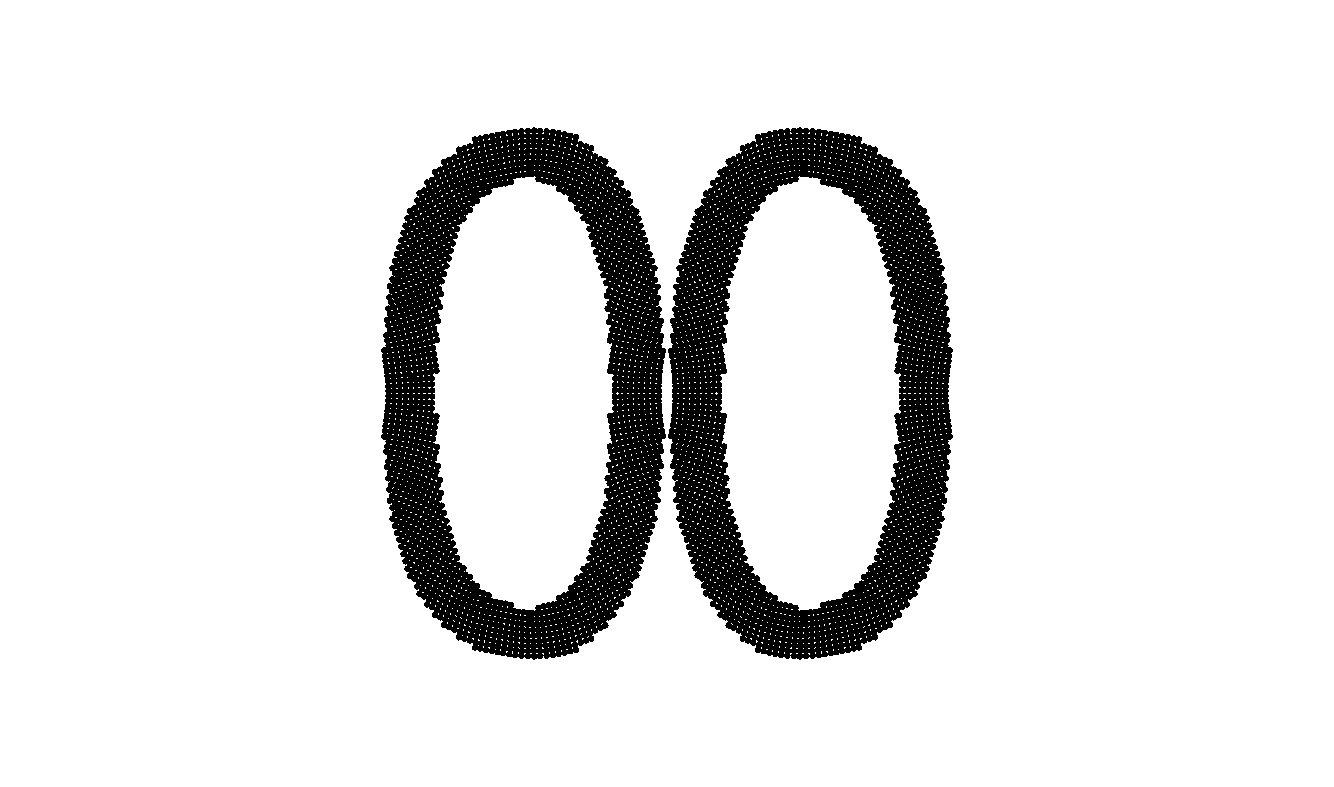} & \includegraphics[width=0.23\textwidth]{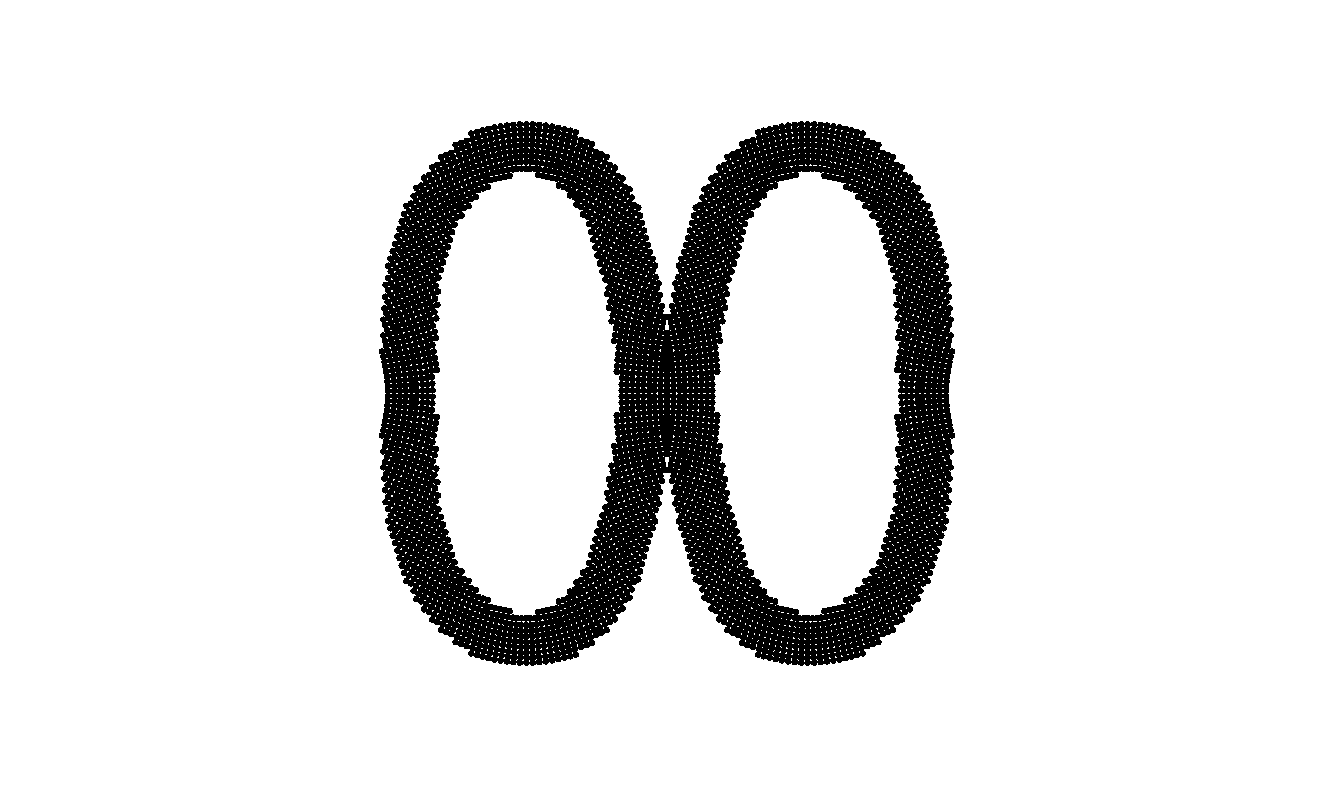} & \includegraphics[width=0.23\textwidth]{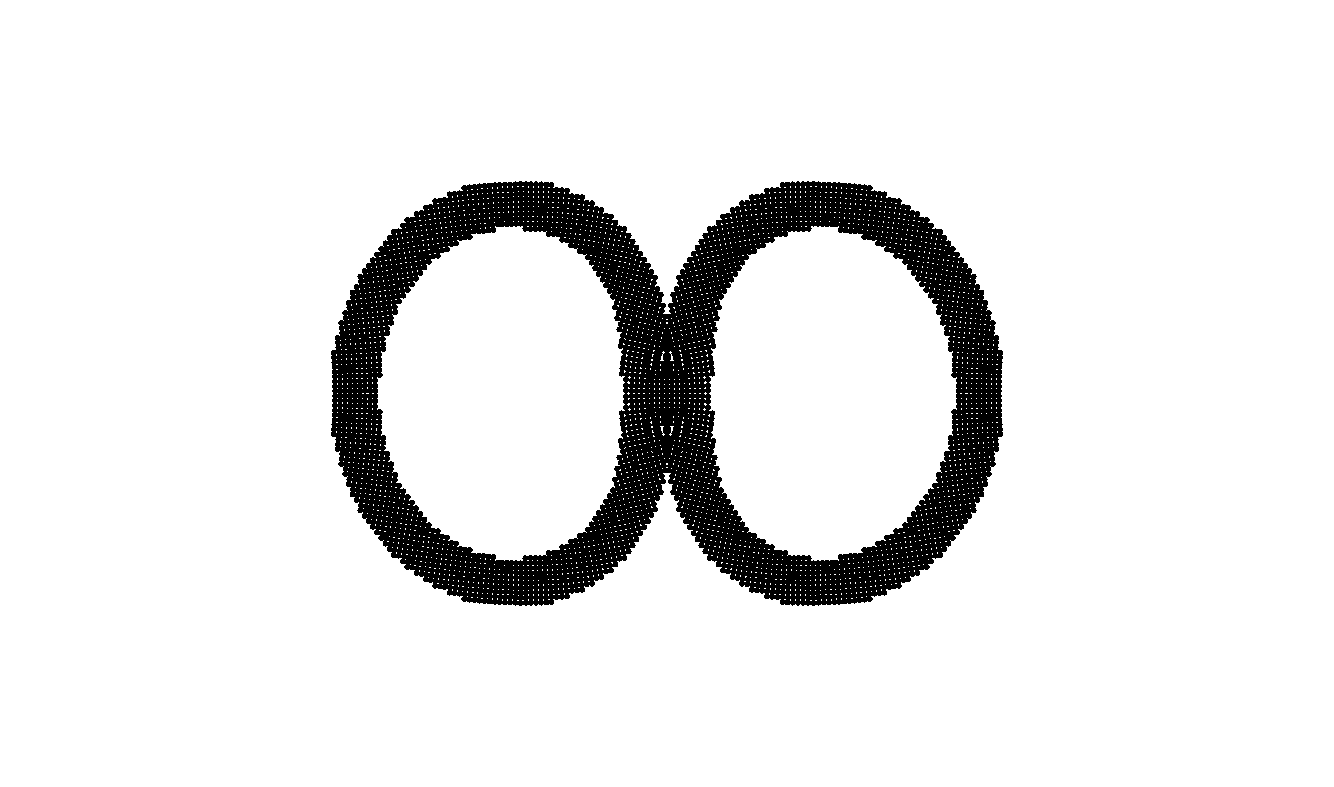} \tabularnewline
UL FEM  & TL FEM & Monaghan & Godunov \tabularnewline
\includegraphics[width=0.23\textwidth]{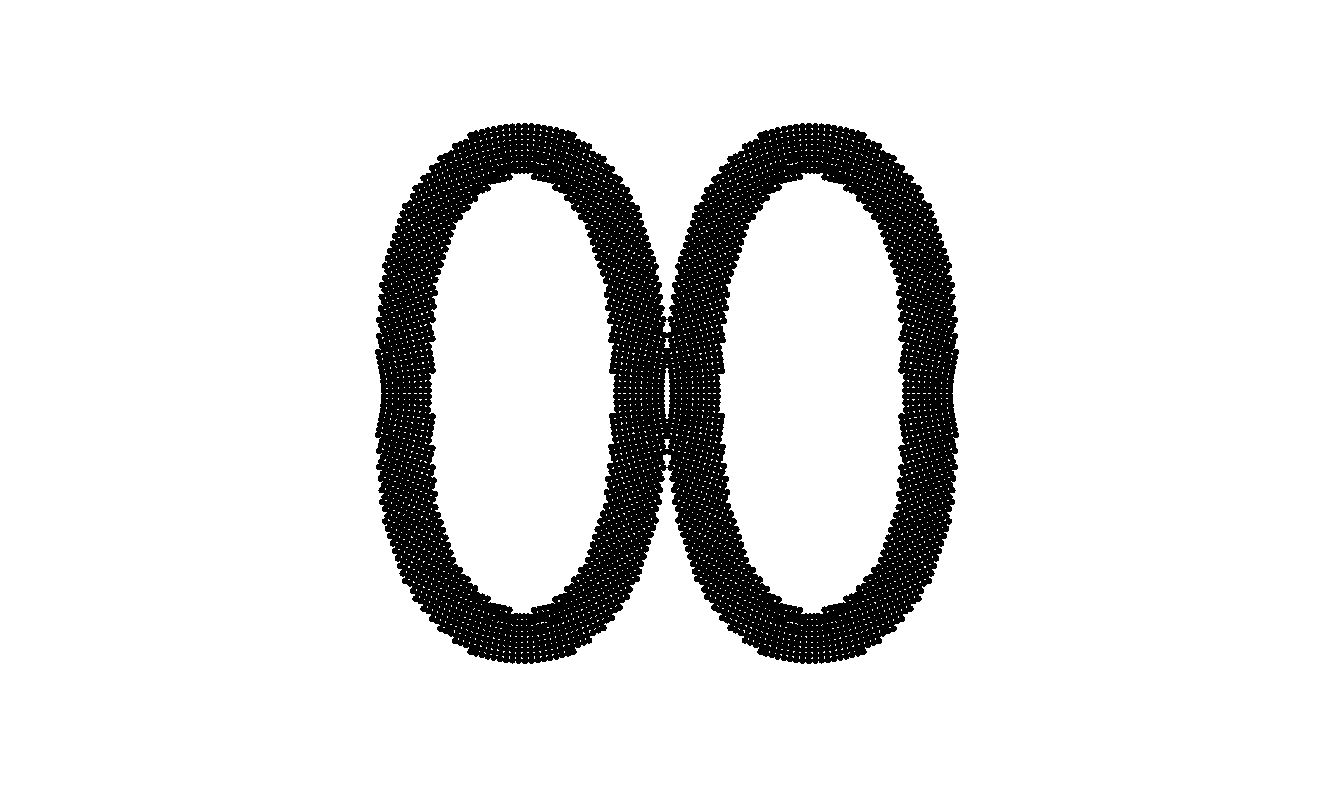}  & \includegraphics[width=0.23\textwidth]{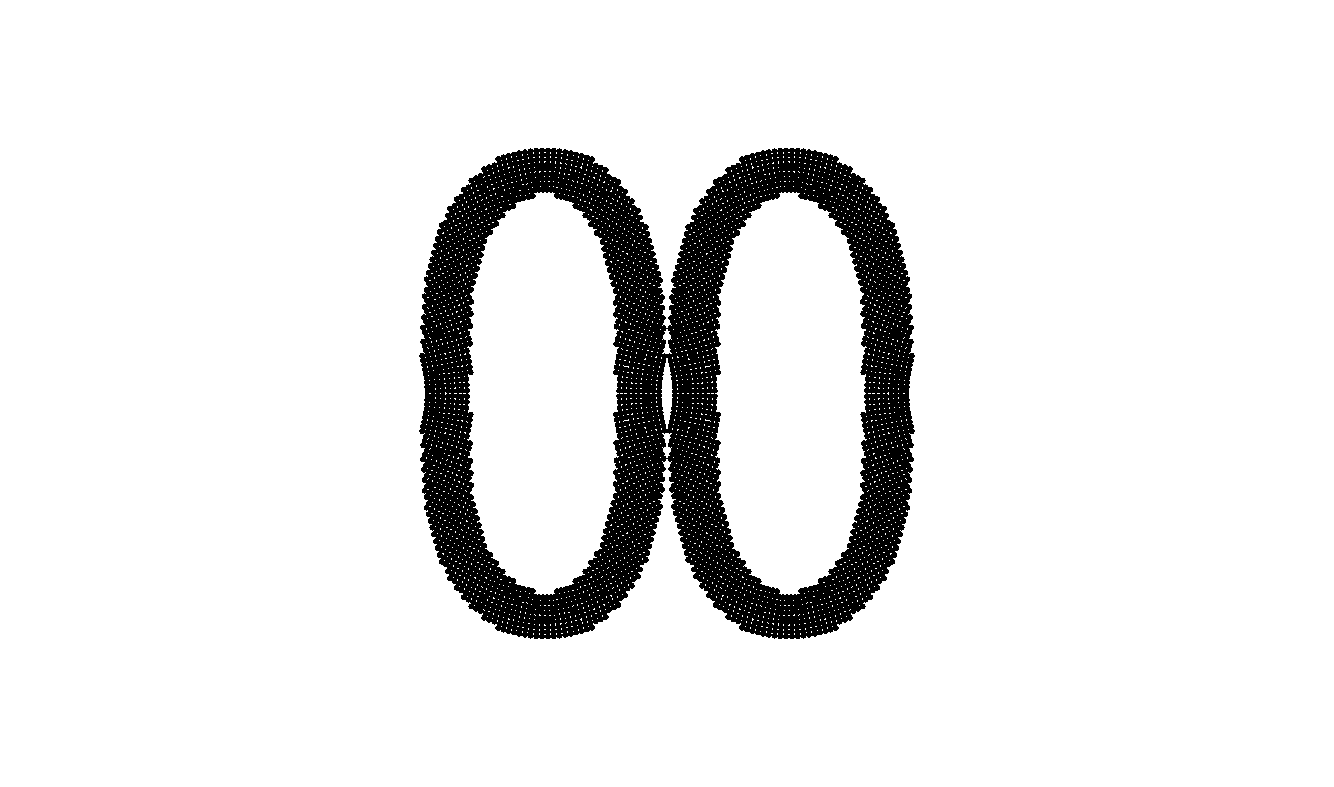} & \includegraphics[width=0.23\textwidth]{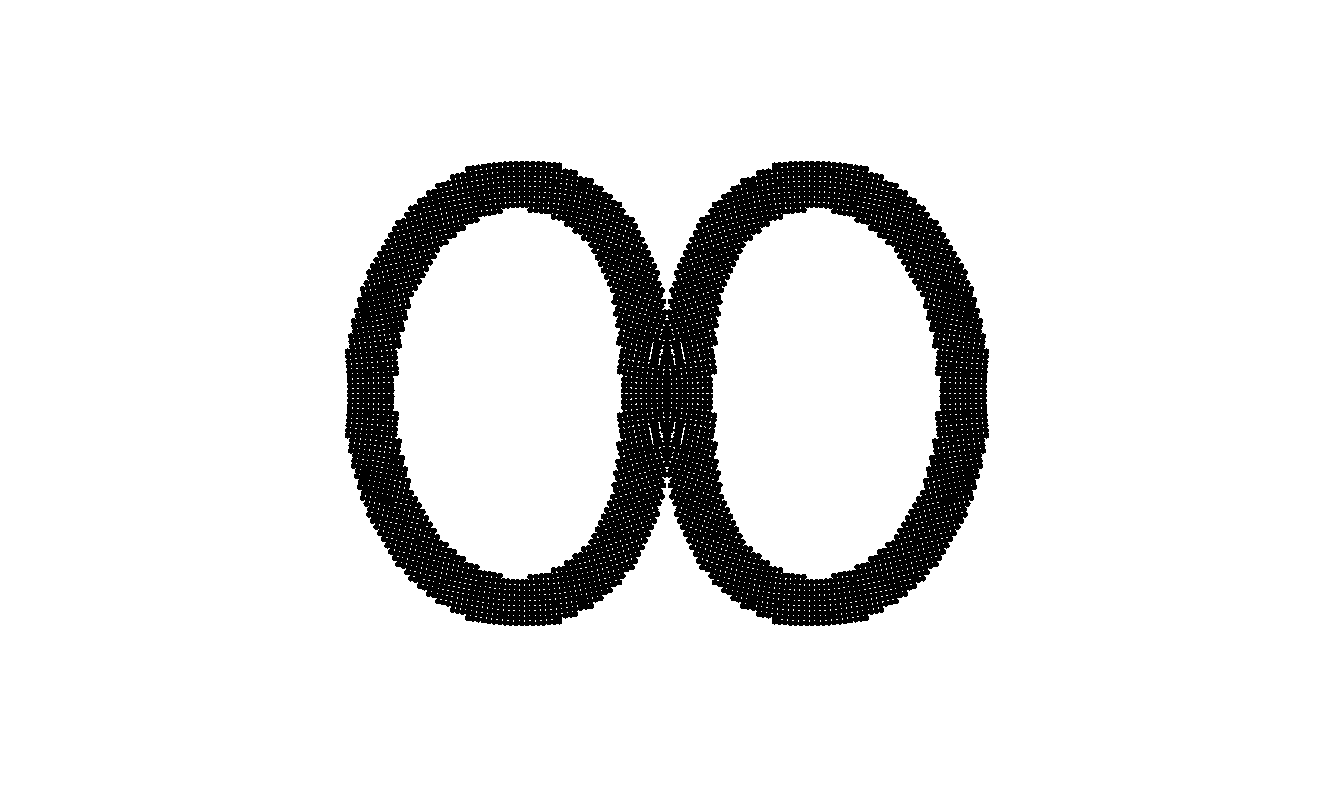} & \includegraphics[width=0.23\textwidth]{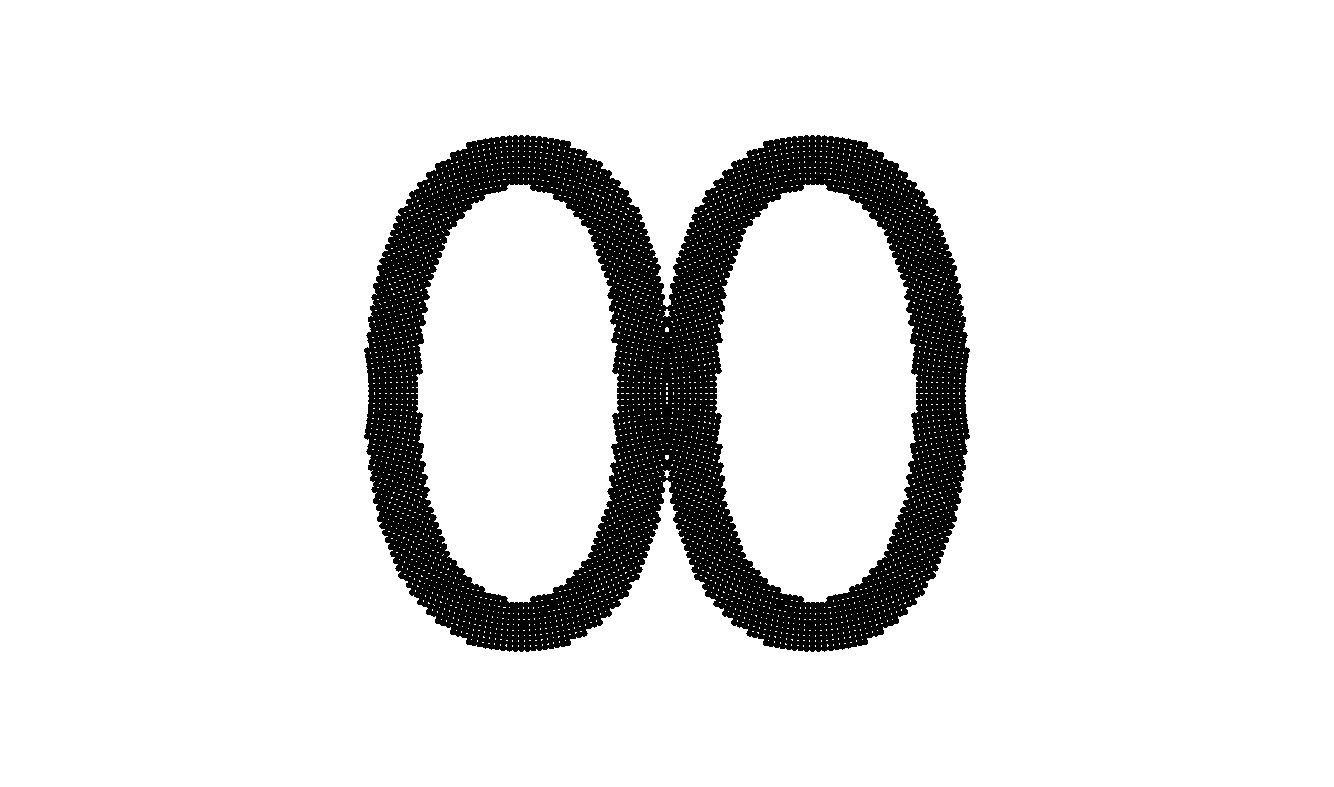} \tabularnewline
Reveles  & Co-rotated & UL weak & TL weak \tabularnewline
\end{tabular}

\caption{Particle distribution at moment of maximum horizontal compression for each case}
\label{fig:rings_compressed}
\end{figure}

Figure \ref{fig:rings_compressed} shows the particle distribution for each method 
at the instant of maximum horizontal compression. The pictures highlight some of
the points mentioned above, like the dissipation of energy in the cases of the 
weak form algorithm in the updated Lagrangian frame and especially the Godunov type
algorithm.

\subsection{Runtimes}

\begin{figure}[htp]
\noindent \centering{}%
\includegraphics[width=\textwidth]{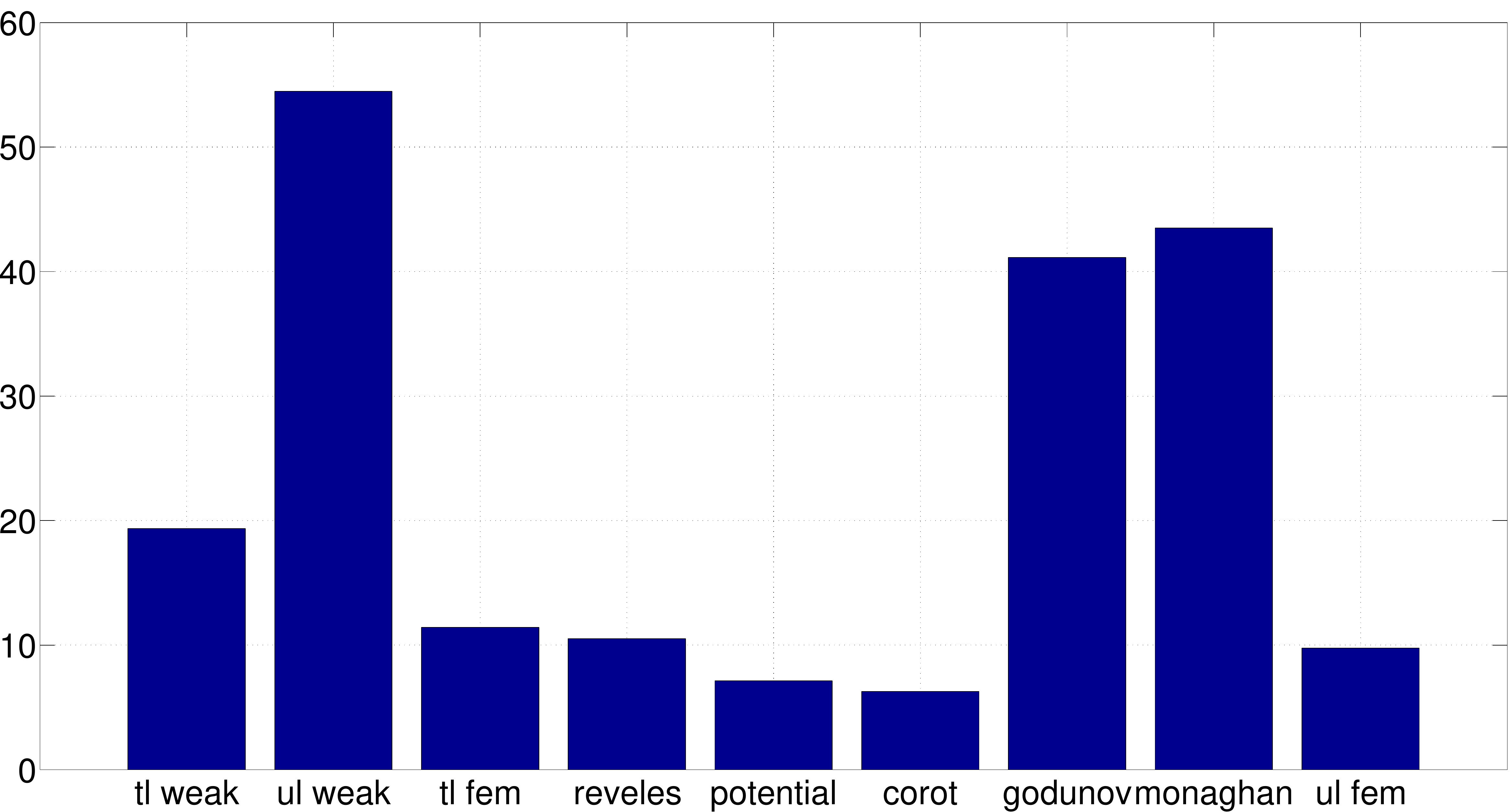}
\caption{Run times of all methods for the first 10'000 time steps for the rotating disk benchmark}
\label{fig:runtimes}
\end{figure}

Even though this paper is not aimed at providing the complete runtime characteristics
of each of the algorithms investigated it is still worthwhile to do a quick comparison.
It has to be noted that none of the algorithms at hand can be called ``optimized''
in any way. The focus of the implementation at hand is at algorithmic clarity and
the runtime of each of the algorithms could be improved greatly by optimizing the
code towards this end. Thus, the following comparison is only valid under the assumption that the runtime would scale equially should the algorithms at hand be tuned for optimal runtime. Admittedly this is hardly realistic. Nonetheless, the runtime of the first 10'000 steps for the
spinning disk benchmark has been recorded for each algorithm and is presented in 
graph \ref{fig:runtimes} to give a rough idea of their runtime characteristics.

The meshless updated Lagrangian methods are the most expensive by a long shot. 
This makes sense since they need to recompute the kernel functions in each time
step. This also entails updating cell lists and turning them into Verlet lists
in the implementation at hand. Next up is the weak form algorithm in the total
Lagrangian frame. This too makes sense since it features the most discretization points.
Beside the particles there is roughly a fourfold of Gauss points to consider.
Not surprisingly the algorithms leveraging ideas from the computer graphics community,
i.e. the potential based and co-rotated one perform fastest. 

\section{Conclusions and Outlook}

As evident from the previous section there is no algorithm that outperforms all
the others. There is no ``silver bullet'' in terms of a meshless
algorithm for large deformation elastodynamics, but rather one has to chose a suitable
algorithm for the application at hand. A quick recap about each algorithms strength
and weaknesses follows:

\begin{itemize}
\item The strong form updated Lagrangian algorithms (Monaghan, Godunov) are numerically quite expensive.
They are able to resolve dynamic impact problems, albeit in the case of the Godunov 
type scheme with excess numerical dissipation. They fail at rotations or tensile tests. See also the discussion in \ref{sec:app0}.

\item Regarding the total Lagrangian framework on the strong form a more mixed picture
is painted. All algorithms are numerically efficient since they do not need to recompute
the kernel functions in each time step nor do they need to consider quadrature points. 
Yet, the computer graphics inspired algorithms have a clear
edge over Reveles' algorithm regarding the runtime. Reveles' algorithm was shown to be very close to FEM
solutions for tension, while the computer graphics algorithms converged to another solution.
The potential based approach was not able to resolve a dynamic impact problem while
Reveles' algorithm needed somewhat elevated levels of artificial viscosity to complete
the benchmark. The co-rotated approach overestimated the stiffness of the material
when compared to other methods in the impact case. Finally, in the rotating disk benchmark
the computer graphics algorithms showed no loss of angular momentum while Reveles' 
algorithm diverged.

\item The updated Lagrangian weak form algorithm conserved angular momentum, converged to the
same solution as the computer graphics algorithms in tension, i.e. disagreed with 
the reference FEM solution and was numerically dissipative in the impact case, though
not as heavily as the Godunov solution. It featured the highest runtime measured. 

\item The total Lagrangian weak form algorithm conserved angular momentum, was consistent
with FEM in tension but was shown to exhibit some amplitude decay in the impact problem.
It's runtime is an interesting compromise between the strong form total Lagrangian 
algorithm and the updated Lagrangian methods.
\end{itemize}

Future work is aimed at testing some of these methods in physically more involved
problems involving plasticity, damage, thermal effects and frictional contact. 

\section{Acknowledgements}

The authors would hereby like to thank the Swiss National Science Foundation for the financial support under Grant No. 200021-149436.

\section{Appendix} \label{sec:app}

\subsection{Discussion of Tensile Benchmark for Monaghan, Godunov Algorithms} \label{sec:app0}

The Godunov algorithm immediately fails at the boundaries for the benchmark discussed
in section \ref{sec:tens}. Since a tensile benchmark has not been published yet for this
algorithm, it is not completely clear whether this is expected behaviour or some problem
specific to the implementation at hand. In fact, no solid mechanics benchmark requiring
the application of Dirichlet type boundaries employing Godunov type SPH has been published
so far. The authors speculate that a new scheme consistent with boundary conditions in 
a Riemann problem would need to be devised. 

\begin{figure}[htp]
\noindent \centering{}%
\includegraphics[width=\textwidth]{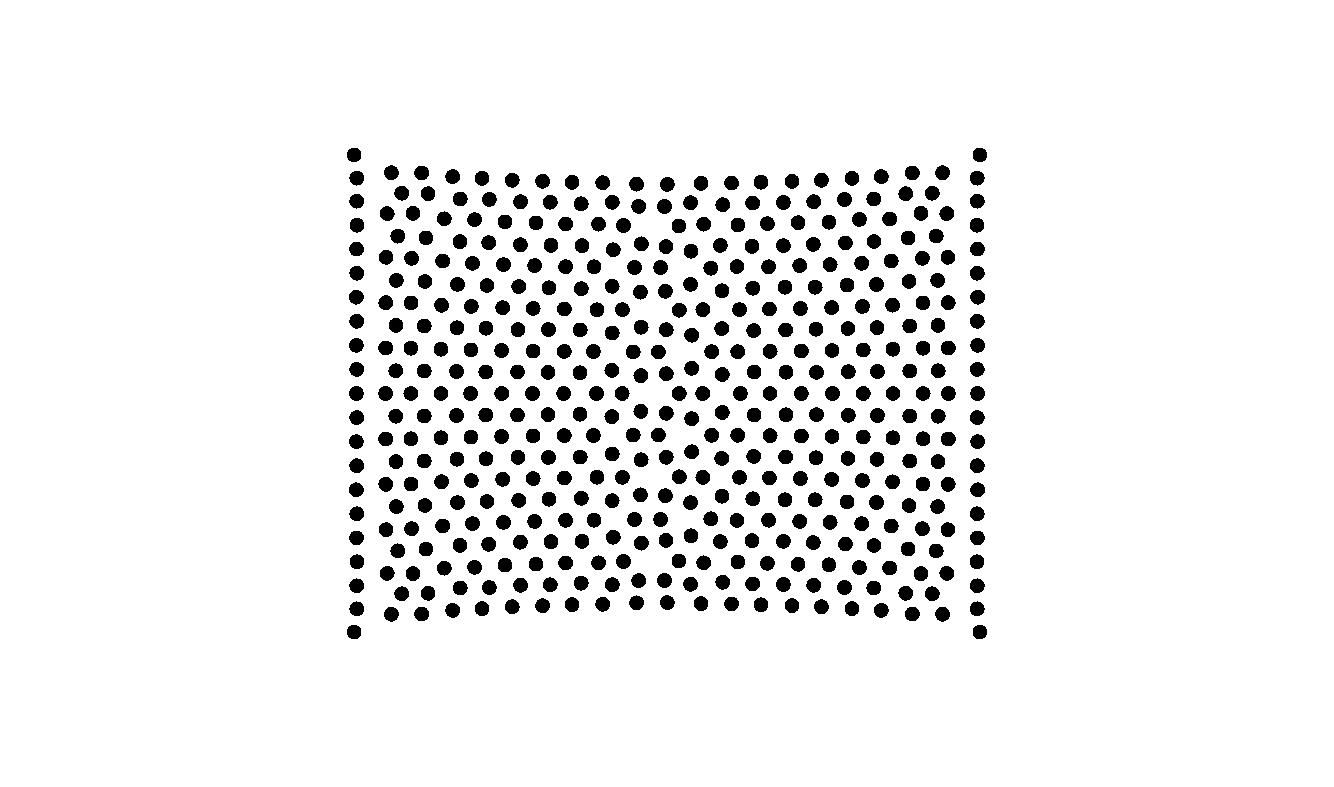}
\caption{Oscillatory failure mode in elastic Monaghan tensile benchmark.}
\label{fig:elast_fail}
\end{figure}

\begin{figure}[htp]
\noindent \centering{}%
\begin{tabular}{c}
\includegraphics[width=0.5\textwidth]{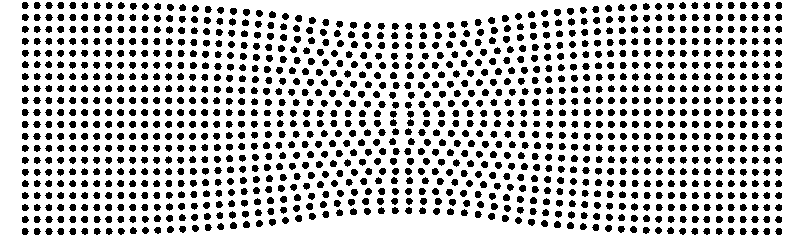} \tabularnewline
\includegraphics[width=0.5\textwidth]{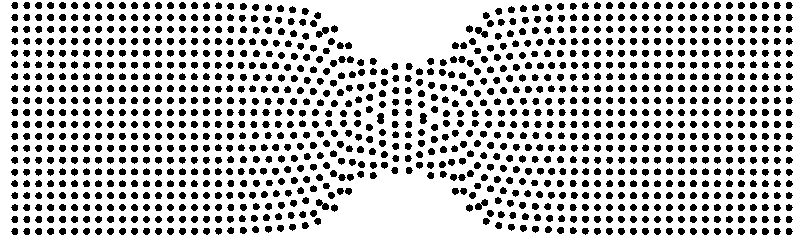} \tabularnewline
\includegraphics[width=0.5\textwidth]{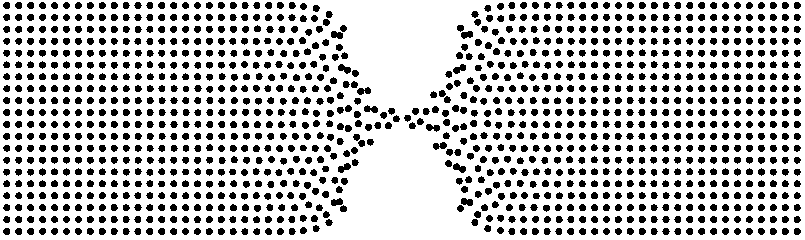} \tabularnewline
\end{tabular}
\caption{Reconstruction of necking and notching given in \cite{gray2001sph}.}
\label{fig:plast_fail}
\end{figure}

The situation is less drastic when considering the Monaghan algorithm. If the Monaghan
algorithm is applied to the tensile test described it fails due to an oscillatory 
displacement mode as can be seen in figure \ref{fig:elast_fail}. The authors in the original paper
do not perform an elastic tensile test but introduce a very simple plasticity model
to their algorithm. This model has been reimplemented to ensure the validity of 
the implementation. The material data is quite obfuscated in the original paper,
employing exotic units like cm/c0, where $c0$ is the reference wave speed for time. For simplicity, Ti6Al4v with material parameters according to \cite{johnson1983constitutive} was assumed
while retaining geometrical proportions \footnote{Note that material parameters only were taken from the paper referenced. The material behaviour is NOT consistent with the Johnson-Cook model but the simple perfect plasticity model from \cite{gray2001sph} was kept.}. Indeed, the necking
and notching presented in the paper are reconstructed closely, see figure \ref{fig:plast_fail}.

\subsection{Proof Regarding Elastic Potential Force} \label{sec:app1}

The intermediate goal is to find an expression for the term $\nabla_{\uline{u}_i}\uuline{\varepsilon}$. In this chapter, the symbol for the displacement field shall be $\uline{\mathcal{U}}$. Albeit slightly unconventional, this is to clearly distinguish between the displacement and it's first component $u$. The derivatives of the first component of the displacement field $\uline{\mathcal{U}}= [u, v, w]$ are given as:
\begin{align}
\nabla u = \begin{bmatrix} u,_{x} \\ u,_{y} \\ u,_{z} \end{bmatrix} \approx \sum_j \nabla \overset{\circ}{W_{ij}}u_j
\end{align} 
where $\overset{\circ}{W_{ij}}$ is the RKPM Kernel function. Note that the implementation
in this paper uses in fact Randles-Libersky corrected Kernel functions. However,
the proof is only slightly more complicated if Randles-Libersky Kernel functions
would be used. \par
For any given particle $i$ we can now take the derivative of that expression with 
respect to the displacement $\uline{\mathcal{U}}_j = (u_j, v_j, w_j)$ at some point 
$\uline{x}_j \neq \uline{x}_i$. 
Note that this is the derivative with respect to the \emph{symbol} $u_j$ 
(and $v_j$, $w_j$, respectively) in the already discretized equation. Also, 
since we look at a single point $j$ the summation over $j$ is dropped (from notation only):
\begin{align}
\frac{\partial}{\partial u_j} \nabla u &= \nabla \overset{\circ}{W_{ij}}  = \uline{d}_j \\
\frac{\partial}{\partial v_j} \nabla u &= \uline{0} \\
\frac{\partial}{\partial w_j} \nabla u &= \uline{0} \\
\end{align}
Analogous results are found for $\nabla v$, $\nabla w$. The three components of 
$\uline{d}_j$ are designated as $d_j^x$, $d_j^y$ and $d_j^z$. With this procedure, 
expressions for all combinations of $\frac{\partial}{\partial \alpha}\beta,_{\gamma}$, 
with $\alpha \in \left\{u,v,w\right\}$, $\beta \in \left\{u,v,w\right\}$ and $\gamma \in \left\{x,y,z\right\}$ 
are obtained. It is now possible to find $\nabla_{\uline{u}_i}\uuline{\varepsilon}$. 
The six independent components of $\uline{\varepsilon}$ are readily expanded as:
\begin{align}
\begin{bmatrix}
   \varepsilon_{xx} \\
   \varepsilon_{yy} \\
   \varepsilon_{zz} \\
   \varepsilon_{xy} \\
   \varepsilon_{yz} \\
   \varepsilon_{zx} \\            
\end{bmatrix}
= 
\begin{bmatrix}
    2u,_{x} + u,_{x}u,_{x} + v,_{x}v,_{x} + w,_{x}w,_{x}\\
    2y,_{y} + u,_{y}u,_{y} + v,_{y}v,_{y} + w,_{y}w,_{y}\\
    2w,_{z} + u,_{z}u,_{z} + v,_{z}v,_{z} + w,_{z}w,_{z}\\
    u,_{y} + v,_{x} + u,_{x}u,_{y} + v,_{x}v,_{y} + w,_{x}w,_{y}\\
    v,_{z} + w,_{y} + u,_{y}u,_{z} + v,_{y}v,_{z} + w,_{y}w,_{z}\\
    w,_{x} + u,_{z} + u,_{z}u,_{x} + v,_{z}v,_{x} + w,_{z}w,_{x}\\    
\end{bmatrix}    
\end{align}
For the first component $\varepsilon_{xx}$:
\begin{align}
\nabla_{\uline{u}_j}\varepsilon_{xx} = \begin{bmatrix} \underbrace{\frac{\partial}{\partial u_j}\varepsilon_{xx}}_{\rm I}, & \underbrace{\frac{\partial}{\partial v_j}\varepsilon_{xx}}_{\rm II}, & \underbrace{\frac{\partial}{\partial w_j}\varepsilon_{xx}}_{\rm III} \end{bmatrix}
\end{align}
\begin{align}
\rm I &= \frac{\partial}{\partial u_j}2u,_{x} + \frac{\partial}{\partial u_j} u,_{x}u,_{x} + \underbrace{\frac{\partial}{\partial u_j}v,_{x}v,_{x}}_{0} + \underbrace{\frac{\partial}{\partial u_j} w,_{x}w,_{x}}_{0} \\
&= 2d_j^x + 2d_j^x \cdot u,_{x} \\
&= 2d_j^x(1 + u,_{x}) \\
\rm II & =\underbrace{\frac{\partial}{\partial v_j} 2u,_{x}}_{0} + \underbrace{\frac{\partial}{\partial v_j} u,_{x}u,_{x}}_{0} + \frac{\partial}{\partial v_j} v,_{x}v,_{x} + \underbrace{\frac{\partial}{\partial v_j} w,_{x}w,_{x}}_0 \\
&= 2d_j^x\cdot v,_{x} \\
\rm III & = \underbrace{\frac{\partial}{\partial w_j}2u,_{x}}_{0} + \underbrace{\frac{\partial}{\partial w_j}u,_{x}u,_{x}}_{0} + \underbrace{\frac{\partial}{\partial w_j}v,_{x}v,_{x}}_0 + \frac{\partial}{\partial w_j}w,_{x}w,_{x} \\
&= 2d_j^x\cdot w,_{x} \\
\end{align}
Thus:
\begin{align}
\nabla_{\uline{u}_j}\varepsilon_{xx} &= \begin{bmatrix} 2(u,x + 1)d_j^x & 2d_j^x v,x & 2d_j^x v,x \end{bmatrix} \\
&= 2\uuline{F}_x\cdot{d_j}^x \\
\end{align}
With analogous results for the derivatives of $\varepsilon_{yy}$ and $\varepsilon_{zz}$:
\begin{align}
\nabla_{\uline{u}_j}\varepsilon_{yy} &= 2\uuline{F}_y\cdot{d_j}^y \\
\nabla_{\uline{u}_j}\varepsilon_{zz} &= 2\uuline{F}_z\cdot{d_j}^z \\
\end{align}
Now, a mixed component shall be derived:
\begin{align}
\nabla_{\uline{u}_j}\varepsilon_{xy} = \begin{bmatrix} \underbrace{\frac{\partial}{\partial u_j}\varepsilon_{xy}}_{\rm I} & \underbrace{\frac{\partial}{\partial v_j}\varepsilon_{xy}}_{\rm II} & \underbrace{\frac{\partial}{\partial w_j}\varepsilon_{xy}}_{\rm III} \end{bmatrix}
\end{align}
\begin{align}
\rm I &= \frac{\partial}{\partial u_j}u,_{y} + \underbrace{\frac{\partial}{\partial u_j}v,_{x}}_{0} + \frac{\partial}{\partial u_j} (u,_{x}u,_{y}) + \underbrace{\frac{\partial}{\partial u_j}v,_{x}v,_{y}}_{0} + \underbrace{\frac{\partial}{\partial u_j} w,_{x}w,_{y}}_{0} \\
&= \frac{\partial}{\partial u_j}u,_{y} + (\frac{\partial}{\partial u_j} u,_{x})u,_{y} + u,_{x}(\frac{\partial}{\partial u_j} u,_{y}) \\
&= d_j^y + d_j^x u,_y + u,_x d_j^y \\
&= d_j^y\cdot(1+u,_x) + d_j^xu,_y \\
\rm II &= \underbrace{\frac{\partial}{\partial v_j}u,_{y}}_{0} + \frac{\partial}{\partial v_j}v,_{x} + \underbrace{\frac{\partial}{\partial v_j} (u,_{x}u,_{y})}_{0} + \frac{\partial}{\partial v_j}v,_{x}v,_{y} + \underbrace{\frac{\partial}{\partial v_j} w,_{x}w,_{y}}_{0} \\
& = \frac{\partial}{\partial v_j}v,_{x} + \frac{\partial}{\partial v_j}v,_{x}v,_{y} \\
& = \frac{\partial}{\partial v_j}v,_{x} + (\frac{\partial}{\partial v_j}v,_{x})v,_{y} + v,_{x}(\frac{\partial}{\partial v_j}v,_{y}) \\
& = d_j^x + d_j^x v,_y + v,_x d_j^y \\
& = d_j^x\cdot(1+v,_y) + d_j^y v,_x \\
\rm III &= \underbrace{\frac{\partial}{\partial v_j}u,_{y}}_{0} + \underbrace{\frac{\partial}{\partial v_j}v,_{x}}_{0} + \underbrace{\frac{\partial}{\partial v_j} (u,_{x}u,_{y})}_{0} + \frac{\partial}{\partial v_j}v,_{x}v,_{y} + \frac{\partial}{\partial v_j} w,_{x}w,_{y} \\
& = \frac{\partial}{\partial v_j} (w,_{x})w,_{y} + w,_{x}\frac{\partial}{\partial v_j}(w,_{y}) \\
& = d_j^x w,_y + w,_x d_j^y
\end{align}
Thus:
\begin{align}
\nabla_{\uline{u}_j}\varepsilon_{xy} &= \begin{bmatrix} d_j^y\cdot(1+u,_x) + d_j^xu,_y & d_j^x\cdot(1+v,_y) + d_j^y v,_x & d_j^x w,_y + w,_x d_j^y \end{bmatrix} \\
&= d_j^y \uuline{F}_x + d_j^x \uuline{F}_y \\
\end{align}
Where $\uuline{F}_x$ is the first row of deformation gradient $\uuline{F}$. With analogous results for the derivatives of $\varepsilon_{xz}$ and $\varepsilon_{zy}$:
\begin{align}
\nabla_{\uline{u}_j}\varepsilon_{xz} &= \uuline{F}_z\cdot{d_j}^x + \uuline{F}_x\cdot{d_j}^z\\
\nabla_{\uline{u}_j}\varepsilon_{zy} &= \uuline{F}_z\cdot{d_j}^y + \uuline{F}_y\cdot{d_j}^z\\
\end{align}
The derivative of the full displacement matrix $\uuline{\varepsilon}$ can now be written as:
\begin{align}
\nabla_{\uline{u}_j}\uuline{\varepsilon} = &2\uuline{F}_x\cdot{d_j}^x + 2\uuline{F}_y\cdot{d_j}^y + 2\uuline{F}_z\cdot{d_j}^z + \\
&2(d_j^y \uuline{F}_x + d_j^x \uuline{F}_y) + 2(\uuline{F}_z\cdot{d_j}^x + \uuline{F}_x\cdot{d_j}^z) + 2(\uuline{F}_z\cdot{d_j}^y + \uuline{F}_y\cdot{d_j}^z)
\end{align}
After some reordering:
\begin{align}
\nabla_{\uline{u}_j} \uuline{\varepsilon} = 2(&\uuline{F}_x\cdot{d_j}^x + \uuline{F}_x\cdot{d_j}^y + \uuline{F}_x\cdot{d_j}^z \\
&\uuline{F}_y\cdot{d_j}^x + \uuline{F}_y\cdot{d_j}^y + \uuline{F}_y\cdot{d_j}^z \\
&\uuline{F}_z\cdot{d_j}^x + \uuline{F}_z\cdot{d_j}^y + \uuline{F}_z\cdot{d_j}^z)\\
\end{align}
Which can be shortened to:
\begin{align}
\nabla_{\uline{u}_j} \uuline{\varepsilon} = 2\cdot \uuline{F}^T\uline{d}_j
\end{align}
Finally, using the symmetry of $\uuline{\sigma}$:
\begin{align}
\uline{\overline{f}}_j &= -\uline{\sigma}\nabla_{\uline{u}_i}\uuline{\varepsilon} \\
&= -\uuline{\sigma} (2\cdot\uuline{F}^T\uline{d}_j) \\
&= -2\uuline{F} \; \uuline{\sigma} \; \uline{d}_j
\end{align}
Note that this is the force per unit volume. To come up with the force that 
particle $i$ induces on particle $j$ one needs to multiply with the volume of $i$:
\begin{align}
\uline{f}_j = -2v_i\uuline{F} \; \uuline{\sigma} \; \uline{d}_j
\end{align}
It remains to determine $\uline{f}_i$. The equilibrium equation for $\uline{f}_i$ is:
\begin{align}
\uline{f}_i = \uuline{\sigma}\nabla_{\uline{u}_i}\uuline{\varepsilon}
\end{align}
The analysis is thus restarted with:
\begin{align}
\nabla u = \begin{bmatrix} u,_{x} \\ u,_{y} \\ u,_{z} \end{bmatrix} = \sum_j \nabla \overset{\circ}{W_{ij}}u_j
\end{align}
But this time the derivative is taken with respect to $\uline{u}_i$. Using RKPM 
the function value of the particle itself is not present in the formula for 
the gradient. Thus, all derivatives are zero:
\begin{align}
\frac{\partial}{\partial u_j} \nabla u &= \uline{0} \\
\frac{\partial}{\partial v_j} \nabla u &= \uline{0} \\
\frac{\partial}{\partial w_j} \nabla u &= \uline{0} \\
\end{align}
Trivially, $\uline{f}_i = 0$. This is in contrast to using a formulation which 
does in fact include $u_i$ in the approximation procedure for the gradient, as 
for example in \cite{muller2004point}.

\section*{References}

\bibliographystyle{plainnat}
\addcontentsline{toc}{section}{\refname}\bibliography{paper.bib}

\end{document}